\documentclass[fleqn,usenatbib]{mnras}

\usepackage{newtxtext,newtxmath}
\usepackage{natbib}

\usepackage[T1]{fontenc}
\usepackage[fleqn]{nccmath}

\DeclareRobustCommand{\VAN}[3]{#2}
\let\VANthebibliography\thebibliography
\def\thebibliography{\DeclareRobustCommand{\VAN}[3]{##3}\VANthebibliography}

\usepackage{graphicx}	
\usepackage{amsmath}	

\def\aap{AA}
\def\prl{Phys. Rev. Lett.}

\def\apjl{ApJL}
\def\na{NewA}

\def\mnras{MNRAS}
\def\apj{ApJ}
\def\apjs{ApJS}

\def\prd{PhysRevD}
\def\jcap{JCAP}
\def\physrep{Phys. Rept.}

\def\qsub{{{\bf{q}}}}
\def\Data{{{\bf{O}}}}

\def\msub{{\boldsymbol{m}_{\rm{sub}}}}

\title[Lensed arcs and flux ratios]{Turbocharging constraints on dark matter substructure through a synthesis of strong lensing flux ratios and extended lensed arcs}
\author[Gilman et al.]{
	Daniel Gilman$^{1, 2}$\thanks{E-mail: gilmanda@uchicago.edu},
	Simon Birrer$^{3}$,
	Anna Nierenberg$^{4}$,
	Maverick S. H. Oh$^{4}$
	\\ \\
	$^1$ Department of Astronomy $\&$ Astrophysics, University of Chicago, Chicago, IL 60637, USA \\
	$^2$ Brinson Prize Fellow \\
	$^3$ Department of Physics and Astronomy, Stony Brook University, Stony Brook, NY 11794, USA \\
	$^4$ University of California, Merced, 5200 N Lake Road, Merced, CA 95341, USA
}

\date{Accepted XXX. Received YYY; in original form ZZZ}

\pubyear{2024}

\begin{document}
	\label{firstpage}
	\pagerange{\pageref{firstpage}--\pageref{lastpage}}
	\maketitle
	
	\begin{abstract}
		Strong gravitational lensing provides a purely gravitational means to infer properties of dark matter halos and thereby constrain the particle nature of dark matter. Strong lenses sometimes appear as four lensed images of a background quasar accompanied by spatially-resolved emission from the quasar host galaxy encircling the main deflector (lensed arcs). We present methodology to simultaneously reconstruct lensed arcs and relative image magnifications (flux ratios) in the presence of full populations of subhalos and line-of-sight halos. To this end, we develop a new approach for multi-plane ray tracing that accelerates lens mass and source light reconstruction by factors of $\sim 100-1000$. Using simulated data, we show that simultaneous reconstruction of lensed arcs and flux ratios isolates small-scale perturbations to flux ratios by dark matter substructure from uncertainties associated with the main deflector mass profile on larger angular scales. Relative to analyses that use only image positions and flux ratios to constrain the lens model, incorporating arcs strengthens likelihood ratios penalizing warm dark matter (WDM) with a suppression scale $m_{\rm{hm}} / M_{\odot}$ in the range $\left[10^7 - 10^{7.5}\right]$, $\left[10^{7.5} - 10^{8}\right]$, $\left[10^8 - 10^{8.5}\right]$, $\left[10^{8.5} - 10^{9}\right]$ by factors of $1.3$, $2.5$, $5.6$, and $13.1$, respectively, for a cold dark matter (CDM) ground truth. The $95\%$ exclusion limit improves by 0.5 dex in $\log_{10} m_{\rm{hm}}$. The enhanced sensitivity to low-mass halos enabled by these methods pushes the observational frontier of substructure lensing to the threshold of galaxy formation, enabling stringent tests of any theory that alters the properties of dark matter halos. 
	\end{abstract}
	
	
	
	
	\section{Introduction}
	The abundance and internal structure of dark matter halos depends on the particle nature of dark matter \citep{Buckley++18}. As such, characterizing the properties of dark matter substructure, the low-mass ($< 10^{10} M_{\odot}$) halos that surround galaxies and permeate the cosmos, enables tests of fundamental dark matter physics. On sub-galactic scales, differences between the concordance cosmological model of cold dark matter (CDM) diverge from the predictions of other theories. For example, if the dark matter has a sufficiently large ($\gtrsim 1 \rm{kpc}$) free-streaming length, the abundance and central density of dark matter halos become suppressed on scales comparable to that of a low-mass galaxy, relative to CDM \citep{Bond++83,Bode++01,Bose++16,Lovell++20,Stucker++22}. In this class of theory, categorically referred to as warm dark matter (WDM), the free-streaming scale depends on the formation mechanism and mass of the dark matter particle \citep[e.g.][]{Schneider++12,AbazajianKusenko19}. Alternatively, theories with self-interacting dark matter (SIDM) posit that the dark matter behave as CDM on cosmological scales, but experiences self-interactions inside high-density regions, such as dark matter halos \citep{Spergel++00,Tulin++13}. In SIDM, halos initially form a central core, and eventually undergo a process referred to as core collapse that causes an order-of-magnitude increase in their central density of halos \citep{Balberg++02,Gilman++21,Gilman++23,Nadler++23,Yang++23}.
	
	Characterizing the properties of substructure on small scales, below $10^9 M_{\odot}$, would have profound consequences for our understanding of dark matter and cosmology. In the standard picture of cosmological structure formation, dark matter halos emerge from the collapse of primordial density fluctuations. Halo mass scales below $10^9 M_{\odot}$  correspond to wave numbers $k > 10 \ \rm{Mpc^{-1}}$, a relatively unconstrained region of the primordial matter power spectrum that could hide clues related to inflation, the early Universe and dark matter \citep{Zentner++03,Bringmann++12,VanTillburg++18,Ando++22,Gilman++22,Esteban++23}. Moreover, the predictions from dark matter theories such as WDM and SIDM diverge more strongly from CDM predictions as one moves to progressively smaller scales and lower halo masses. 
	
	In search of new physics, observational probes of dark matter structure from dwarf galaxies \citep{Kim++18,Correa++21,Nadler++21,Bechtol++22,Dekker++22,AkitaAndo23,Slone++23,Nadler++24}, stellar streams \citep{Bovy++17,Banik++18,Bonaca++19,Banik++21}, and gravitational lensing \citep{DalalKochanek02, Vegetti++14,Nierenberg++14,Hezaveh++16,Birrer++17,Nierenberg++17,Vegetti++18, Gilman++20, Gilman++21, He++22,Sengul+22,Wagner-Carena++23,Dike++23,Powell++23,Dhanasingham++23,Keeley++23, Gilman++23,Mondino++23,Nightingale++24} characterize the properties of substructure on sub-galactic scales. If the dark matter has no coupling to the standard model besides gravity, these cosmic probes of dark matter constitute the only experiments with which to investigate its properties \citep[see the reviews by][]{DrlicaWagner++22,Bechtol++22}. 
	
	Among the various observational probes that constrain dark matter properties through studies of low-mass halos, strong gravitational lensing provides the unique capability to characterize the properties of substructure across cosmological distances, and across several Gyr of cosmic time (see the recent review by \citet{Vegetti++23}). Strong lensing refers to a phenomenon in which multiple highly-magnified and distorted images of a background source appear due to deflection of light around an intervening cosmic structure, such as a galaxy. As lensing depends only on gravity, it circumvents systematic uncertainties associated with using luminous matter 
	as a tracer for the underlying dark matter, and can characterize both the abundance and internal structure of dark matter halos \citep{Minor++21,Amorisco++22,Gilman++22,Ballard++23}. Through the direct, purely gravitational detection of dark halos, strong lensing can extend the reach of cosmic probes of dark matter structure to scales below the threshold of galaxy formation $\sim 10^7 M_{\odot}$. 
	
	Performing such a measurement requires exquisite data. Over the last decade, radio interferometry \citep{Koopmans++04,McKean++07,Spingola++18}, the Hubble Space Telescope (HST) \citep{Shajib++19,Nierenberg++17,Nierenberg++20}, the W.M. Keck Observatory \citep{Nierenberg++14}, and the James Webb Space Telescope (JWST) \citep{Nierenberg++23} have observed a particular class of strong lens system in which a quasar becomes quadruply-imaged. These systems are ideally suited to probe low-mass dark matter structure because the relative magnifications among lensed images (flux ratios) experience strong perturbation by low-mass halos. The minimum mass sensitivity of the flux ratios is determined by the size of the lensed background source \citep{DoblerKeeton06}, and recent JWST observations of the ``warm dust region'' around the background quasar are expected to provide sensitivity to halos at $\lesssim 10^7 M_{\odot}$. 
	
\begin{figure}
		\centering
		\includegraphics[trim=1.5cm 6cm 1cm
		6cm,width=0.48\textwidth]{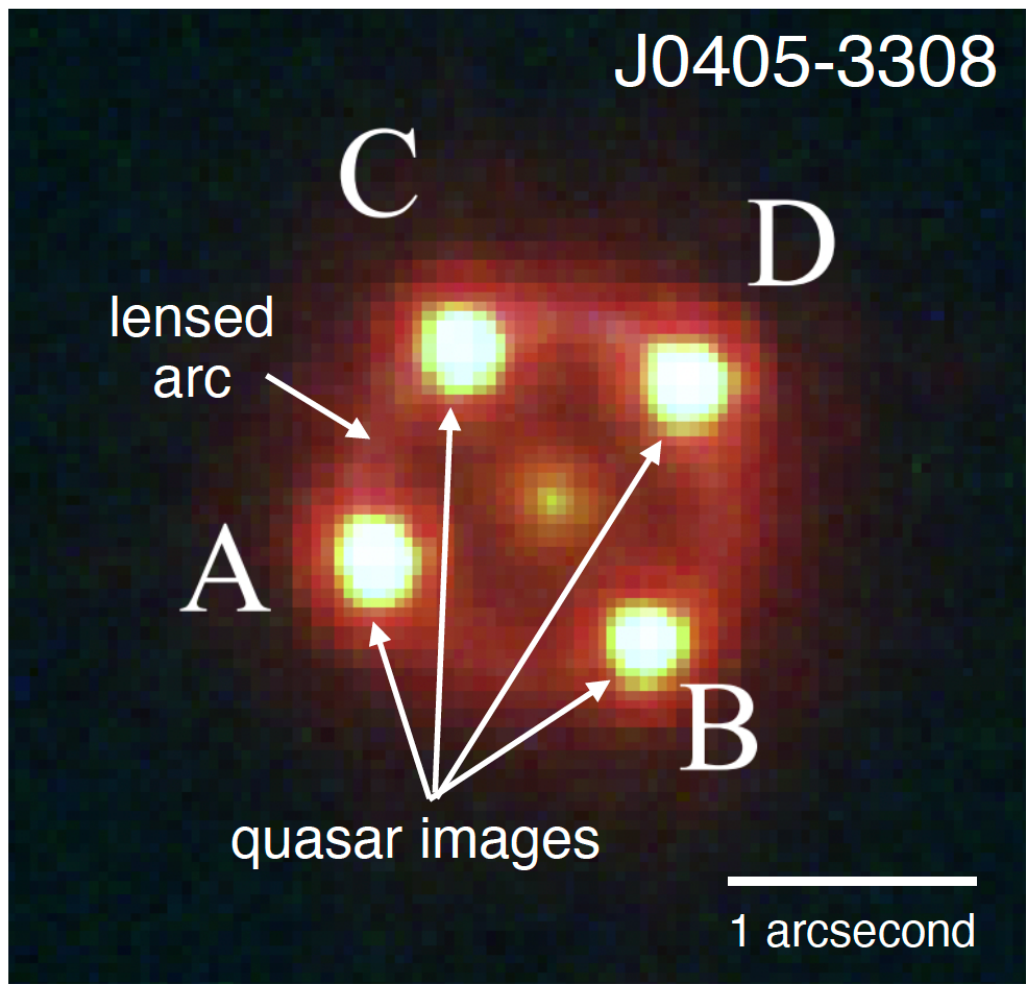}
		\caption{\label{fig:wgdj0405} An HST image of J0405-3308, the type of strong lens system we consider in this work. Four images of a background quasar (labeled A, B, C, D) appear alongside a lensed arc that encircles the main deflector. The methodology presented in this paper develops the formalism to self-consistently reconstruct the lensed images, the relative magnifications, and the lensed arc in the presence of dark matter subhalos and line-of-sight halos. The Figure is adapted from \citet{Shajib++19}.}
	\end{figure}

	As shown in Figure \ref{fig:wgdj0405}, a quadruply-imaged quasar can sometimes appear alongside spatially-resolved lensed emission from the quasar host galaxy, or \textit{lensed arcs}. While the flux ratios provide sensitive localized constraints on the lens model, the lensed arcs that encircle the main deflector impose stringent constraints on the mass profile across larger angular scales \citep[e.g.][]{Shajib++20,Powell++22}. Incorporating constraints from the arcs leads to tighter constraints on the main deflector mass profile, improving the precision of model-predicted flux ratios \citep{Oh++24}. However, due mainly to computational limitations, no existing methodology enables the self-consistent reconstruction of lensed arcs and quasar flux ratios in the presence of potentially tens-of-thousands of dark matter halos. As a result, substructure lensing analyses performed with quadruple-image lenses use only the image positions and flux ratios to constrain the lens model and the properties of dark matter substructure. 
	
	To make best use of existing and future flux ratio measurements, in this paper we introduce a new lens modeling methodology to characterize the properties of substructure in quadruply-imaged quasars with extended lensed arcs. To reduce the computational costs associated with this analysis, we introduce a new approximation for multi-plane ray tracing that accelerates lens mass and source light reconstruction by factors of 100-1000, depending on the number of halos in the lens model. We demonstrate the accuracy of this methodology by performing end-to-end Bayesian inference on simulated datasets. As we will show, the joint reconstruction of lensed arcs and flux ratios leverages complementary information from angular scales that span from the typical Einstein radius ($\sim 1$ arcsecond) down to the milli-arcsecond scales probed by flux ratios. The methods we develop enable more robust constraints on the properties of low-mass dark matter halos and the nature of dark matter. 
	
	This paper is organized as follows: In Section \ref{sec:lensmodeling}, we describe the Bayesian inference problem of inferring substructure properties from strong lens modeling. In Section \ref{sec:multiplanelensmodeling}, we detail the computational challenge that has precluded the joint modeling of lensed quasar flux ratios and lensed arcs, and introduce a methodology for multi-plane ray tracing that alleviates this computational burden. Section \ref{sec:simulationsetup} discusses how we create simulated datasets to evaluate the performance of the lens modeling techniques presented in Section \ref{sec:multiplanelensmodeling} in the context of constraining warm dark matter. Section \ref{sec:results} presents the results of applying the methodology to the simulated datasets, and quantifies the improvement afforded by reconstructing lensed arcs simultaneously with flux ratios relative to analyses that use only the image positions and flux ratios to constrain substructure properties. We summarize our finding and give concluding remarks in Section \ref{sec:conclusions}. 
	
	We perform lensing calculations using the open-source software package {\tt{lenstronomy}}\footnote{https://github.com/lenstronomy/lenstronomy} \citep{BirrerAmara18,Birrer++21}. We generate the populations of dark matter substructure using the open-source software {\tt{pyHalo}}\footnote{https://github.com/dangilman/pyHalo} \citep{Gilman++20}. To perform the forward modeling calculations that interface between {\tt{lenstronomy}} and {\tt{pyHalo}}, we use the new  open-source software {\tt{samana}}\footnote{https://github.com/dangilman/samana}. We assume a flat cosmology with $\Omega_{\rm{m}} = 0.32$, $\sigma_8 = 0.81$ and $H_0 = 67.4 \ \rm{km} \ \rm{s^{-1}} \ \rm{Mpc^{-1}}$ \citep{PlanckCosmo}. 
	
	\section{Dark matter substructure inference}
	\label{sec:lensmodeling}
	We begin in Section \ref{ssec:bayesianinf} by phrasing our objective as a Bayesian inference problem in which we use image positions, flux ratios, and imaging data to constrain a set of hyper-parameters that determine the properties of dark matter substructure. In Section \ref{ssec:inferencesummaries}, we discuss an Approximate Bayesian Computing method to compute the joint likelihood function of lensed image positions, flux ratios, and the imaging data. 
	\begin{figure*}
		\centering
		\includegraphics[trim=1cm 0cm 5cm
		5cm,width=0.9\textwidth]{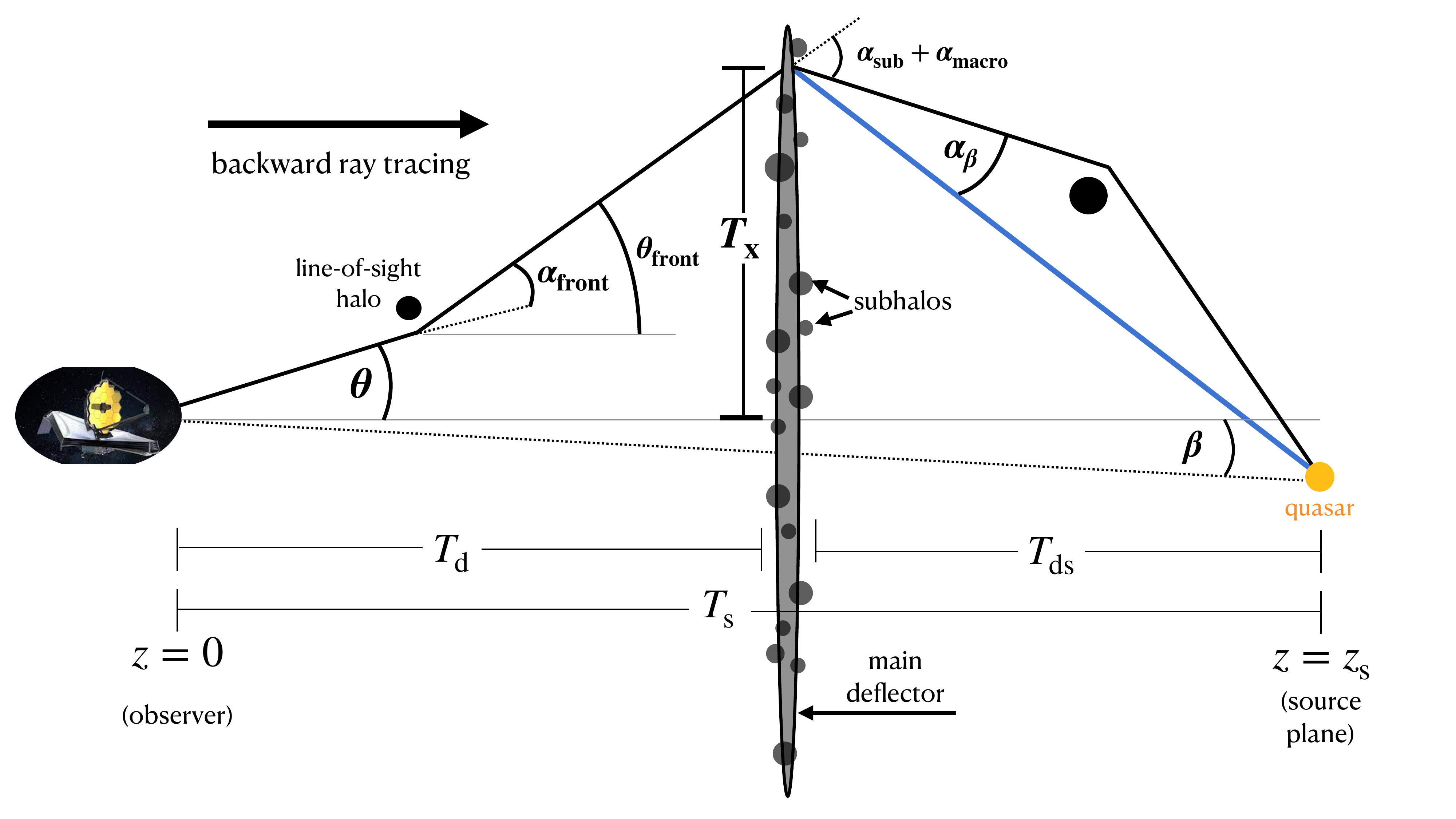}
		\caption{\label{fig:schematic} An illustration of the angles and comoving distances that appear in the derivation of Equation \ref{eqn:lenseqndecoupled}. The figure depicts a backwards ray tracing operation beginning from the observer on the left and ending at the source on the right. By performing one ray tracing calculation through the lens volume to the source plane, we calculate $\boldsymbol{\alpha_{\beta}}$, an effective deflection field from halos at $z_d < z < z_s$ and apply this deflection field across the main deflector lens plane. This simplification preserves the essential properties of the multi-plane deflection field that results from Equation \ref{eqn:lenseqn} while permitting lens mass and source reconstruction with the same computational cost as a single-plane calculation.}
	\end{figure*}
	
	\subsection{The Bayesian inference problem}
	\label{ssec:bayesianinf}
	For a quadruply-imaged quasar with lensed arcs from the quasar host galaxy, such as the example shown in Figure \ref{fig:wgdj0405}, the observables include the relative positions of the four lensed quasar images $\boldsymbol{O_{\rm{pos}}}$, the three flux ratios $\boldsymbol{O_{\rm{f}}}$, and the imaging data $\boldsymbol{O_{\rm{img}}}$ associated with the lensed arcs. Given a set of hyper-parameters $\qsub$ that parameterize a dark matter theory, our aim is to compute the posterior probability distribution
	\begin{eqnarray}
		p\left(\boldsymbol{q} | \boldsymbol{O} \right) &\propto& \pi \left(\qsub\right)\prod_{i} \mathcal{L}\left(\Data_i | \qsub \right) 
	\end{eqnarray}
	where $\boldsymbol{O}=\left(\boldsymbol{O_1}, \boldsymbol{O_2}, .... \right)$ represents the data from a sample of lenses, $\pi\left(\qsub\right)$ represents the prior on $\qsub$, and where $\mathcal{L}\left(\Data_i | \qsub \right)$ represents the likelihood function for a single lens with data $\Data_i \equiv \left(\boldsymbol{O_{\rm{pos}}}, \boldsymbol{O_{\rm{img}}}, \boldsymbol{O_{\rm{f}}}\right)$. To connect the hyper-parameters $\qsub$ with $\boldsymbol{O_i}$, we must sample from an enormous volume of parameter space that specifies the masses, positions, density profiles of a typically very large $\mathcal{O}\left(10^3\right)$ number of dark matter subhalos and field halos. Hereafter, we will refer to $\msub$ as a {\textit{realization}} of substructure that we generate from the dark matter model specified by $\qsub$. The likelihood follows from marginalizing over many realizations
		\begin{equation}
			\label{eqn:likelihoodsingle}
			\mathcal{L}\left(\Data_i | \qsub \right) = \frac{1}{w\left(\qsub | \boldsymbol{O}_{i}\right)} \int p\left(\Data_i | \msub,\boldsymbol{\mathcal{N}}\right) p\left(\msub  | \qsub \right) p\left(\boldsymbol{\mathcal{N}}\right) d\boldsymbol{\mathcal{N}} d \msub.
		\end{equation}
		In the preceding equation, we have introduced a vector of nuisance parameters, $\boldsymbol{\mathcal{N}}$, which include quantities that parameterize the main deflector mass profile (hereafter the {\textit{macromodel}}), the spatial extent and structure of the lensed background quasar, and the surface brightness profile of the quasar host galaxy; $p\left(\boldsymbol{\mathcal{N}}\right)$ represents a prior on these parameters. We have included importance weights $w\left(\qsub | \boldsymbol{O_{{i}}}\right)$, a topic we will discuss in Section \ref{ssec:imgdatalike}. 
		
		The first term in the integrand, $p\left(\Data_i | \msub,\boldsymbol{\mathcal{N}}\right)$, is the product of an astrometric term computed with the lensed image positions $ p\left(\boldsymbol{O}_{\rm{pos}} | \msub,\boldsymbol{\mathcal{N}}\right)$, a term that depends on the imaging data, $p\left(\boldsymbol{O}_{\rm{img}} | \msub,\boldsymbol{\mathcal{N}}\right)$, and the flux ratios $ p\left(\boldsymbol{O}_{\rm{fr}} | \msub,\boldsymbol{\mathcal{N}}\right)$. We perform this integral through Monte-Carlo sampling of the parameter space, and define a likelihood through an Approximate Bayesian Computing approach, as discussed in the next section.
		
		We compute observables $\boldsymbol{O_i}$ using the multi-plane lens equation \citep{Blandford++86}
		\begin{equation}
			\label{eqn:lenseqn}
			{\boldsymbol{\theta_{\rm{K}}}}= \boldsymbol{\theta} - \underbrace{ \frac{1}{D_{\rm{s}}} \sum_{n=1}^{K-1} D_{\rm{ns}}{\boldsymbol{\alpha_{\rm{n}}}} \left(D_{\rm{n}} \boldsymbol{\theta_{\rm{n}}}\right)}_{ \boldsymbol{\alpha_{\rm{eff}}}\left(\boldsymbol{\theta},\msub,\boldsymbol{\mathcal{N}}\right)},
		\end{equation}
		which describes backwards ray propagation through lens planes indexed by $n$. For later use, we have also introduced an effective multi-plane deflection angle $\boldsymbol{\alpha_{\rm{eff}}} \left(\boldsymbol{\theta}, \msub, \boldsymbol{\mathcal{N}}\right)$. The quantity $\boldsymbol{\alpha_n}$ represents the deflection field acting in each lens plane, $D_{\rm{i}}$ represents an angular diameter distance to the $i$th lens plane, and $D_{\rm{ij}}$ represents an angular diameter distance between planes $i$ and $j$. The subscripts $s$ and $d$ denote the source plane and the lens plane of the main deflector. The angle $\boldsymbol{\theta}$ represents an angle on the sky as seen by an observer. 
		
		\subsection{Efficient sampling methods and summary statistics}
		\label{ssec:inferencesummaries}
		The high dimension of the data vector that includes image positions, flux ratios, and imaging data, $\boldsymbol{O_i}$, poses challenges related to exploring the vast parameter space of possible lens model configurations, most of which do not fit all of the available data. Current analysis methods deal with this issue by reducing the dimension of the data vector before computing the likelihood function. For example, in an analysis of eight quadruply-imaged quasars that used only the image positions and flux ratios, \citet{Gilman++20} perform a non-linear optimization of the macromodel such that each proposed lens model satisfies the lens equation for the observed image positions in the presence of substructure. 
		
		In this work, we use a similar strategy to focus computational resources in regions of parameters space that contribute most significantly to the integral in Equation \ref{eqn:likelihoodsingle}. The task at hand involves the simultaneous reconstruction of the main deflector mass and source light profile in the presence of a fixed population of dark matter subhalos and line-of-sight halos, $\msub$. Using a particle swarm optimization (PSO) implemented in {\tt{lenstronomy}}, we simultaneously reconstruct the imaging data and background source while applying a non-linear solver to a portion of the lens macromodel to satisfy the lens equation for the quasar image positions. We guide the PSO towards viable regions of parameters space by punishing poor fits to the imaging data on a pixel-by-pixel basis in the image plane. In Section \ref{ssec:approximation}, we describe an approximation for full multi-plane ray tracing that we use to accelerate the PSO and any subsequent multi-plane ray tracing calculations in the presence of the full population of halos described by $\msub$. 
		
		The lens models that result from PSO optimizations provide better fits to the imaging data than the overwhelming majority lens models proposed only on the basis of matching the image positions. Once we have constructed these lens models, we proceed to compute the astrometric, flux ratio and imaging data likelihoods, as described in the next three sub-sections. 
		
		\subsubsection{The astrometric likelihood}
		As discussed in the previous section, during the particle swarm optimization we use the approach discussed by \citet{Birrer++15} and apply a non-linear solver to a portion of the lens macromodel such that the lens equation is automatically satisfied for each proposed lens model. To handle observational measurement uncertainties in the lensed quasar image positions, we add astrometric perturbations to the image positions prior to performing the PSO, and later on we will evaluate the flux ratios at these new perturbed coordinates. Thus, at this stage we have accounted for $\mathcal{L}\left(\boldsymbol{O_{\rm{pos}}} | \msub, \boldsymbol{\mathcal{N}}\right)$ by selecting lens models that only reproduce the observed image positions to high precision. 
		
		\subsubsection{The flux ratio likelihood}
		Following existing methods \citep{Gilman++19}, we compute the flux ratio likelihood using an Approximate Bayesian Computing (ABC) rejection algorithm based on a summary statistic $S$ computed with respect to the three observed flux ratios, $\boldsymbol{O_{\rm{f}}}$, and $\boldsymbol{f}\left(\msub, \boldsymbol{\mathcal{N}}\right)$, the model-predicted flux ratios 
		\begin{equation}
			\label{eqn:frstat}
			S\left(\boldsymbol{O_{\rm{f}}, \boldsymbol{\msub}, \mathcal{N}}\right) \equiv \sqrt{\sum_{i=1}^3 \left[{O_{\rm{f(i)}}} - {f_{(\rm{i})}}\left(\msub, \boldsymbol{\mathcal{N}}\right)\right]^2}.
		\end{equation}
		We accept a realization and the corresponding parameters drawn from $\pi\left(\boldsymbol{q}\right)$ if $S <  \epsilon$, where $\epsilon$ represents a tolerance threshold. For a given $\epsilon$, the number of accepted samples between two regions of parameter space approximates the relative likelihood between the two regions of parameter space. ABC rejection algorithms converge to exact (relative) likelihoods as $\epsilon\rightarrow 0$ while the number of samples tends to infinity. 
		
		In our analysis, we generate $\sim 10^6$ realizations per lens, and accept the top $N_{\rm{accept}}=3,000$ samples corresponding to the lowest values of $S$. This results in values of $\epsilon$ that range between 0.01 and 0.1 among the (mock) lenses in our sample. As in previous work, we handle observational measurement uncertainties by adding them post-processing to the model-predicted flux ratios before computing $S$. The convergence tests performed for ABC rejection algorithms applied to image flux ratios by \citet{Gilman++20} motivate our choice of the tolerance threshold on $\epsilon$. Finally, we obtain a continuous approximation of the flux-ratio sampling distribution marginalized over $\boldsymbol{\mathcal{N}}$, $p\left(\boldsymbol{O_{\rm{f}}}| \bf{q}_{\rm{sub}} \right)$, by applying a Gaussian kernel density estimator to the accepted samples drawn from the prior on $\qsub$.
		
		\subsubsection{The imaging data likelihood}
		\label{ssec:imgdatalike}
		We account for the imaging data constraints by applying an ABC rejection cut to a second summary statistic, $L$, which we define as the probability of observing the imaging data $L \equiv p\left(\boldsymbol{O_{\rm{img}}} | \msub, \boldsymbol{\mathcal{N}}\right)$. In principle, we could incorporate information directly through importance sampling with weights equal to $L$, but this causes numerical stability issues when the relative likelihood among the accepted samples fluctuate by large factors of $\sim e^{5}$ due to the high dimension of the data vector. We define the acceptance threshold on $L$ as the value of $L$ that corresponds to the top $2\%$ of the imaging data likelihoods. This choice represents a compromise between selecting models that best fit the imaging data, and accepting realizations that minimize the $S$ statistic computed with respect to the flux ratios. The optimum choice for this threshold should depend on the constraining power of the imaging data, which in turn depends on the specifics of the lens system of in question and in particular the brightness of the lensed arc.
		
		As we show in Section \ref{sec:results}, the imaging data enables stronger conclusions regarding substructure properties by imposing tighter constraints on the mass profile of the main deflector than one obtains from analyzing only the image positions and flux ratios. However, our method of incorporating imaging data exposes the likelihood function to a systematic bias associated with reconstruction of the source light profile with substructure in the lens model. This bias manifests as a systematic preference for lens models with less substructure. Moreover, in rare cases a dark matter halo imparts a strong perturbation to the surface brightness of a lensed arc, a feature of the data that the inference methodology we present in this work is not intended to properly capture in the likelihood. For reasons we elaborate on further in Appendix \ref{app:A}, we conclude that the imaging data likelihood will not give a reliable inference of substructure properties on its own when calculated according to the methodology we present in this work. 
		
		To mitigate the effects of systematics associated with the imaging data in the presence of substructure, we demand that our posterior distribution have the property
		\begin{equation}
			\label{eqn:posteriorcriterion}
			\lim_{\epsilon \rightarrow \infty} p\left(\qsub | \boldsymbol{O_{\rm{img}}, \boldsymbol{O_{\rm{pos}}, \boldsymbol{O_{\rm{f}}}}}\right) = \pi \left(\qsub \right).
		\end{equation}
		That is, the posterior should equal the prior on $\qsub$ when computed \textit{only} using the imaging data (as $\epsilon \rightarrow \infty$ the flux ratio ratio likelihood term becomes uninformative). We can arrange that our posterior meets this requirement by an appropriate definition of the importance sampling weights in Equation \ref{eqn:likelihoodsingle}. We define the importance weights, $w\left(\qsub | \boldsymbol{O_{i}}\right)$, as
		\begin{eqnarray}
			\label{eqn:importanceweights2}
			w \left(\qsub |  \boldsymbol{O_{\rm{img}}}, \boldsymbol{O_{\rm{pos}}} \right) &=& \boldsymbol{\mathcal{L}} \left(\boldsymbol{O_{\rm{img}}}, \boldsymbol{O_{\rm{pos}}} | \qsub\right) \\ 
			\nonumber &=&\int p\left(\boldsymbol{O_{\rm{img}}}, \boldsymbol{O_{\rm{pos}}} | \msub,\boldsymbol{\mathcal{N}}\right) \\
			\nonumber &&\times p\left(\msub  | \qsub \right)  p\left(\boldsymbol{\mathcal{N}}\right) d\boldsymbol{\mathcal{N}} d \msub.
		\end{eqnarray}
		In other words, we divide the likelihood computed for all available data, $\Data_i$, by the imaging data and astrometric likelihood, ignoring flux ratio information. As one can easily verify, applying the importance weights in Equation \ref{eqn:importanceweights2} and performing a dark matter inference without incorporating constraints from the flux ratios yields a posterior distribution equal to $\pi\left(\qsub\right)$. 
		
		One can interpret this term as a prior on the dark matter model $\qsub$ applied to each lens that depends on the parameterization of the source light profile in $\boldsymbol{\mathcal{N}}$. The source reconstruction process involves solving for a configuration of light in the source plane that maximizes the likelihood of the lensed image in the image plane. As discussed in Section \ref{ssec:lightmodels}, we model the source light profile as a parametric component plus a shapelet basis expansion, and we optimize the coefficients of the shapelet sets \citep{Birrer++15} simultaneously with the parametric components. Properties of substructure can, to some degree, be absorbed by the source light profile, and the degree to which this occurs depends on various choices regarding the source light model and the process of reconstructing the extended light in the source plane \citep[e.g.][]{Ballard++23}. The importance weights in Equation \ref{eqn:importanceweights2} reflect our knowledge that choices regarding the source light reconstruction will affect our beliefs regarding the probability of a dark matter model $\boldsymbol{\qsub}$. The importance weights $w\left(\qsub |  \boldsymbol{O_{\rm{img}}}, \boldsymbol{O_{\rm{pos}}} \right)$ quantify this effect, and their incorporation in the likelihood corrects for the resulting bias.
		
		\section{Accelerating multi-plane lens modeling with the ``Decoupled multi-plane'' formalism}
		\label{sec:multiplanelensmodeling}
		The approach outlined in the previous section presents a viable strategy for performing a dark matter inference using image positions, imaging data, and flux ratios to constrain the lens model. As discussed in the first paragraph of Section \ref{ssec:inferencesummaries}, a crucial step in the inference method involves a non-linear optimization of the lens mass and source light profile with respect to the imaging data through a particle swarm optimization, and a simultaneous non-linear solver applied to the lens macromodel such that the lens equation is satisfied for the quasar image positions, for each realization. However, with exact ray tracing methods, performing this calculation with thousands of dark matter halos along the line of sight is computationally intractable. 
		
		In this section, we begin in \ref{ssec:multiplanelensing} by reviewing the methodology of lens mass and source reconstruction using existing lens modeling techniques, and explain how the computational intractability of lens mass and source light reconstruction with line-of-sight halos derives from the recursive nature of Equation \ref{eqn:lenseqn}. In Section \ref{ssec:approximation}, we introduce a new approximation for mutli-plane lensing that preserves the non-linear effects associated with Equation \ref{eqn:lenseqn} while accelerating calculations by factors of 100-1000. In Section \ref{ssec:validity}, we show that this formalism predicts effectively indistinguishable flux ratio likelihoods from full multi-plane ray tracing.  
		
		\subsection{Multi-plane lens modeling}
		\label{ssec:multiplanelensing}
		To understand computational challenges posed by Equation \ref{eqn:lenseqn}, we will walk through the step-by-step procedure of performing a lens mass and source light reconstruction with substructure included along the line of sight and in the main lens plane. Figure \ref{fig:schematic} depicts the backwards ray tracing procedure described by Equation \ref{eqn:lenseqn}, and will serve as a useful guide for understanding the lens modeling procedures discussed below, as well as the methods introduced in the next subsection. 
		
		For what follows, we assume that we have generated a realization of substructure $\msub$ from the dark matter model $\qsub$. The masses, positions, and density profiles of the halos are held fixed for the remainder of the calculation. Our objective is to simultaneously reconstruct the mass distribution of the main deflector and the lensed background source in the presence of the realization specified by $\msub$. The only unknown parameters are those that describe the mass profile of the main deflector, or more generally, the deflection associated with the main deflector $\boldsymbol{\alpha_{\rm{macro}}}$. Our task is determine $\boldsymbol{\alpha_{\rm{macro}}}$ subject to the requirement that these deflection angles satisfy the lens equation for the four image positions while simultaneously fitting the imaging data. A standard lens modeling procedure proceeds as follows: 
		\begin{enumerate}
			\item Using Equation \ref{eqn:lenseqn}, we ray trace from the viewer to the pixel location in the plane of the main deflector an angle $\boldsymbol{\theta}$. We record the direction of a light ray as it intersects the main lens plane, $\boldsymbol{\theta_{\rm{front}}} = \boldsymbol{\theta} + \boldsymbol{\alpha_{\rm{front}}}$, and the comoving transverse distance of a light ray where it hits the main lens plane $\boldsymbol{T_{\rm{x}}}\left(\boldsymbol{\theta},\boldsymbol{\alpha_{\rm{front}}}\right)$. Once the light ray reaches the main lens plane, we compute the deflection angles produced by main deflector subhalos, $\boldsymbol{\alpha_{\rm{sub}}\left(\boldsymbol{T_{\rm{x}}}\right)}$. For a given realization of halos, we only need to perform the calculation of $\boldsymbol{T_{\rm{x}}}$ and $\boldsymbol{\alpha_{\rm{sub}}\left(\boldsymbol{T_{\rm{x}}}\right)}$ once for a given angle $\boldsymbol{\theta}$ because the ray tracing thus far does not depend on the unknown  $\boldsymbol{\alpha_{\rm{macro}}}$. 
			\item Propose a set of deflection angles $\boldsymbol{\alpha_{\rm{macro}}}\left(\boldsymbol{T_{\rm{x}}}\right)$ produced by the main deflector. This deflection field is unknown, and we will try to optimize it subject to the constraints imposed by the data. Methods discussed by \citep{Gilman++19} generate proposals for $\boldsymbol{\alpha_{\rm{macro}}}$ by selecting lens models that satisfy the lens equation. In this work, we use a particle swarm optimization to generate proposals for $\boldsymbol{\alpha_{\rm{macro}}}$ that satisfy the lens equation while also reproducing the lensed arcs. 
			\item Using the proposed $\boldsymbol{\alpha_{\rm{macro}}}$, we ray trace with Equation \ref{eqn:lenseqn} until reaching the source plane. The deflection angles produced by halos at $z > z_{\rm{d}}$ depend on $\boldsymbol{\alpha_{\rm{macro}}}$ due to the recursive nature of Equation \ref{eqn:lenseqn}, so this step involves thousands of function evaluations per pixel in the lensed image.
			\item We evaluate the surface brightness of the source light in the souce plane, and then cast this light back to the image plane to produce a lensed image. We then evaluate the imaging data likelihood. 
			\item Repeat steps $ii-iv$ until obtaining a maximum likelihood estimation (or in some cases, a Markov Chain) for the parameters of interest. During each iteration of Steps $ii-iv$, we must re-evaluate the deflections by halos at $z > z_{\rm{d}}$ because they are coupled to $\boldsymbol{\alpha_{\rm{macro}}}$. 
		\end{enumerate}
		Because step $iii$ involve backwards ray tracing operations through all lens planes between the main deflector and the source plane, the computation time scales in proportion with the number of halos in this region. For a typical configuration with a lens at $z_{\rm{d}} = 0.5$ and a source at $z_{\rm{s}} = 2.0$, CDM predicts $\sim 1750$ halos in the mass range $10^6 - 10^{10} M_{\odot}$ between the main deflector and source\footnote{This number assumes we render halos in a volume shaped like a double cone that opens towards the lens with an opening angle of six arcseconds, and closes at the source position.}. Thus, lens mass and source light reconstruction with line-of-sight halos between the main deflector and the source takes approximately 1750 times as long as a single-plane reconstruction. The likelihood function in Equation \ref{eqn:likelihoodsingle} dictates millions of such calculations per lens. Assuming the particle swarm optimization takes $\sim 1$ minute for a single plane reconstruction, performing the lens modeling for 1,000,000 realizations of substructure would take $\sim 1,750 \times 10^{6} \ \rm{CPU} \ \rm{minutes}$, or $\sim 3,000$ CPU years per lens. In practical terms, to perform this analysis on a sample of ten lenses with a computing allocation of $\sim 10^6 \ \rm{CPU}$ hours, we require an increase in speed by a factor of at least $\sim 250$. 
		
		In summary, the computational difficulties that preclude the joint reconstruction of image flux ratios and lensed arcs in substructure lensing analyses stem from the recursive nature of the multi-plane lens equation. In the next section, we introduce an approximation for full multi-plane ray tracing that circumvents the repeated evaluation of step $iii$ in the procedure described earlier in this section, accelerating calculations by factors of up to $\sim 1,000$. 
		\begin{figure}
			\centering
			\includegraphics[trim=0cm 0cm 0cm
			0cm,width=0.48\textwidth]{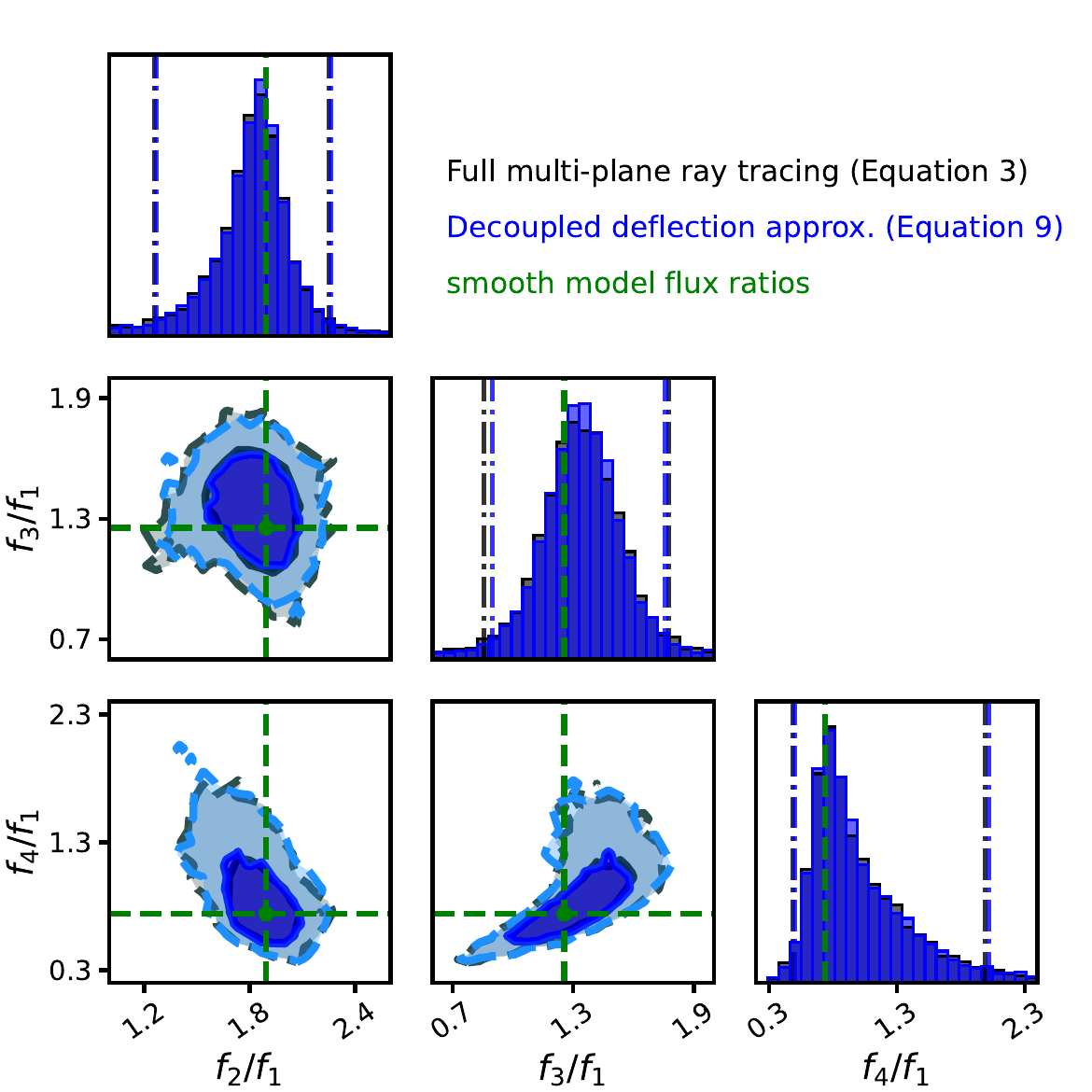}
			\caption{\label{fig:frtest} The joint distribution of flux ratios for a simulated mock lens system generated with exact multi-plane ray tracing (black) and with decoupled multi-plane formalism (blue). The flux ratios of the mock lens system that does not contain substructure are marked in green. The decoupled multi-plane formalism predicts the same distribution of flux ratios as exact ray tracing.}
		\end{figure}
		
		\subsection{The decoupled multi-plane approximation for multi-plane lensing}
		\label{ssec:approximation}
		We begin with a reasonable estimate for the deflection field produced by the main deflector, $\boldsymbol{\hat{\alpha}_{\rm{macro}}}$. As a reasonable estimate for this deflection field, we choose one that satisfies the lens equation for the image positions without substructure included in the lens model. Using Equation \ref{eqn:lenseqn}, we can perform Step $i-iii$ in the lens modeling procedure described in the previous section, using our initial guess for $\boldsymbol{\hat{\alpha}_{\rm{macro}}}$ to ray-trace through the entire lens system to the source plane. Inserting our choice of $\boldsymbol{\hat{\alpha}_{\rm{macro}}}$ into Equation \ref{eqn:lenseqn} for a given realization of substructure, we ray-trace through the lens system and reach an angular coordinate $\boldsymbol{\hat{\beta}}$ given by
		\begin{equation}
			\boldsymbol{\hat{\beta}}\left(\boldsymbol{\theta}, \msub, \boldsymbol{  \hat{\alpha}_{\rm{macro}} }\right) = \boldsymbol{\theta} - \boldsymbol{\alpha_{\rm{eff}}}\left(\boldsymbol{\theta}, \msub, \boldsymbol{  \hat{\alpha}_{\rm{macro}} }\right)
		\end{equation}
		where $\boldsymbol{\alpha_{\rm{eff}}}$ is the effective multi-plane deflection angle defined in Equation \ref{eqn:lenseqn}, and where we have explicitly included our proposal for the macromodel deflections $\boldsymbol{\hat{\alpha}_{\rm{macro}}}$ in place of the nuisance parameters $\boldsymbol{\mathcal{N}}$. 
		
		Using $\boldsymbol{\hat{\beta}}$ we define a deflection field associated with line-of-sight halos behind the main deflector, $\boldsymbol{\alpha_{\rm{\beta}}}$, as illustrated by the blue line in Figure \ref{fig:schematic}. This deflection field is a function of the following quantities: First, $\boldsymbol{T_{\rm{x}}}$ represents the comoving position where a light ray strikes the main lens plane; second, an angle $\boldsymbol{\theta_{\rm{front}}}$, which is the observed angle on the sky $\boldsymbol{\theta}$ plus the cumulative deflections from foreground halos; third, $\boldsymbol{\alpha_{\rm{sub}}}$, the deflection field associated with main deflector subhalos; and fourth, the coordinate on the source plane $\boldsymbol{\hat{\beta}}$ we compute with Equation \ref{eqn:lenseqn} and $\boldsymbol{\hat{\alpha}_{\rm{macro}}}$. Expressing $\boldsymbol{\alpha_{\rm{\beta}}}$ in terms of these quantities gives
		\begin{equation}
			\label{eqn:alphabeta}
			\boldsymbol{\alpha_{\beta}}= \frac{1}{T_{\rm{ds}}}\boldsymbol{T_{\rm{x}}} - \frac{T_{\rm{s}}}{T_{\rm{ds}}} \boldsymbol{\hat{\beta}}  + \boldsymbol{\theta_{\rm{front}}} - \boldsymbol{\alpha_{\rm{sub}}} - \boldsymbol{\hat{\boldsymbol{\alpha}}_{\rm{macro}}}.
		\end{equation} 
		The approximation we make is to apply the deflection field $\boldsymbol{\alpha_{\beta}}$ across the main lens plane for any subsequent ray tracing operation for a given population of dark matter halos. This results in a lens equation for a coordinate on the source plane $\boldsymbol{\beta}$
		\begin{equation}
			\label{eqn:lenseqndecoupled}
			\boldsymbol{\beta}= \boldsymbol{\hat{\beta}} + \frac{T_{\rm{ds}}}{T_{\rm{s}}} \left(\boldsymbol{\hat{\alpha}_{\rm{macro}}} - \boldsymbol{\alpha_{\rm{macro}}}\right).
		\end{equation}
		Note that we must still evaluate the main deflector deflection angles $\boldsymbol{\alpha_{\rm{macro}}}$ at the positions $\boldsymbol{T_{\rm{x}}}$ defined in step $i$ in the previous subsection. This version of the multi-plane lens equation resembles a single plane lens equation that is linear in the macromodel deflections $\boldsymbol{\alpha_{\rm{macro}}}$. 
		
		The deflection field $\boldsymbol{\alpha_{\beta}}$ is computed with Equation \ref{eqn:lenseqn}, so we expect it will capture the small-scale lensing distortions associated with multi-plane ray tracing. However, after performing a lens modeling operation with Equation \ref{eqn:lenseqndecoupled}, we cannot go back to interpret the physical properties (mass, location, density profile, etc.) of individual dark matter halos behind the main deflector. Physically, our assumption implies that the deflection field produced by the background population of substructures depends primarily on the intrinsic dark matter characteristics of that population and other fixed geometrical effects, and that flux ratio statistics we obtain from considering many realizations of substructure do not depend strongly on the coupling to the main deflector deflection field. This formalism shares some similarities with the perturbative formalism presented by \citet{Fleury++21}, but differs in that we compute $\boldsymbol{\alpha_{\beta}}$ with exact ray tracing. 
		
		Equation \ref{eqn:lenseqndecoupled} increases the speed of lens mass and source light reconstruction with line-of-sight halos included in the lens model because it decouples deflections produced by halos behind the main deflector from deflections produced in the main lens plane. Thus, using Equation \ref{eqn:lenseqndecoupled} requires only one full ray tracing calculation to the source plane per coordinate in the image plane, circumventing the repeated evaluation of step $iii$ for each new proposal of $\boldsymbol{\alpha_{\rm{macro}}}$. A lens mass and source light reconstruction with substructure performed according to this procedure takes less than ten minutes, corresponding to an increase in speed by a factor between 100-1000, depending on the number of line-of-sight halos behind the main deflector. We have added this decoupled multi-plane approximation for optional use in {\tt{lenstronomy}} \footnote{This \href{https://github.com/lenstronomy/lenstronomy-tutorials/blob/main/Notebooks/LensModeling/modeling_with_decoupled_multiplane.ipynb}{notebook} illustrates the functionality in combination with {\tt{pyHalo}}.}.
		
		\subsection{Validity of the decoupled multi-plane approximation}
		\label{ssec:validity}
		We can check the validity of the approximation discussed in the previous section for predicting image positions and flux ratio statistics through comparisons with the statistics obtained by exact multi-plane ray tracing. First, using exact ray tracing techniques discussed by \citet{Gilman++19} we compute $p_{{\rm{eqn3}}}\left(\boldsymbol{O_{\rm{pos}}}, \boldsymbol{O_{\rm{f}}} | \boldsymbol{q_{\rm{cdm}}}\right)$, the probability of observing a given set of flux ratios with subhalos and line-of-sight halos included in the lens model with properties as prescribed by CDM (represented by $\boldsymbol{q_{\rm{cdm}}}$). We then repeat this procedure using the ray tracing approximation discussed in the previous section to compute the likelihood $p_{{\rm{eqn9}}}\left(\boldsymbol{O_{\rm{pos}}}, \boldsymbol{O_{\rm{f}}} | \boldsymbol{q_{\rm{cdm}}}\right)$. For details regarding the substructure models we use for these tests, we refer to Section \ref{ssec:darkmattermodels}. We perform these calculations using the image positions of a reference mock lens model created without substructure (hereafter, the ``smooth model''). This reference model has a cross image configuration with lens (source) redshifts $z_{\rm{d}} = 0.5$ ($z_{\rm{source}} = 1.5$), and a mass profile parameterized as an elliptical power law with Einstein radius of $1 \  \rm{arcsec}$, an axis ratio of $0.75$, a logarithmic mass profile slope of $-2.0$, and external shear strength of $0.09$. We have also performed a similar test for a fold image configuration. 
		
		Figure \ref{fig:frtest} shows the joint likelihoods for the three flux ratios of the mock lens, with the flux ratios of the reference smooth lens model marked with the green points. The likelihood computed through exact ray tracing is shown in black, while the likelihood computed with the approximation introduced in the previous sub-section is represented in blue. The likelihoods peak at the smooth model flux ratio values, which is simply a statement that substructure introduces perturbations around the flux ratios predicted by the smooth lens model. A dark matter inference with flux ratios requires the characterization of this multi-dimensional probability density. Dark matter substructure deforms the structure of this probability distribution in a distinct way from other sources of small-scale perturbation, such as multipole terms added to the main deflector mass profile.
		
		The requirement of the approximation we introduce is that it predict identical flux ratio statistics for a given dark matter model. Kolmogorov-Smirnov tests performed on the marginal distributions of the flux ratios shown in Figure \ref{fig:frtest} return p-values greater than 0.99, indicating that the distributions of model-predicted flux ratios are statistically indistinguishable. We have also tested the approximation for an identical set of 10,000 realizations, as opposed to 10,000 unique realizations generated from $\qsub$. We  compute the model-predicted flux ratios with and without the decoupled multiplane approximation, and initialize the solver applied to the macromodel parameters from the same random seed. Predicting the same flux ratios for an identical realization of halos is not formally a requirement that the method needs to satisfy to predict consistent flux ratio statistics, but nonetheless we find that for identical realizations, 95$\%$ of the samples obtained from exact ray tracing and the approximation we present differ by less than $5\%$. We can conclude that the approximation presented in the previous section does not introduce a detectable bias in the model-predicted image flux ratios for a CDM dark matter model while decreasing the time per flux ratio calculation by a factor of   $5-10$.
		
		Unlike the flux ratios and image positions, it is computationally intractable to compare our approximation with exact ray tracing when reconstructing imaging data for the reasons discussed in Section \ref{ssec:multiplanelensing}. Instead, we perform tests with simulated datasets to verify that we obtain unbiased inferences of dark matter substructure properties when using this approximation to reconstruct imaging data with substructure included in the lens model. We describe the details of these simulations in the following section.  
		
		\begin{figure}
			\centering
			\includegraphics[trim=0.5cm 1.5cm 0.5cm
			2cm,width=0.45\textwidth]{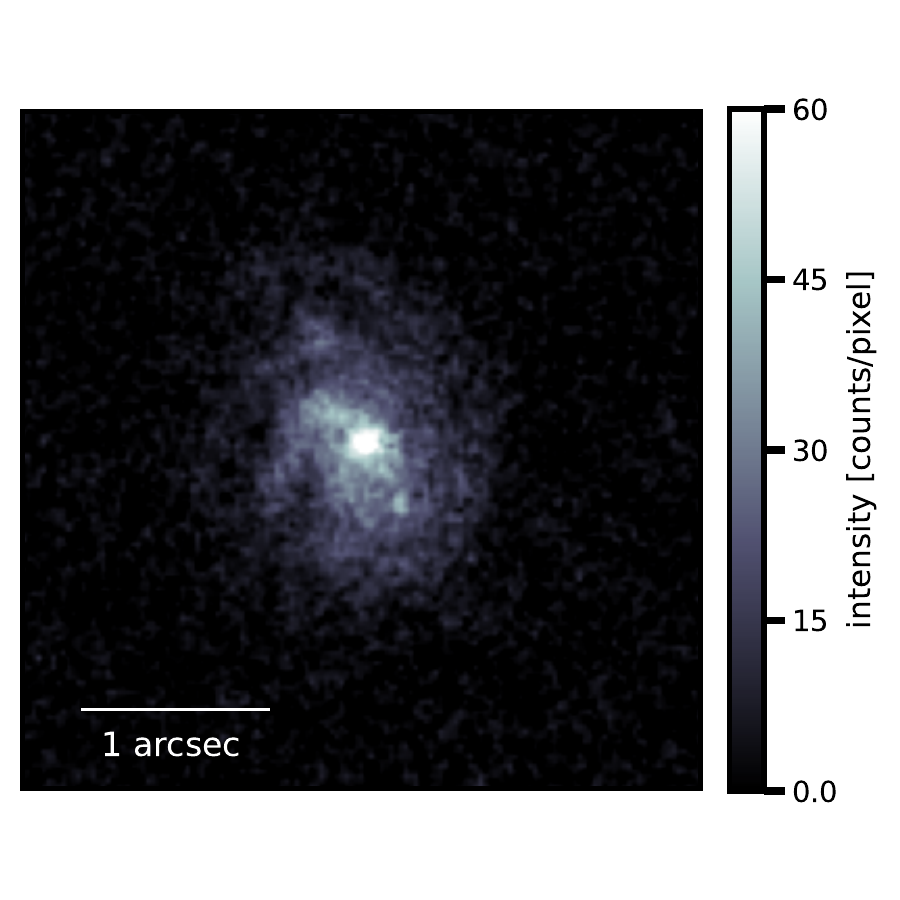}
			\caption{\label{fig:source} The spiral galaxy used in the mock lens systems as the lensed quasar host galaxy. This image was extracted from the COSMOS catalog \citep{Wagner-Carena++23}.}
		\end{figure}
		\section{Tests on simulated data}
		\label{sec:simulationsetup}
		This section details the how we create simulated datasets with which to test the lens modeling methodology discussed in the previous section. Section \ref{ssec:darkmattermodels} discusses the warm dark matter model on which we test the methodology. Section \ref{ssec:massprofilemaindeflector} discusses the properties of the simulated main deflectors we use in our simulations. Section \ref{ssec:lightmodels} details the parameterization of the lens and source light models used to create and model the simulated datasets. Section \ref{ssec:obs} details the modeling assumptions related to the observing conditions and point spread function. Throughout this section we use $\mathcal{U}$ to represent a uniform prior and $\mathcal{N}$ to represent a Gaussian prior, not be confused with $\mathcal{N}$, the vector of nuisance parameters appearing in Equation \ref{eqn:likelihoodsingle}.
		\begin{figure*}
			\includegraphics[trim=0.25cm 2cm 0.25cm
			1cm,width=0.45\textwidth]{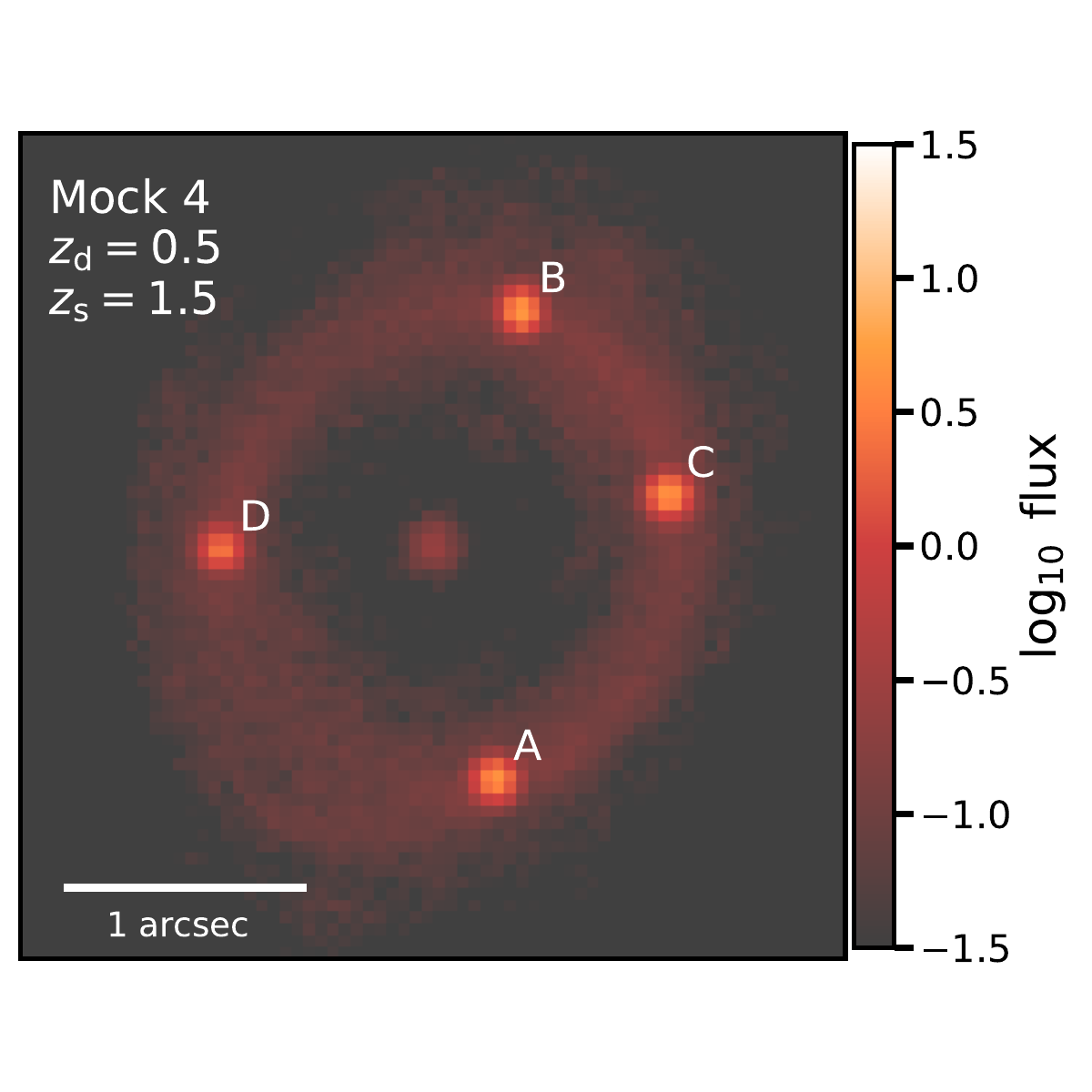}
			\includegraphics[trim=0.25cm 2cm 0.25cm
			1cm,width=0.45\textwidth]{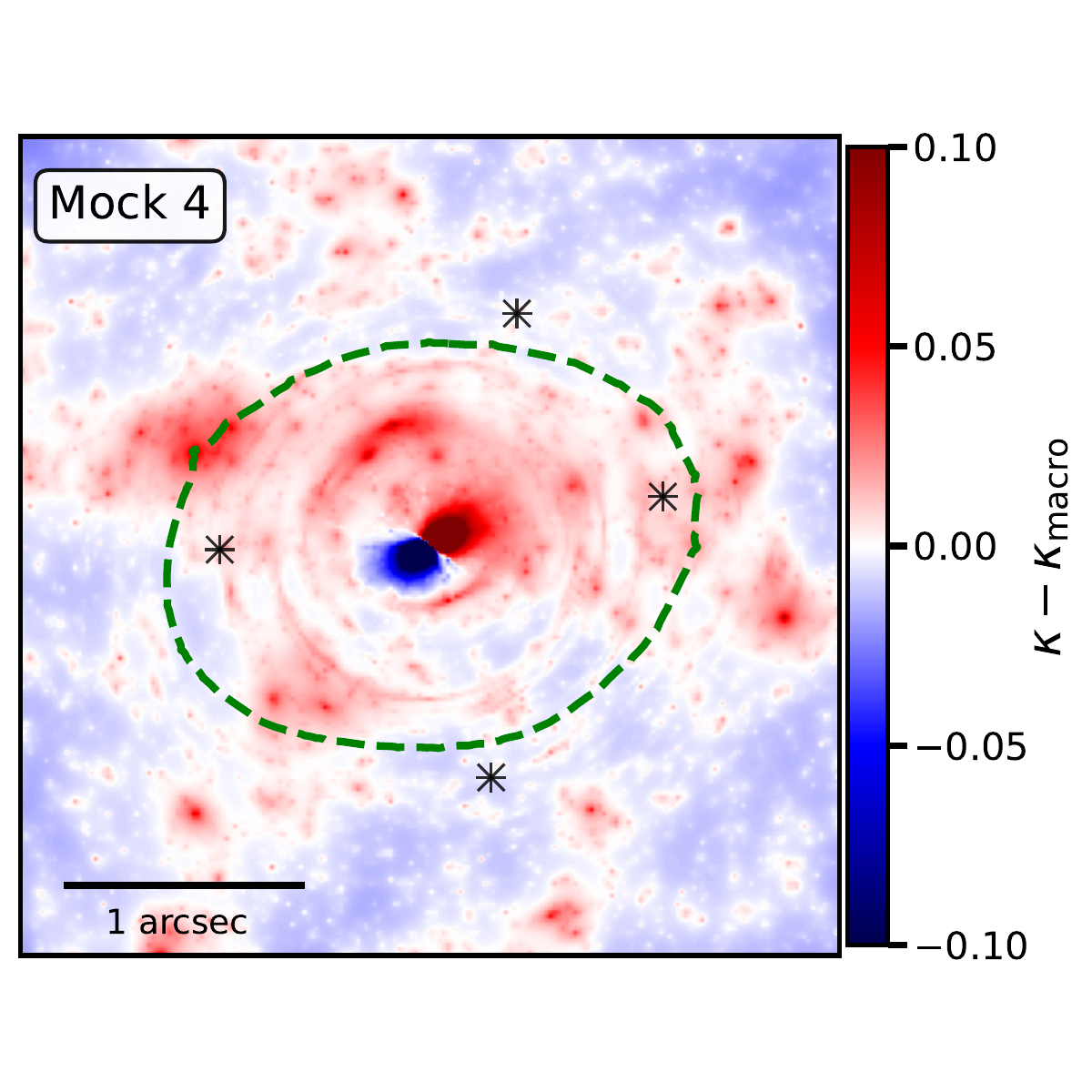}
			\caption{\label{fig:mock4image} The imaging data (left) and the true convergence in dark matter substructure (right) for mock lens $\#$4. This system has flux ratios consistent with those predicted by a smooth lens model.}
		\end{figure*}
		\begin{figure*}
			\includegraphics[trim=0.25cm 2cm 0.25cm
			1cm,width=0.45\textwidth]{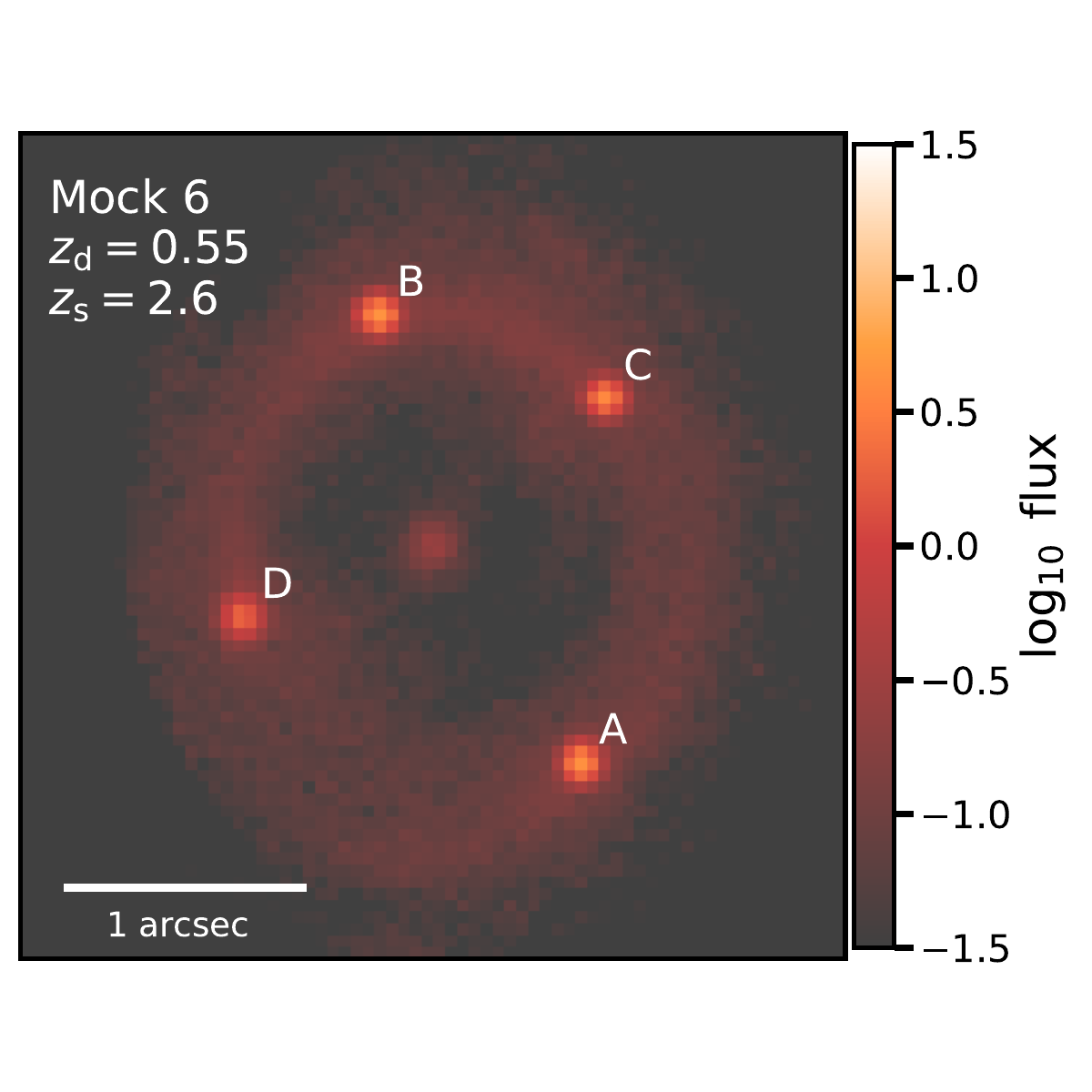}
			\includegraphics[trim=0.25cm 2cm 0.25cm
			1cm,width=0.45\textwidth]{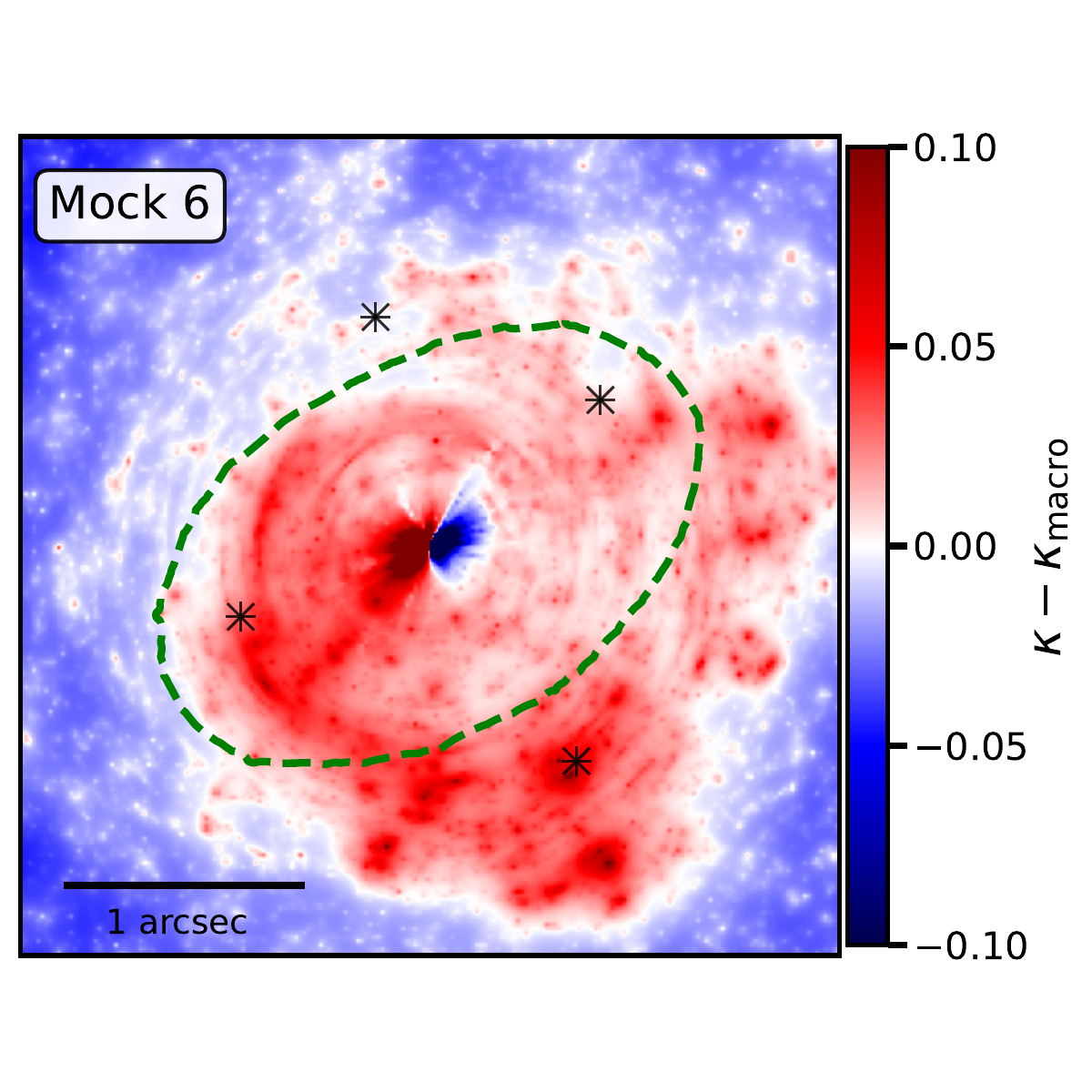}
			\caption{\label{fig:mock6image}The imaging data (left) and the true convergence in dark matter substructure (right) for mock lens $\#$6. Several dark matter halos and line-of-sight halos near image A impart a strong perturbation to the magnification of this image.}
		\end{figure*}
		\begin{figure*}
			\includegraphics[trim=0.25cm 2cm 0.25cm
			1cm,width=0.45\textwidth]{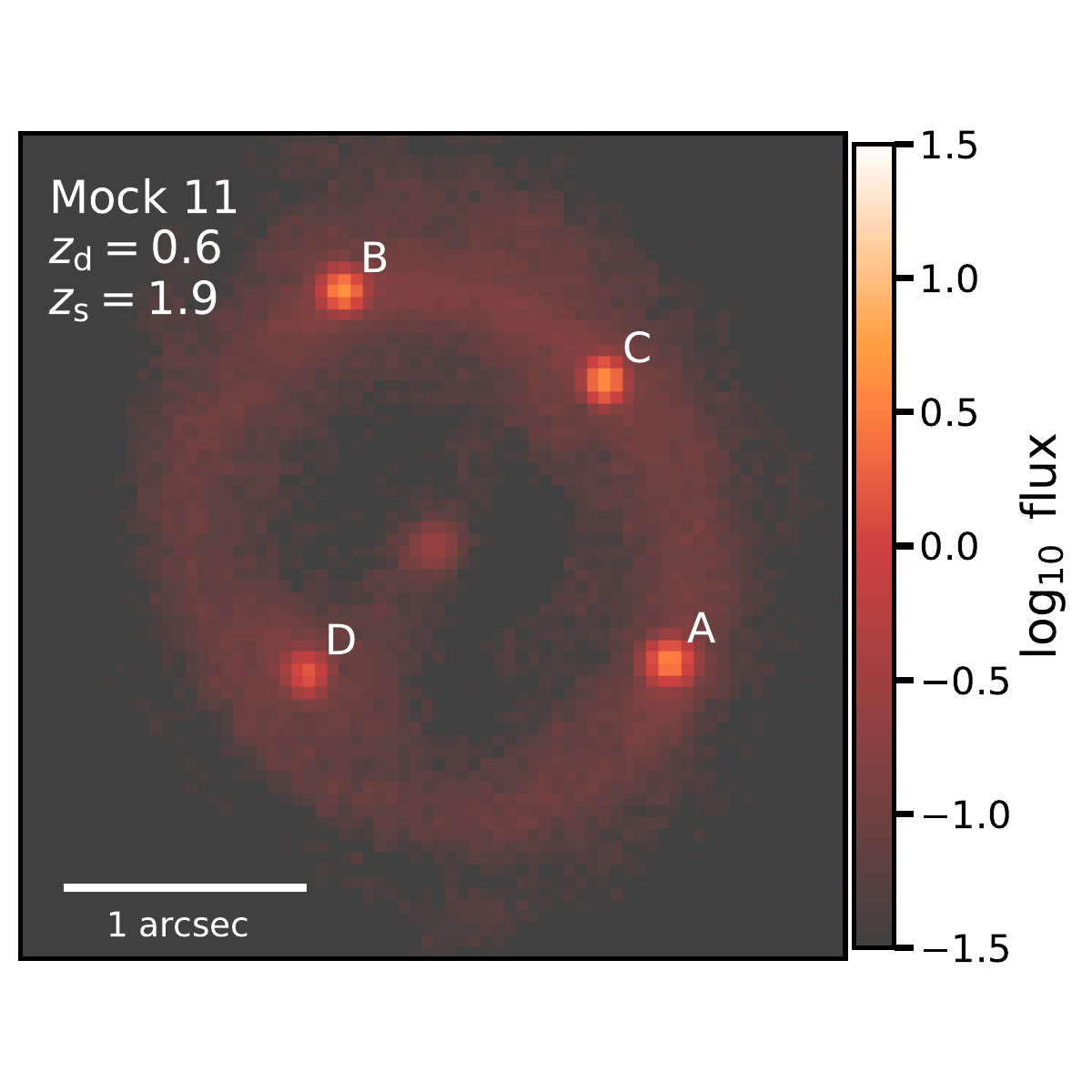}
			\includegraphics[trim=0.25cm 2cm 0.25cm
			1cm,width=0.45\textwidth]{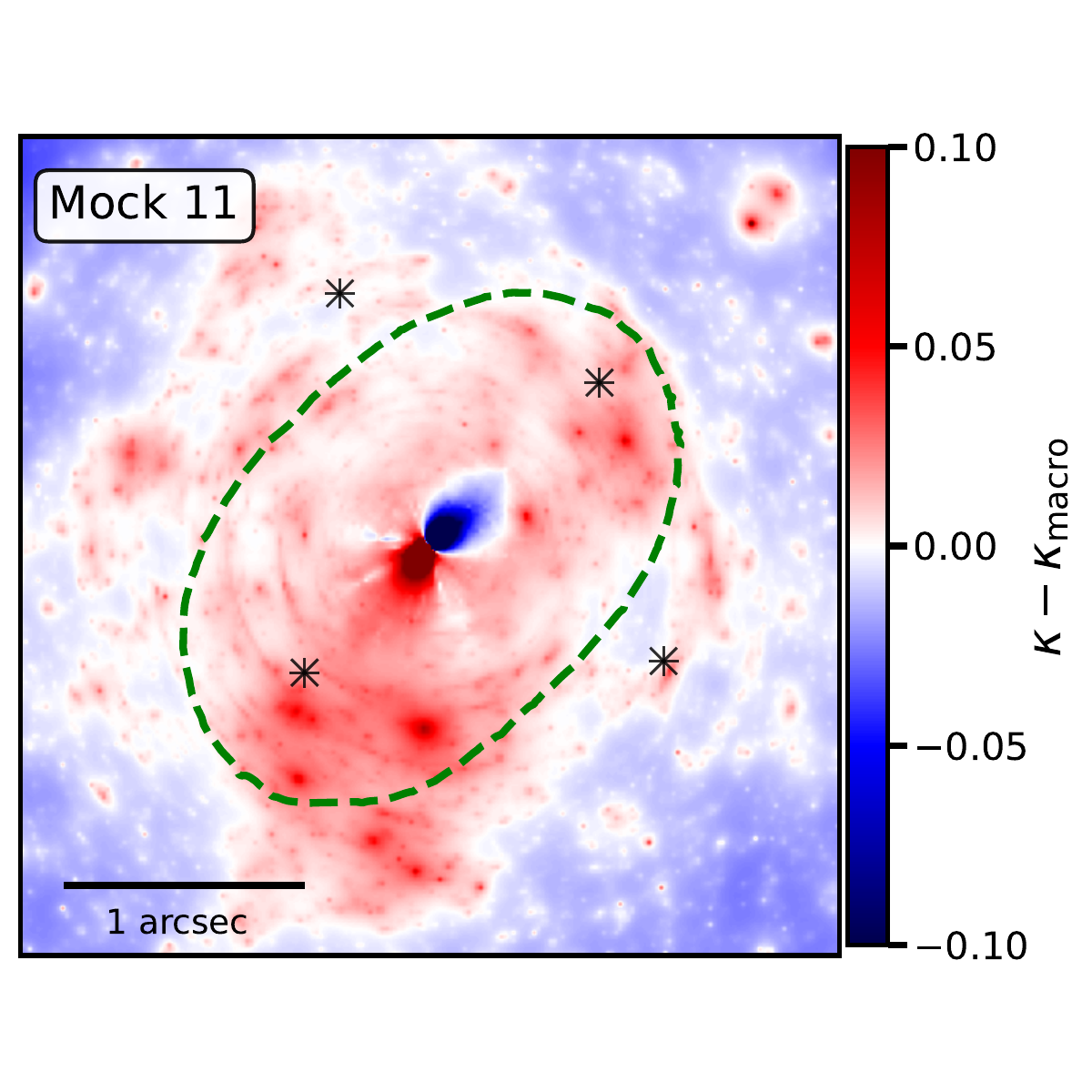}
			\caption{\label{fig:mock11image} The imaging data (left) and the true convergence in dark matter substructure (right) for mock lens $\#$11. Like Mock 4, this system has flux ratios consistent with those predicted by a smooth lens model.}
		\end{figure*}
		\begin{figure*}
			\includegraphics[trim=0.25cm 2cm 0.25cm
			1cm,width=0.45\textwidth]{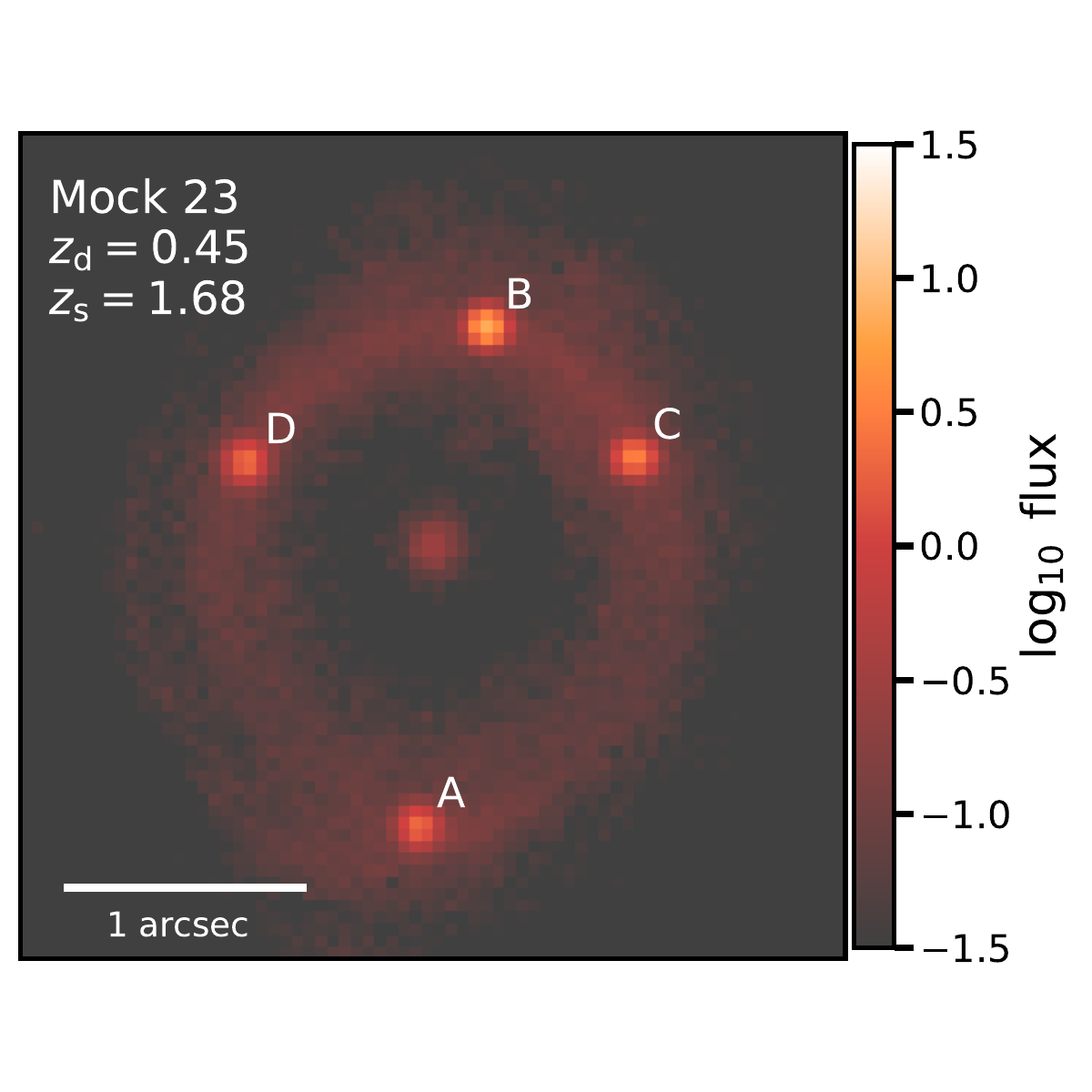}
			\includegraphics[trim=0.25cm 2cm 0.25cm
			1cm,width=0.45\textwidth]{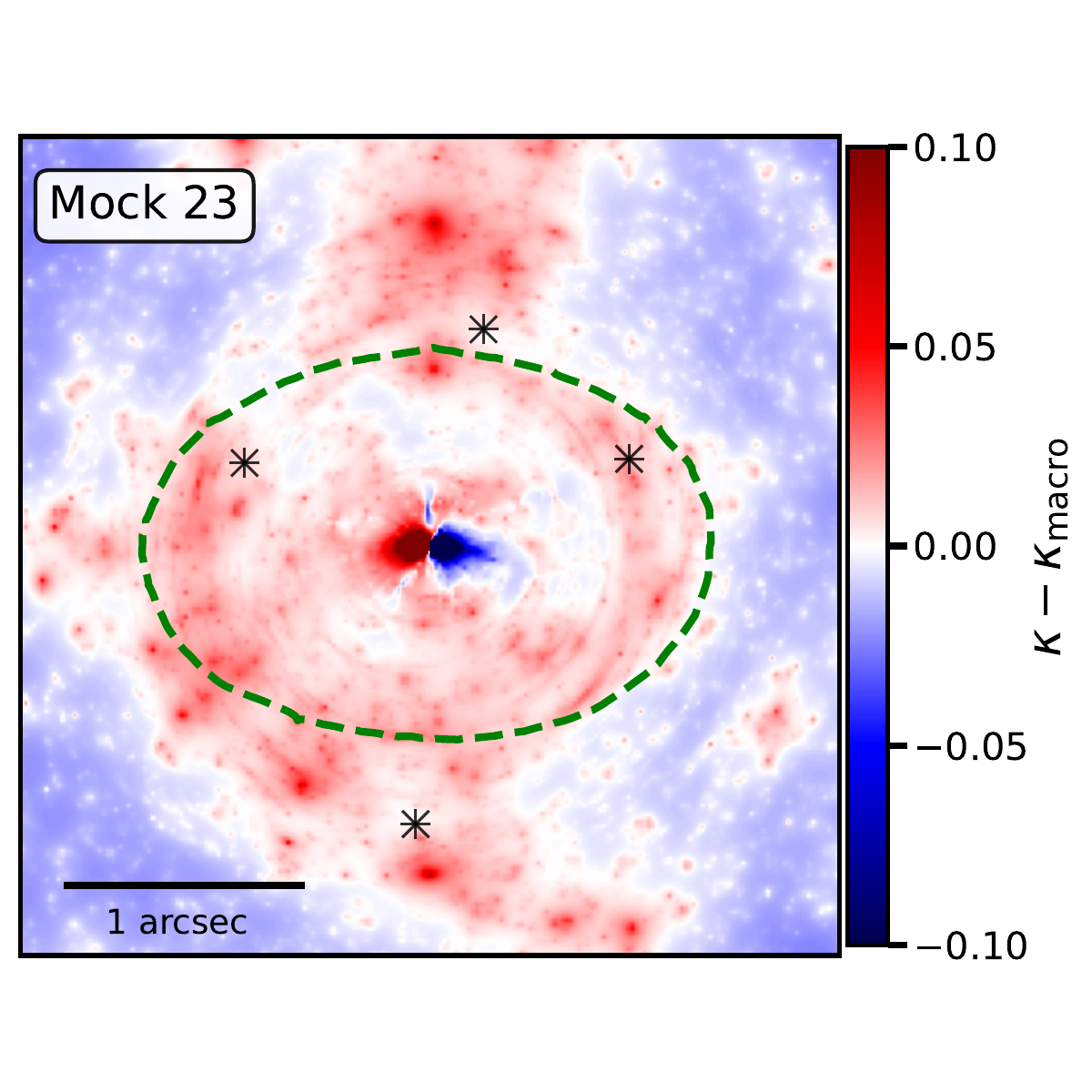}
			\caption{\label{fig:mock23image} The imaging data (left) and the true convergence in dark matter substructure (right) for mock lens $\#$23. Like Mock 6, the flux ratios in this system experience perturbation by halos.}
		\end{figure*}
		\subsection{Modeling of dark matter subhalos and line-of-sight halos}
		\label{ssec:darkmattermodels}
		We test the methodology presented in Section \ref{sec:multiplanelensmodeling} on a warm dark matter (WDM) model, although the methodology we present is applicable to any theory that predicts the abundance and internal structure of halos. Warm dark matter refers to a class of theory in which the abundance and density profiles of dark matter halos become suppressed below a certain mass threshold determined by the free-streaming length. We model the halo mass function in WDM using the parametric form 
		\begin{equation}
			\label{eqn:suppression}
			\frac{dN_{\rm{WDM}}}{dm} = \frac{dN_{\rm{CDM}}}{dm} \left(1 + \left(a\frac{m_{\rm{hm}}}{m}\right)^b\right)^{c},
		\end{equation}
		with $a = 2.3$, $b = 0.8$, and $c = -1.0$ \citep{Lovell++20}, and $dN_{\rm{CDM}}/  dm$ represents the (sub)halo mas function in CDM. We draw line-of-sight halos from the Sheth-Tormen \citep{ST01} halo mass function, and generate subhalos from a mass function of the form
		\begin{equation}
			\label{eqn:shmf}
			\frac{d^2N}{dm dA} = \frac{\Sigma_{\rm{sub}}}{m_0} \left(\frac{m}{m_0}\right)^{-\alpha}.
		\end{equation}
		We use a pivot scale $m_0 = 10^{8} M_{\odot}$, and a logarithmic profile slope $\alpha = 1.9$ \citep{Springel++08}. The normalization is related to the projected mass in substructure in the range $10^6 - 10^{10}$ solar masses by $f_{\rm{proj}} = 10^7\left(\Sigma_{\rm{sub}} / 0.01 \  \rm{kpc^{-2}}\right) \ \rm{M_{\odot}} \ \rm{kpc^{-2}}$, and N-body simulations of massive ellipticals predict typical values in the range $\sim 2-3 \times 10^7 M_{\odot} / \rm{kpc^{-2}}$ \citep[e.g][]{Fiacconi++16}, although these numbers are uncertain due to various numerical uncertainties inherent to the simulations, such as the artificial disruption of subhalos and subhalo finders. To encompass a broad range of theoretical uncertainties associated with $\Sigma_{\rm{sub}}$, we use a log-uniform prior between $\log_{10} \Sigma_{\rm{sub}} = -2.5 $ and $\log_{10}\Sigma_{\rm{sub}} = -1.0$. We apply the same suppression function in Equation \ref{eqn:suppression} to both the line-of-sight and subhalo mass functions. 
		
		We model halo density profiles as tidally-truncated Navarro-Frenk-White profiles \citep{NFW,Baltz++09}
		\begin{equation}
			\rho\left(r,r_s,r_t\right) =  \frac{\rho_s}{\left(r/r_s\right)\left(1+r / r_s\right)^2} \frac{r_t^2}{r_s^2 + r_t^2}
		\end{equation}
		where $\rho_s$ is a characteristic central density, $r_s$ is the scale radius, and $r_t$ implements a tidal radius. For field halos, we set $r_t$ equal to $r_{\rm{50}}$, which is comparalbe to the splash-back radius of a halo \citep{DiemerKravtsov14,Adhikari++14,More++15}. We truncated subhalo density profiles based on the mass and three-dimensional position inside the host (see \citet{Gilman++20}). 
		
		The central density $\rho_s$ determines the lensing efficiency of a halo of a fixed total mass.  The delayed onset of structure formation in WDM models suppresses the central density of halos with mass below $m_{\rm{hm}}$ \citep{Bose++16,Ludlow++16}. The concentration-mass relation establishes the connection between $\rho_s$ and halo mass through the concentration parameter, $c$. We use the concentration-mass relation presented by \citet{DiemerJoyce19}, and implemented with the software {\tt{colossus}} \citep{Diemer18}, to assign concentrations to CDM halos. We compute the concentrations in WDM according to \citep{Bose++16}
		\begin{equation}
			\label{eqn:suppressionc}
			c_{\rm{WDM}}\left(m,z\right) = c_{\rm{CDM}}\left(m,z\right) \left(1+60\frac{m_{\rm{hm}}}{m}\right)^{-0.17} \left(1+z\right)^{\beta\left(z\right)},
		\end{equation}
		where $\beta\left(z\right) = 0.026 - 0.04 z$. 
		
		We generate line-of-sight halos (subhalos) with (infall) masses in the range $10^6 - 10^{10} M_{\odot}$. Halos more massive than $10^{10} M_{\odot}$ would likely contain a luminous galaxy, in which case we would include these objects in the lens macromodel. Halos less massive than $10^6 M_{\odot}$ lie below the sensitivity threshold of our data given our assumptions regarding the background source size (see the next sub-section). 
		
		In the dark matter inference, we sample $\Sigma_{\rm{sub}}$ from a log-uniform prior $\log_{10} \Sigma_{\rm{sub}} \in\mathcal{U}\left(-2.5, -1.0\right)$. This is a broad and uninformative prior that we use to reveal any degeneracies between $\Sigma_{\rm{sub}}$ and $m_{\rm{hm}}$, and to understand to what degree imaging data can break these degeneracies. We sample $\log_{10} m_{\rm{hm}} \in \mathcal{U} \left(4.0, 10.0\right)$. For both the halo mass function and concentration-mass relation, values of $m_{\rm{hm}} < 10^{5} M_{\odot}$ result in mass functions and concentration-mass relations that deviate from the CDM prediction below the estimated sensitivity threshold of our data; thus, we can consider these realizations as consistent with CDM. We fix the values of all other parameters introduced in this section to the stated values. 
		
		We create two sets of simulated data with which to test our inference methodology. First, we create a sample of 25 lenses with a CDM ground truth using $\Sigma_{\rm{sub}} = 0.05 \  \rm{kpc^{-2}}$ and $m_{\rm{hm}} = 0$. The chosen value of $\Sigma_{\rm{sub}}$ roughly corresponds to the amount of substructure inferred by \citet{Gilman++20} and is consistent with N-body to within an $\mathcal{O}\left(10\right)$ factor \citep{Fiacconi++16}. Second, we create a sample of 25 lenses with a WDM ground truth with  $\Sigma_{\rm{sub}} = 0.04 \  \rm{kpc^{-2}}$ and $m_{\rm{hm}} = 10^{7.5} M_{\odot}$. Both sets of mocks have the same main deflector mass models and background sources, but the image positions and flux ratios between them differ slightly due to the different populations of halos in the lens models. 
		
		\begin{figure*}
			\includegraphics[trim=0.5cm 1cm 0.5cm
			1cm,width=0.95\textwidth]{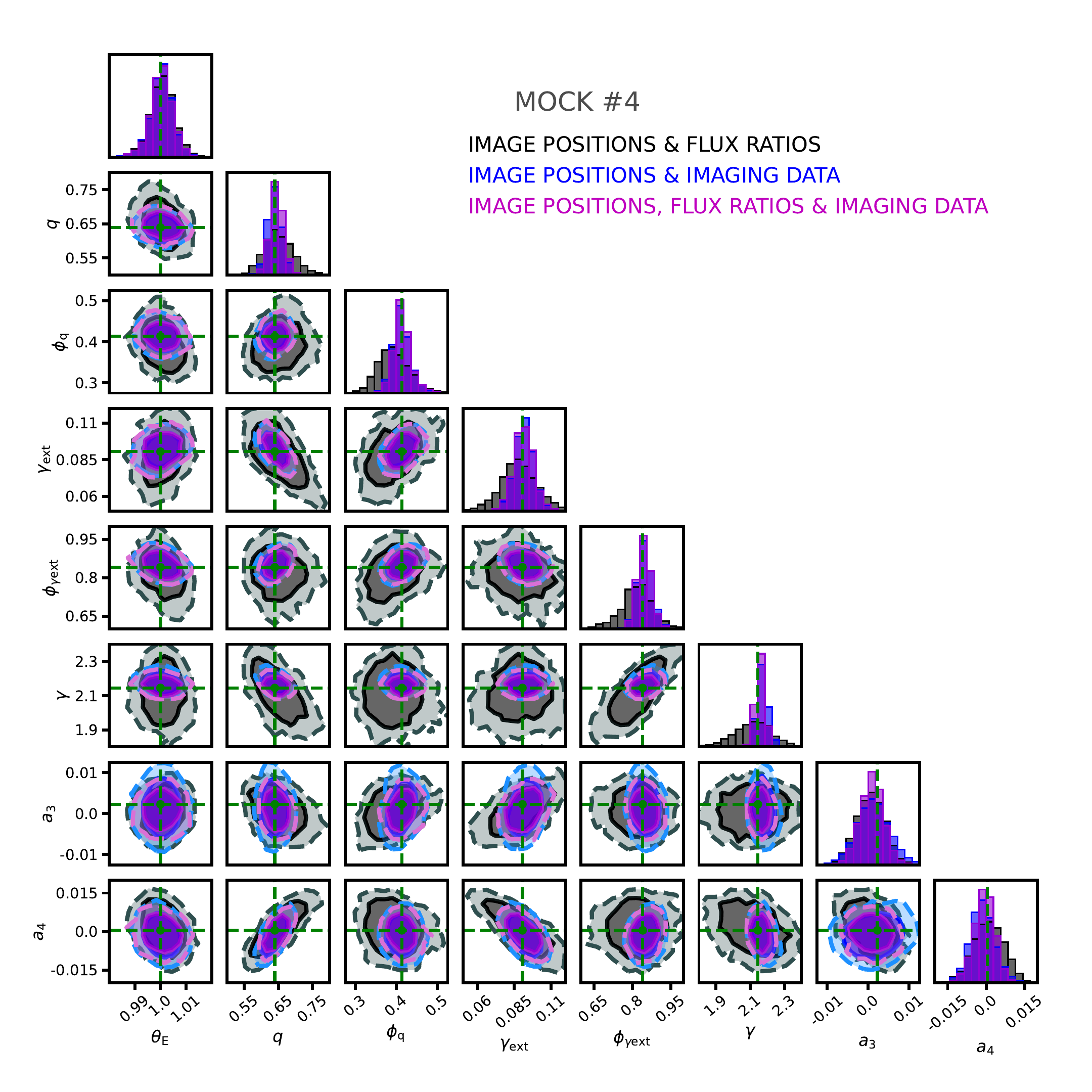}
			\caption{\label{fig:mock4macro} The inference on the macromodel parameters in Mock 4 in the CDM ground truth sample. The black distribution shows the inference obtained from using only image positions and flux ratios to constrain the lens model.The blue distribution results from using image positions and imaging data to constrain the lens model, and the magenta results from using image positions, flux ratios, and imaging data. These distributions are marginalized over the properties of substructure in each system. From left, the x-axis labels correspond to the normalization of the main deflector mass profile $\theta_{\rm{E}}$, the axis ratio $q$, the position angle of the main deflector ellipticity $\phi_{\rm{q}}$, the external shear strength $\gamma_{\rm{ext}}$, the position angle of the external shear $\phi_{\rm{\gamma\rm{ext}}}$, the logarithmic profile slope of the main deflector $\gamma$, and the multipole moments $a_3$ and $a_4$. Contours enclose $68 \%$ and $95 \%$ credible intervals, and the true lens model parameters for the mock lens are marked with the green crosshairs.}
		\end{figure*}
		\begin{figure*}
			\includegraphics[trim=0.5cm 1cm 0.5cm
			1cm,width=0.95\textwidth]{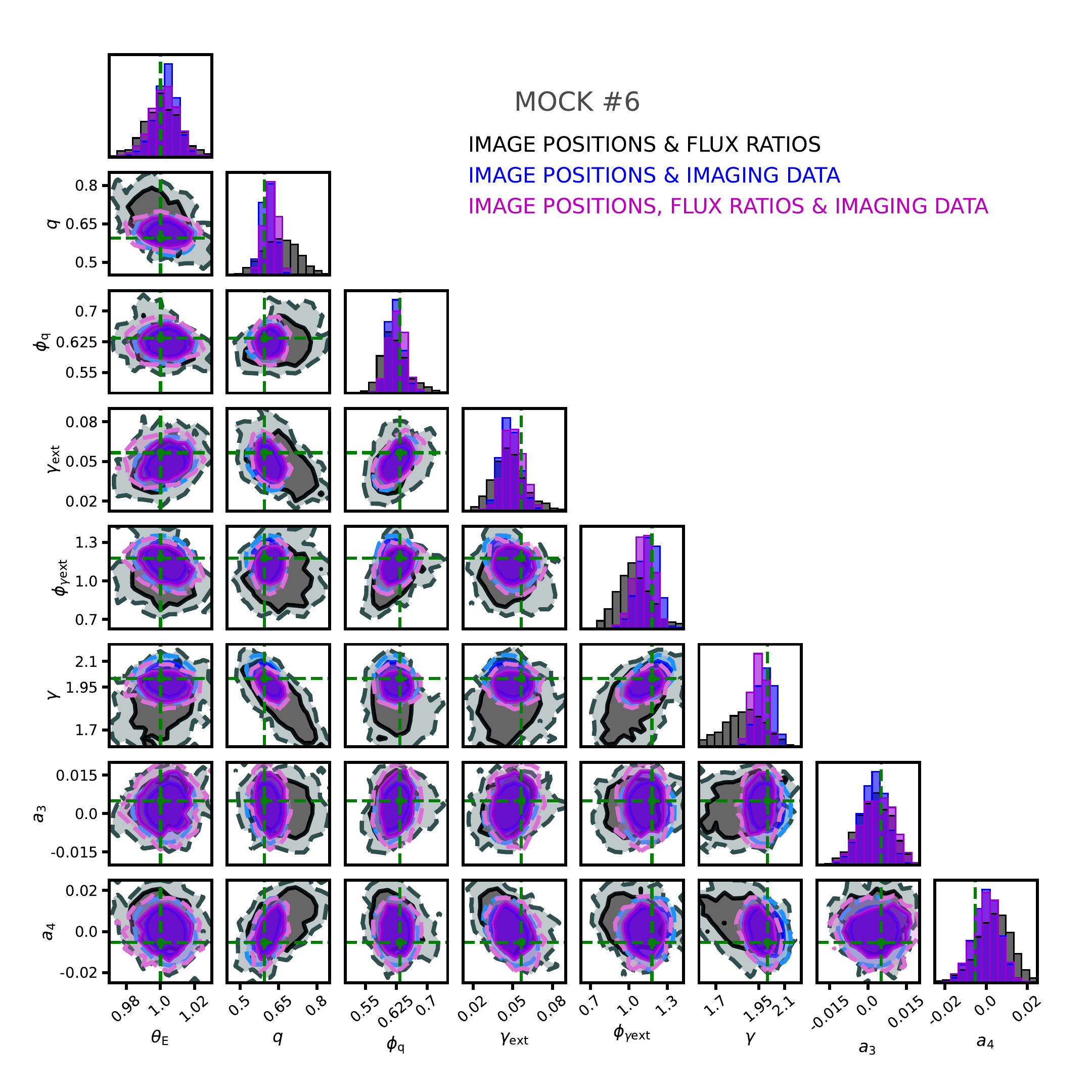}
			\caption{\label{fig:mock6macro} The inference on the macromodel parameters in Mock 6 in the CDM ground truth sample obtained from only the image positions and flux ratios (black), from the image positions and imaging data (blue), and from the image positions, flux ratios, and imaging data (magenta).}
		\end{figure*}
		\begin{figure*}
			\includegraphics[trim=0.5cm 1cm 0.5cm
			1cm,width=0.95\textwidth]{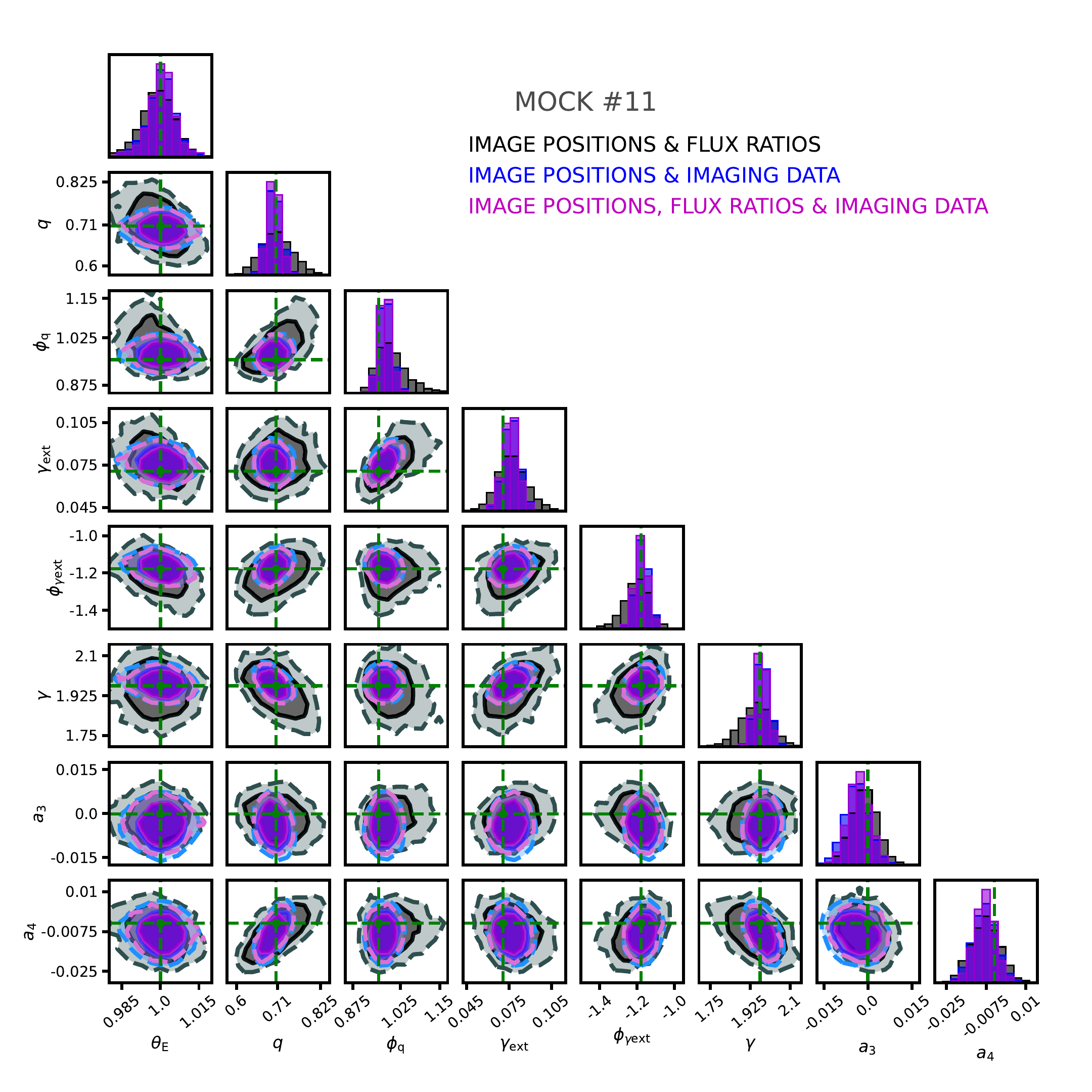}
			\caption{\label{fig:mock11macro} The inference on the macromodel parameters in Mock 11 in the CDM ground truth sample obtained from only the image positions and flux ratios (black), from the image positions and imaging data (blue), and from the image positions, flux ratios, and imaging data (magenta).}
		\end{figure*}
		\begin{figure*}
			\includegraphics[trim=0.5cm 1cm 0.5cm
			1cm,width=0.95\textwidth]{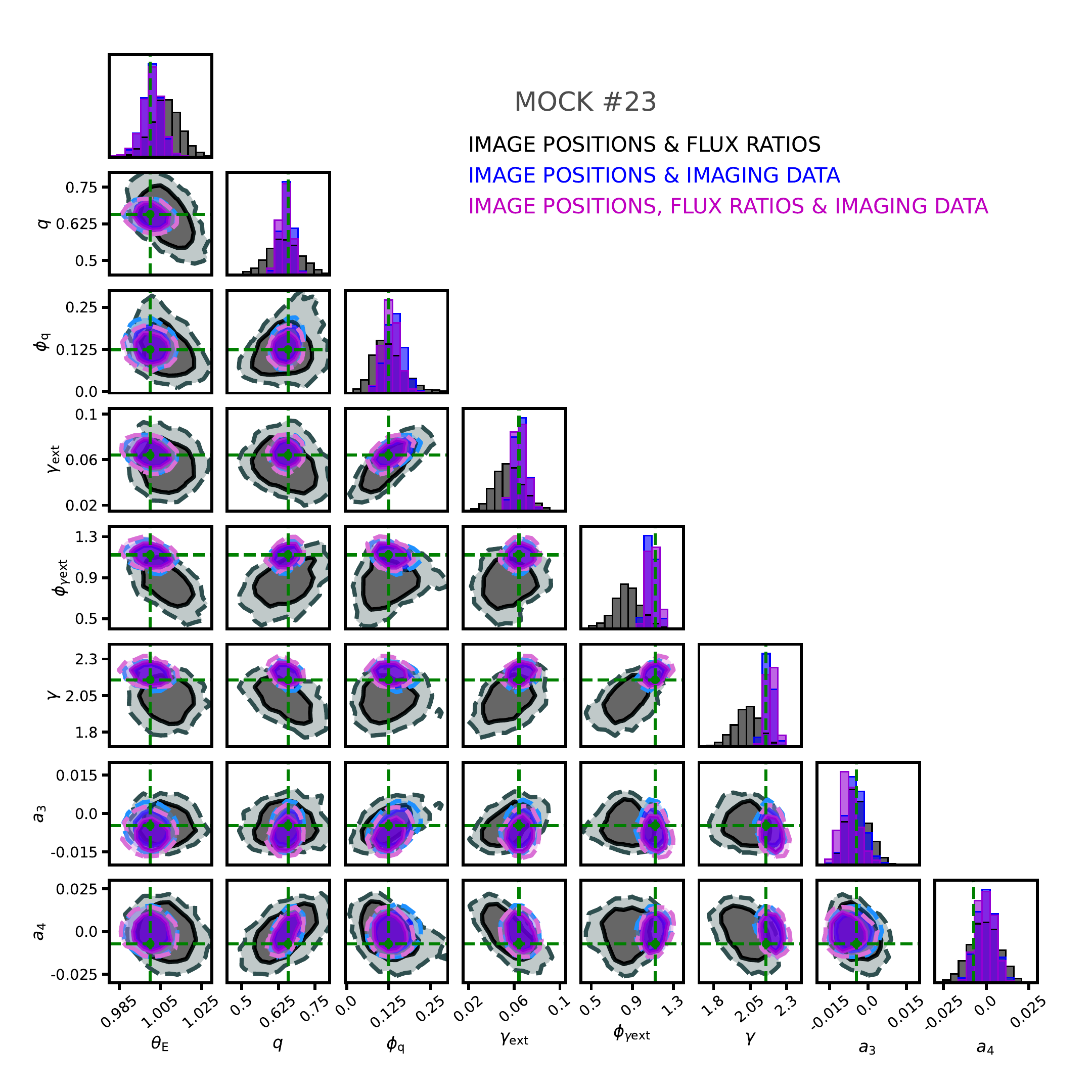}
			\caption{\label{fig:mock23macro} The inference on the macromodel parameters in Mock 23 in the CDM ground truth sample obtained from only the image positions and flux ratios (black), from the image positions and imaging data (blue), and from the image positions, flux ratios, and imaging data (magenta).}
		\end{figure*}
		\begin{figure*}
			\includegraphics[trim=0.25cm 0.1cm 0.25cm
			0.4cm,width=0.4\textwidth]{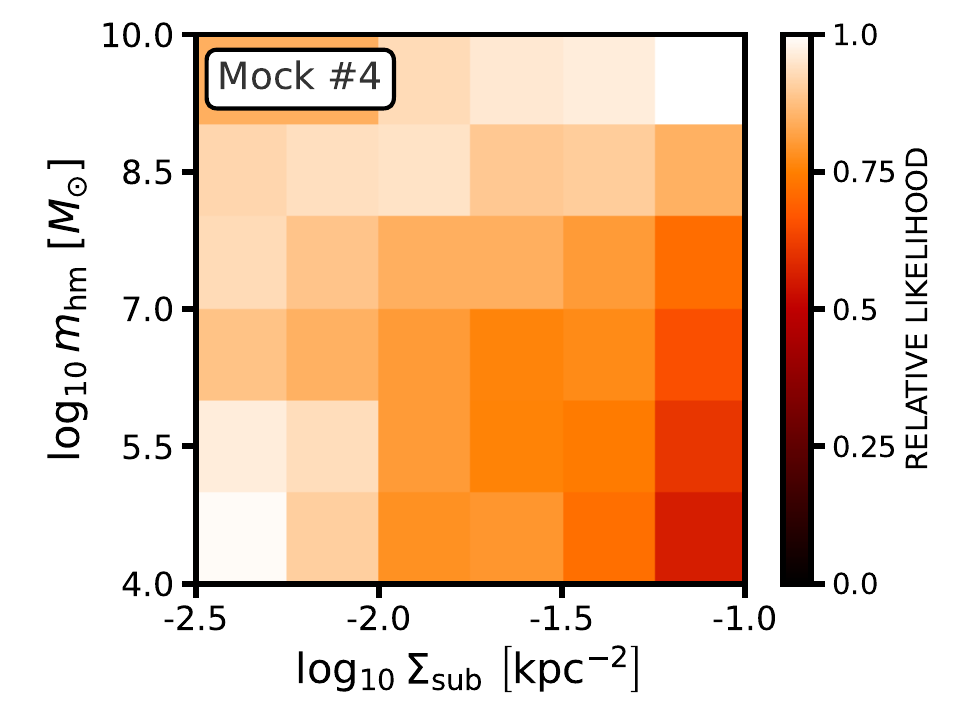}
			\includegraphics[trim=0.25cm 0.1cm 0.25cm
			0.4cm,width=0.4\textwidth]{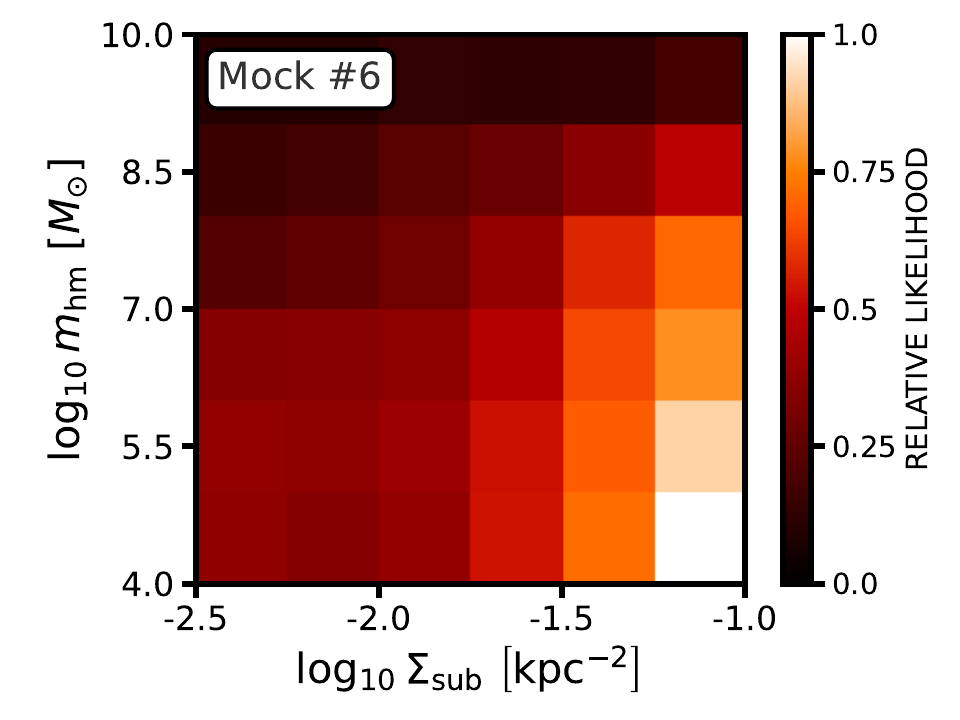}
			\includegraphics[trim=0.25cm 0.1cm 0.25cm
			0.4cm,width=0.4\textwidth]{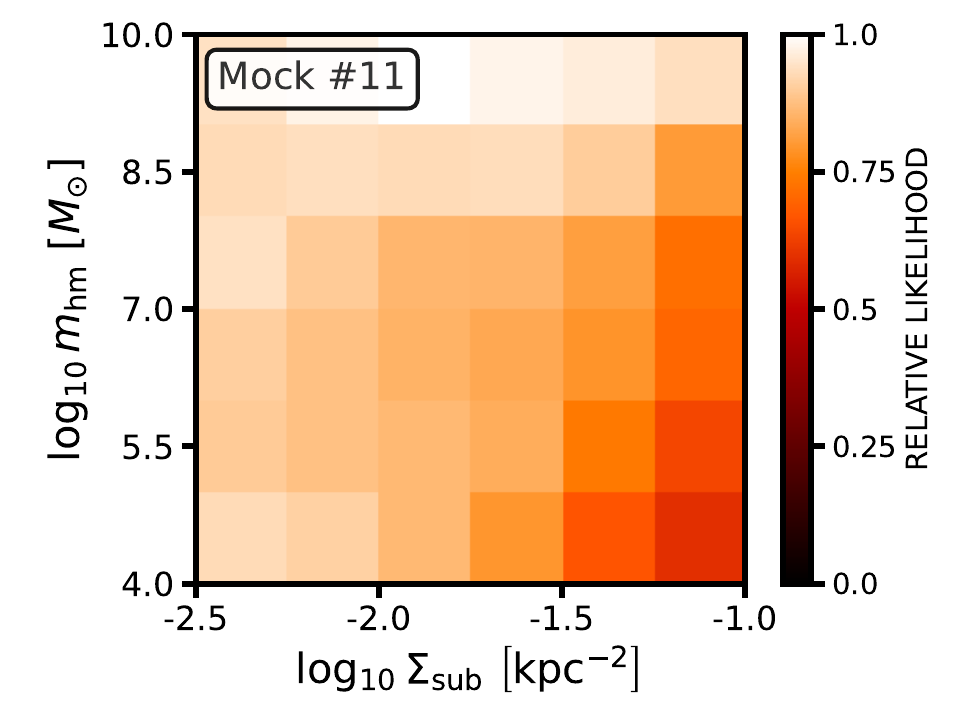}
			\includegraphics[trim=0.25cm 0.1cm 0.25cm
			0.4cm,width=0.4\textwidth]{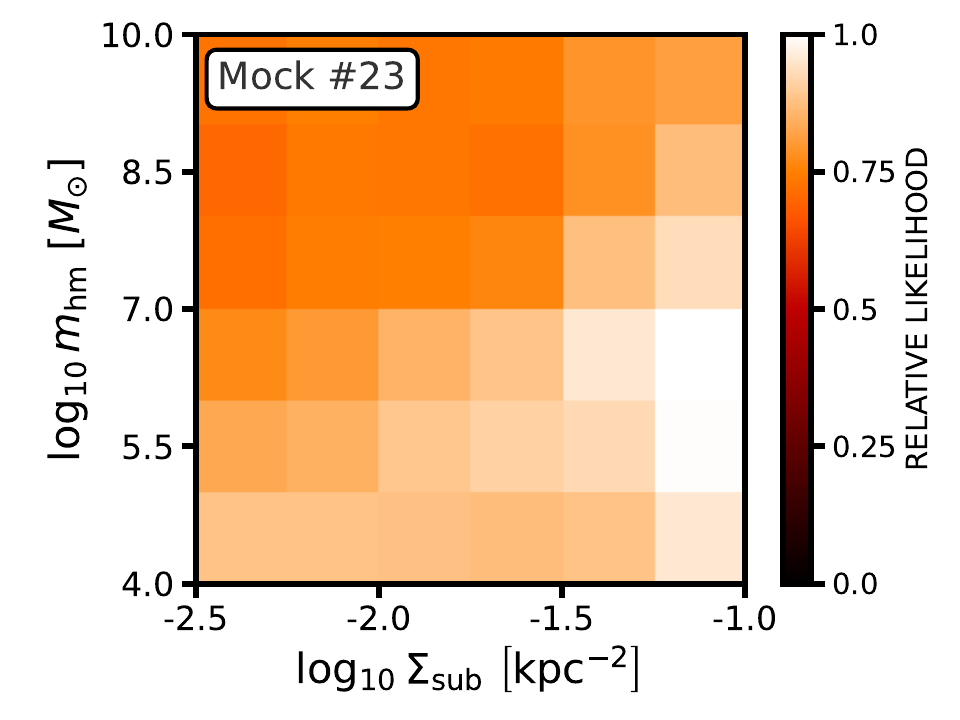}
			\caption{\label{fig:likelihoods} The likelihood functions for the normalization of the subhalo mass function, $\Sigma_{\rm{sub}}$, and the turnover scale of a WDM mass function, $m_{\rm{hm}}$, for Mocks 4, 6, 11, and 23 in the CDM ground truth sample. These likelihoods result from the down-selection on flux ratio and imaging data summary statistics described in Section \ref{ssec:inferencesummaries}. The color scale indicates relative likelihood between different points in parameter space.}
		\end{figure*}
		
		\begin{figure*}
			\includegraphics[trim=0.2cm 0.2cm 0.25cm
			2.5cm,width=0.33\textwidth]{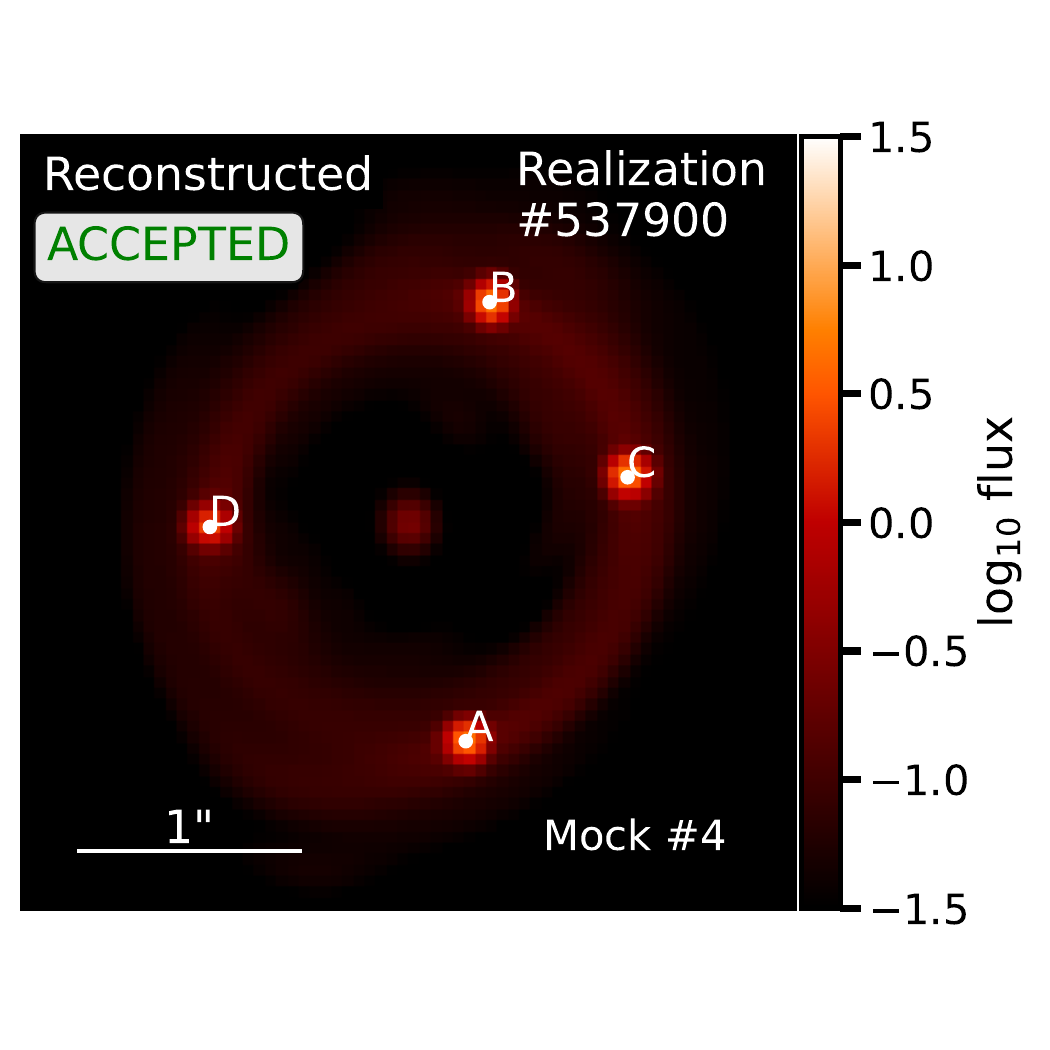}
			\includegraphics[trim=0.2cm 0.2cm 0.25cm
			2.5cm,width=0.33\textwidth]{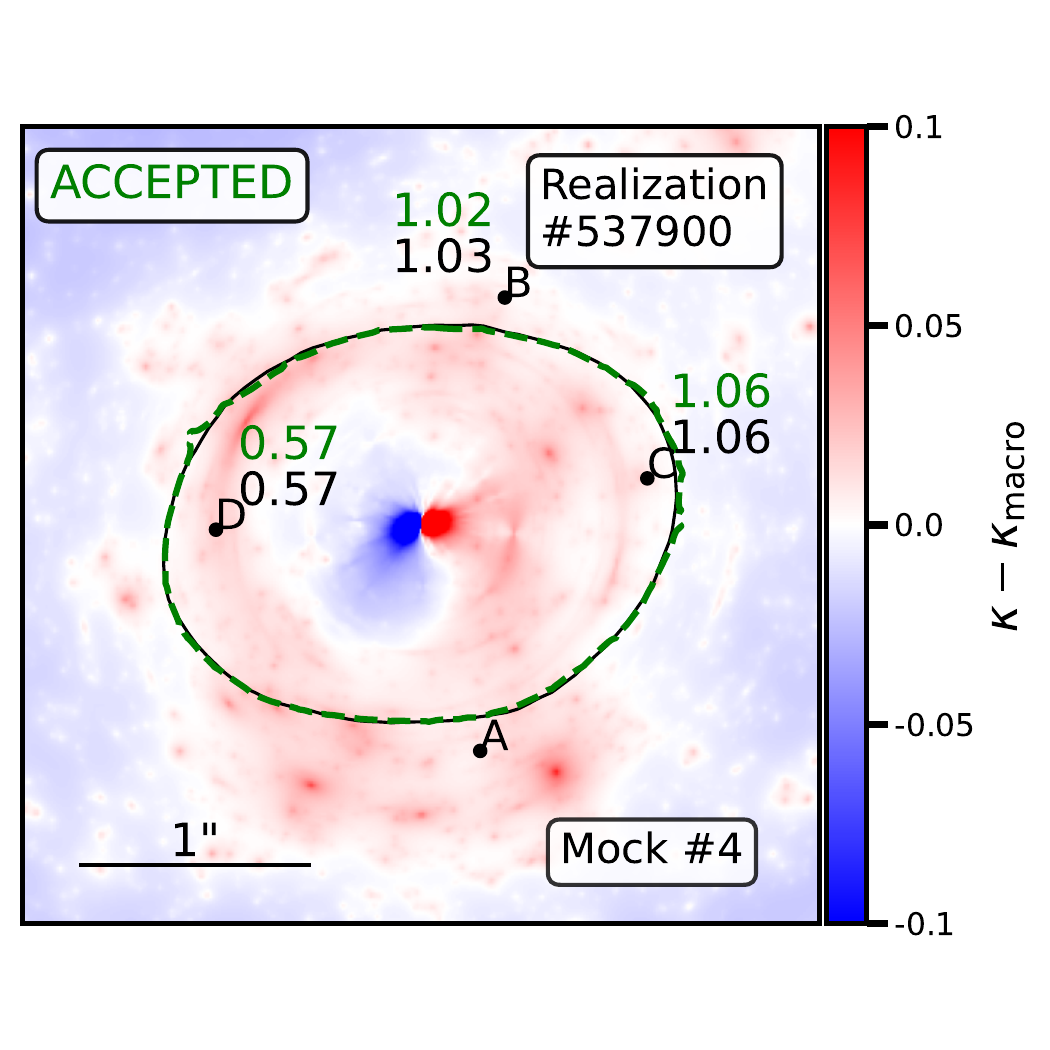}
			\includegraphics[trim=0.2cm 0.2cm 0.25cm
			2.5cm,width=0.33\textwidth]{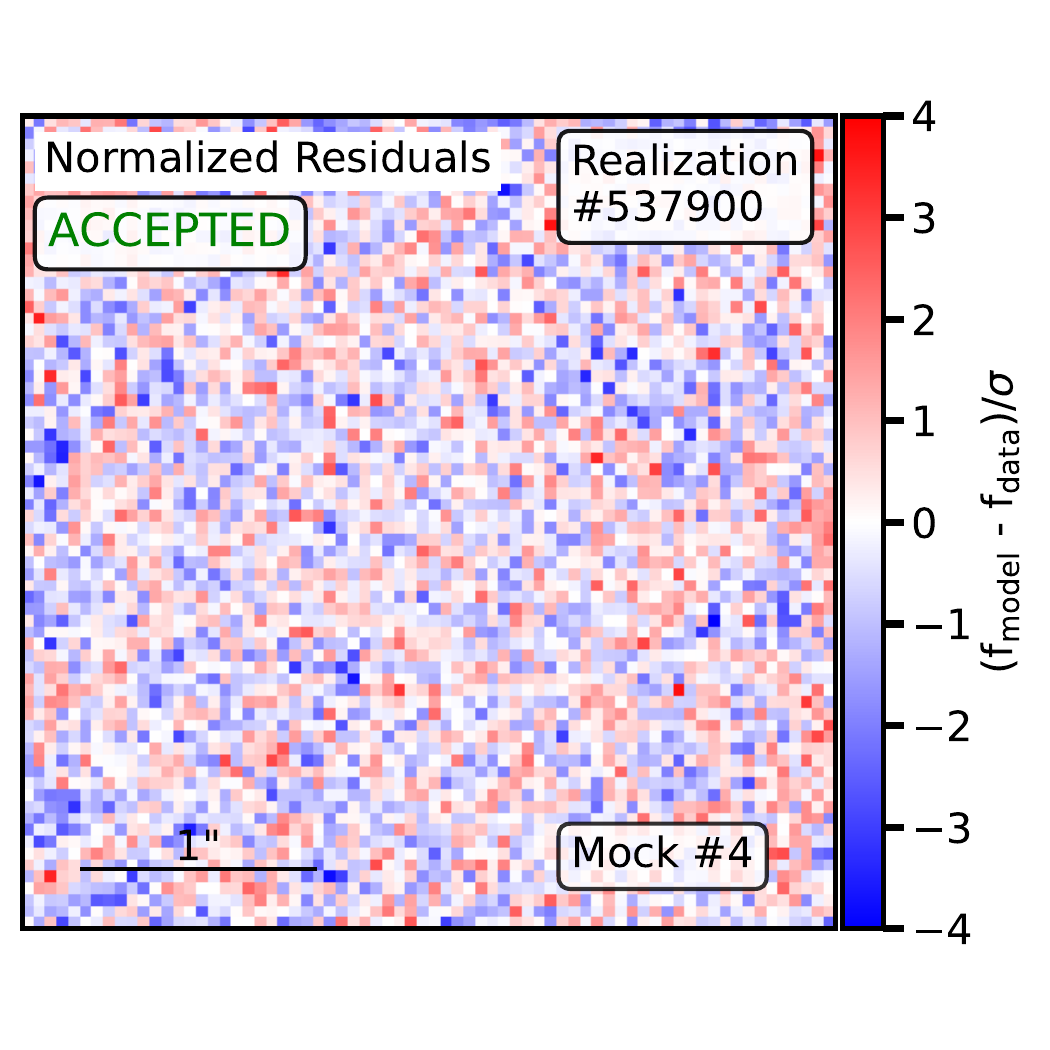}
			\includegraphics[trim=0.2cm 0.2cm 0.25cm
			1.5cm,width=0.33\textwidth]{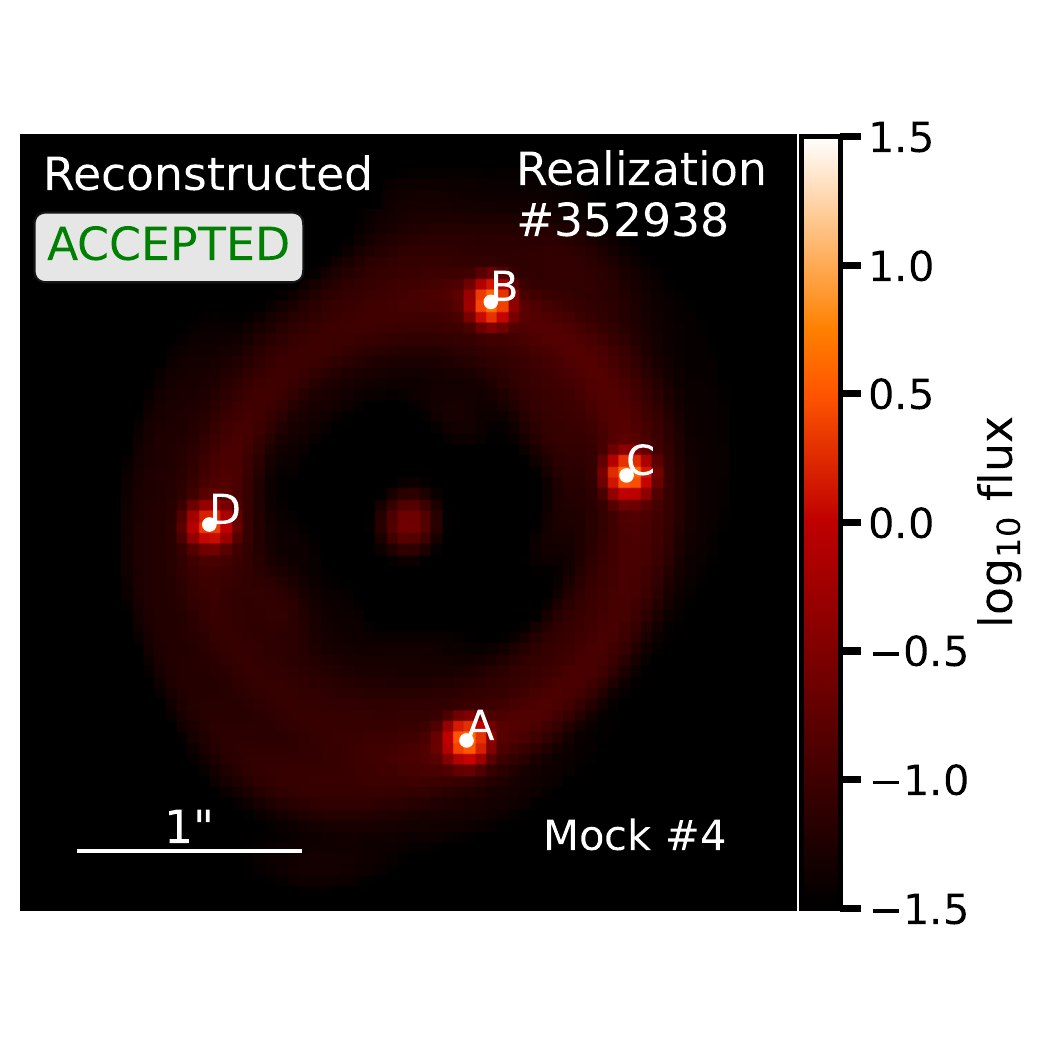}
			\includegraphics[trim=0.2cm 0.2cm 0.25cm
			1.5cm,width=0.33\textwidth]{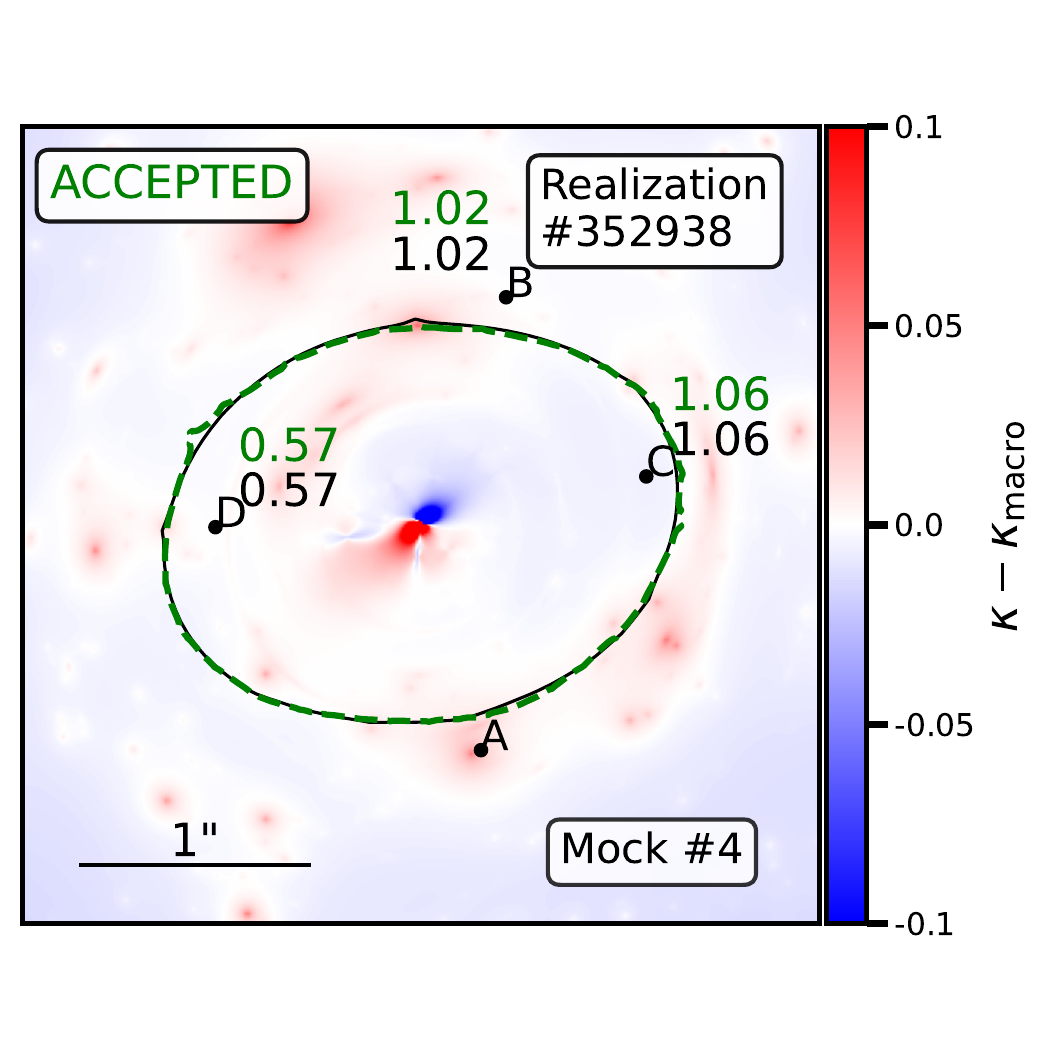}
			\includegraphics[trim=0.2cm 0.2cm 0.25cm
			1.5cm,width=0.33\textwidth]{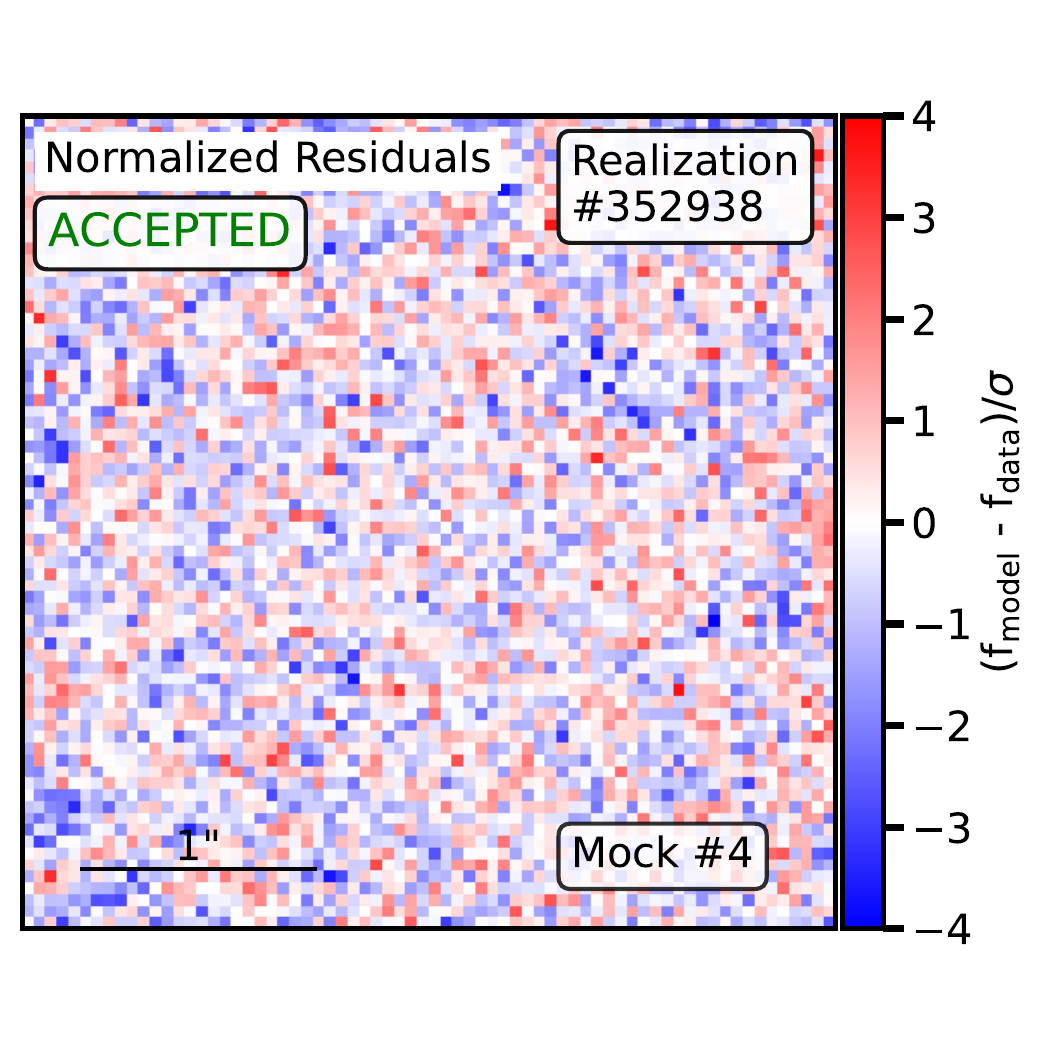}
			\includegraphics[trim=0.2cm 0.2cm 0.25cm
			1.5cm,width=0.33\textwidth]{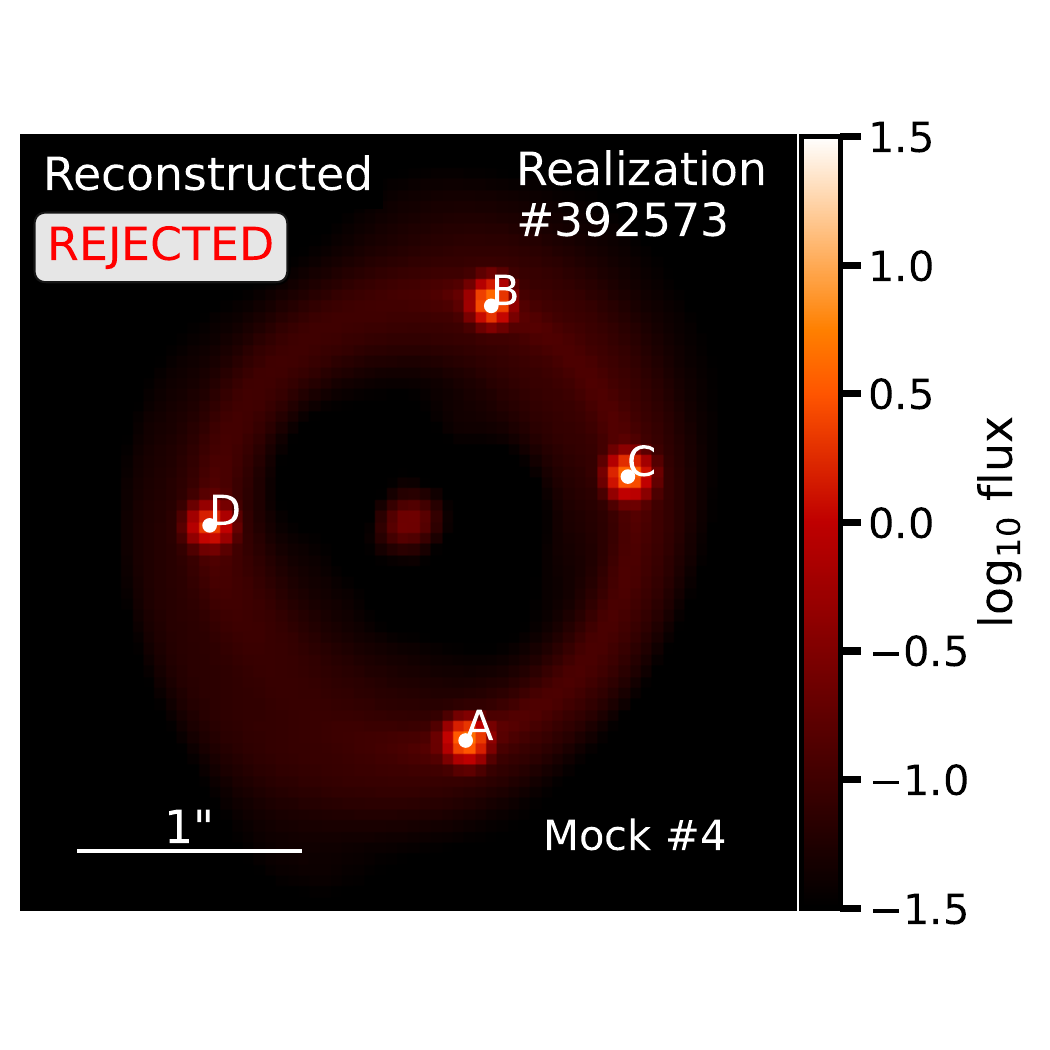}
			\includegraphics[trim=0.2cm 0.2cm 0.25cm
			1.5cm,width=0.33\textwidth]{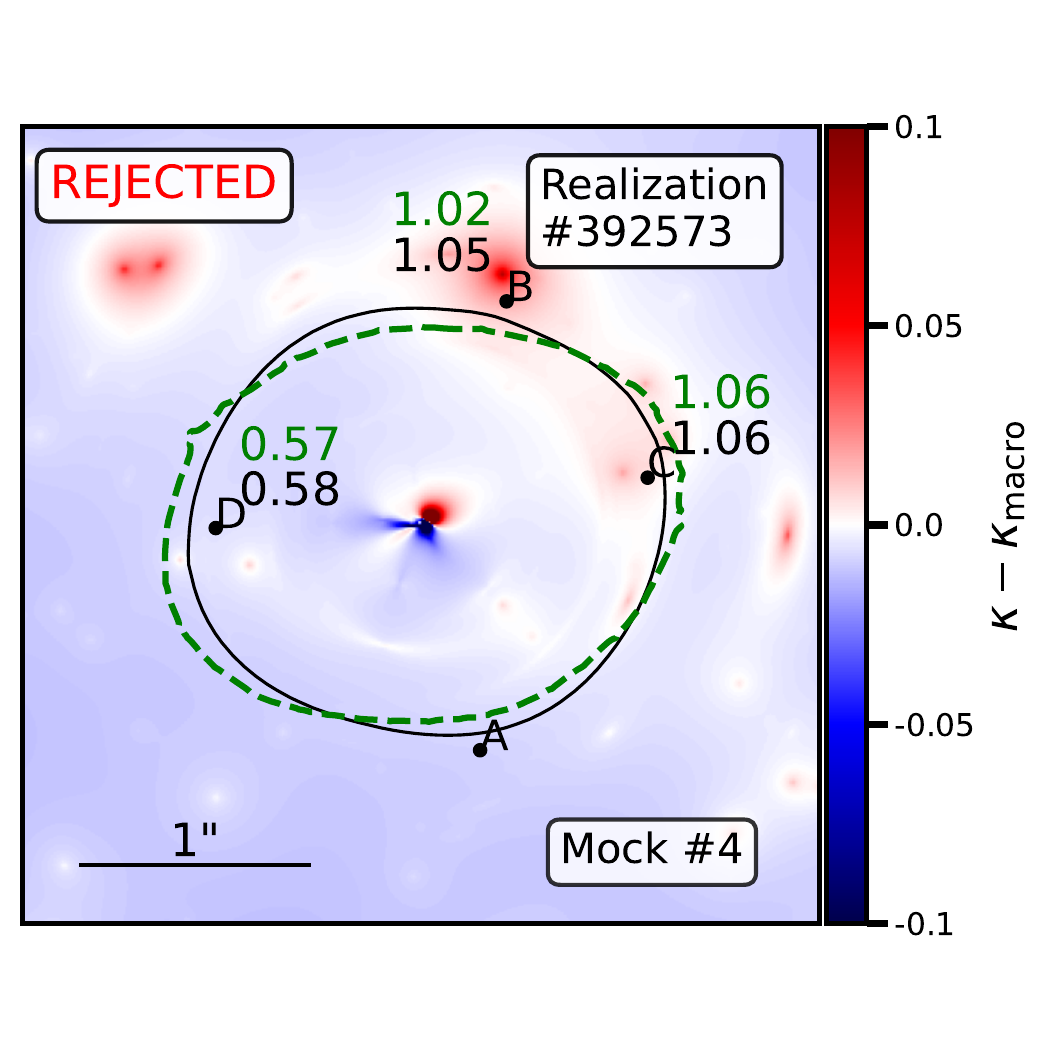}
			\includegraphics[trim=0.2cm 0.2cm 0.25cm
			1.5cm,width=0.33\textwidth]{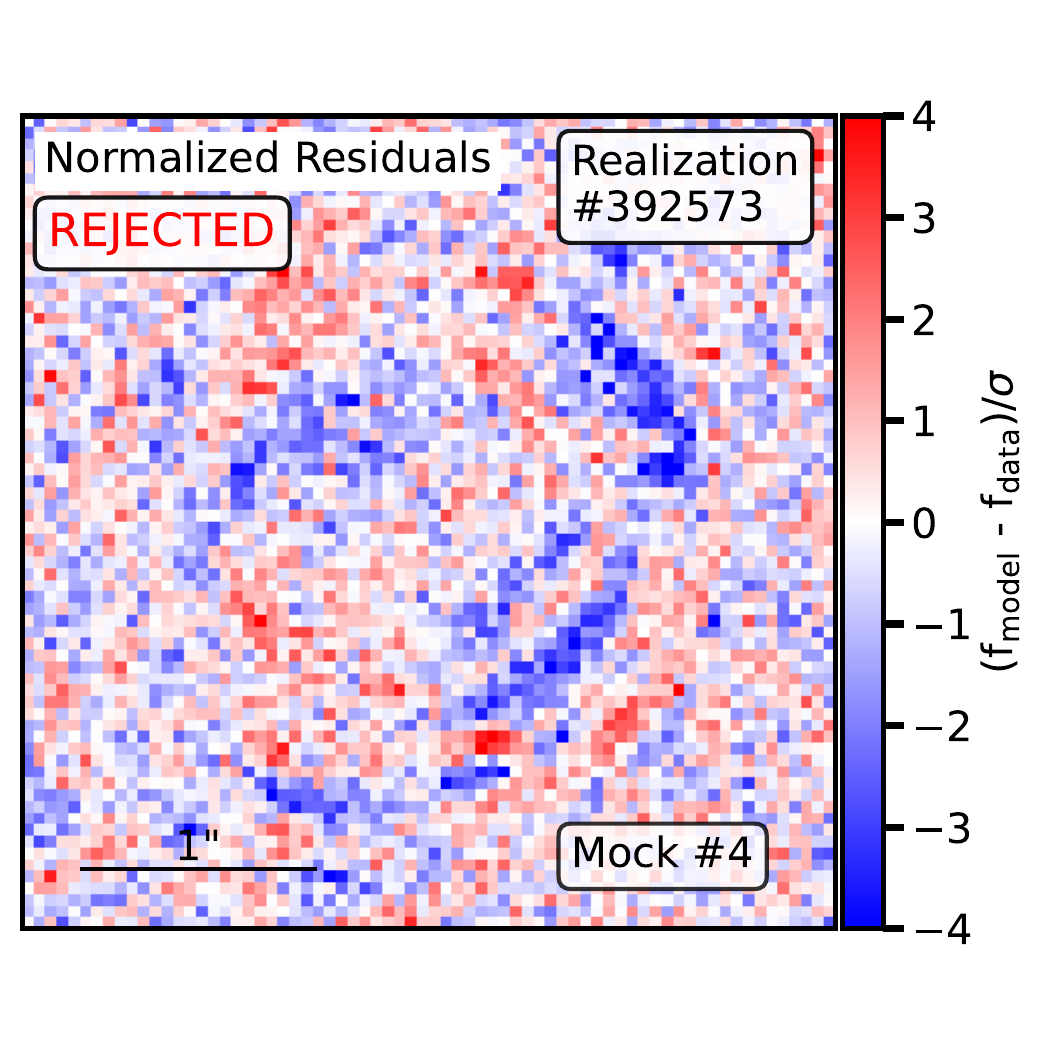}
			\includegraphics[trim=0.2cm 2.5cm 0.25cm
			1.5cm,width=0.33\textwidth]{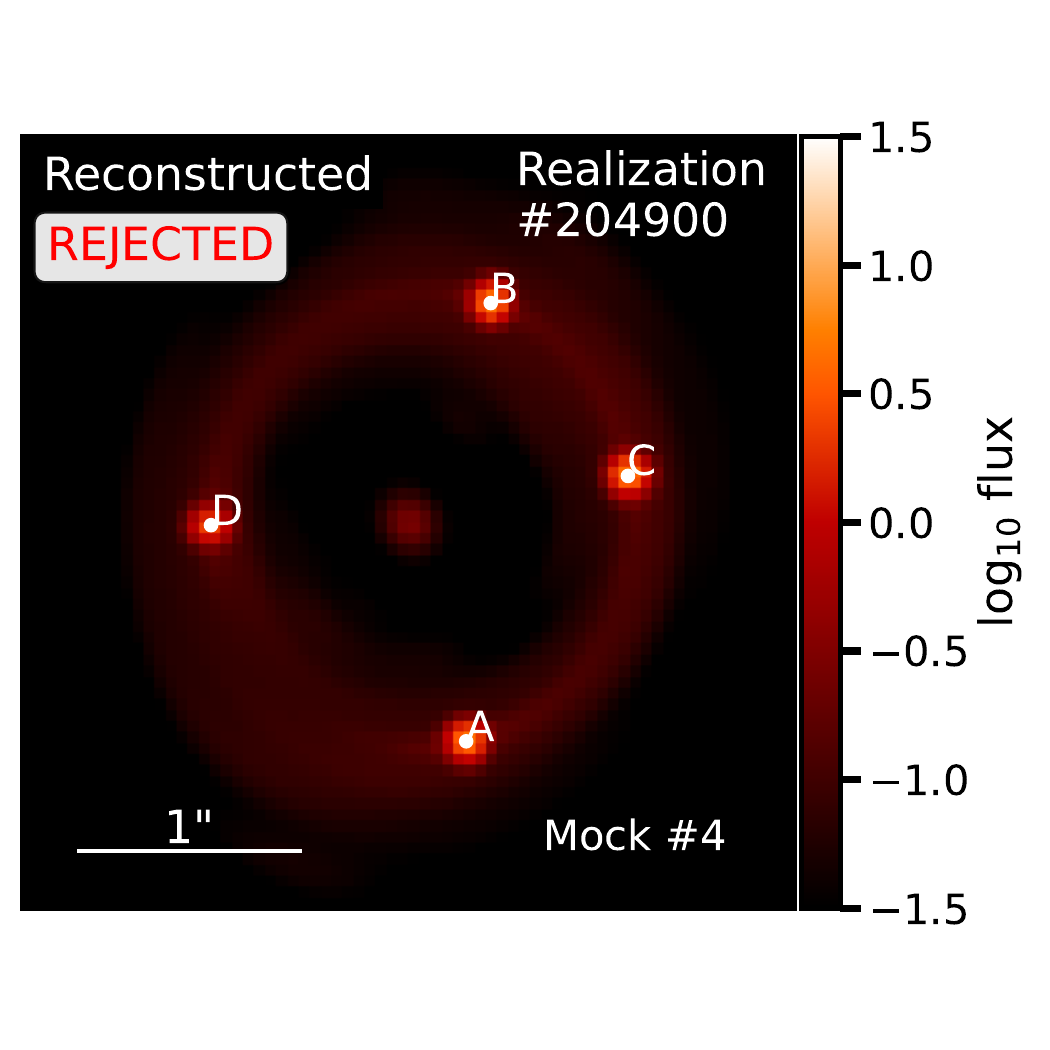}
			\includegraphics[trim=0.2cm 2.5cm 0.25cm
			1.5cm,width=0.33\textwidth]{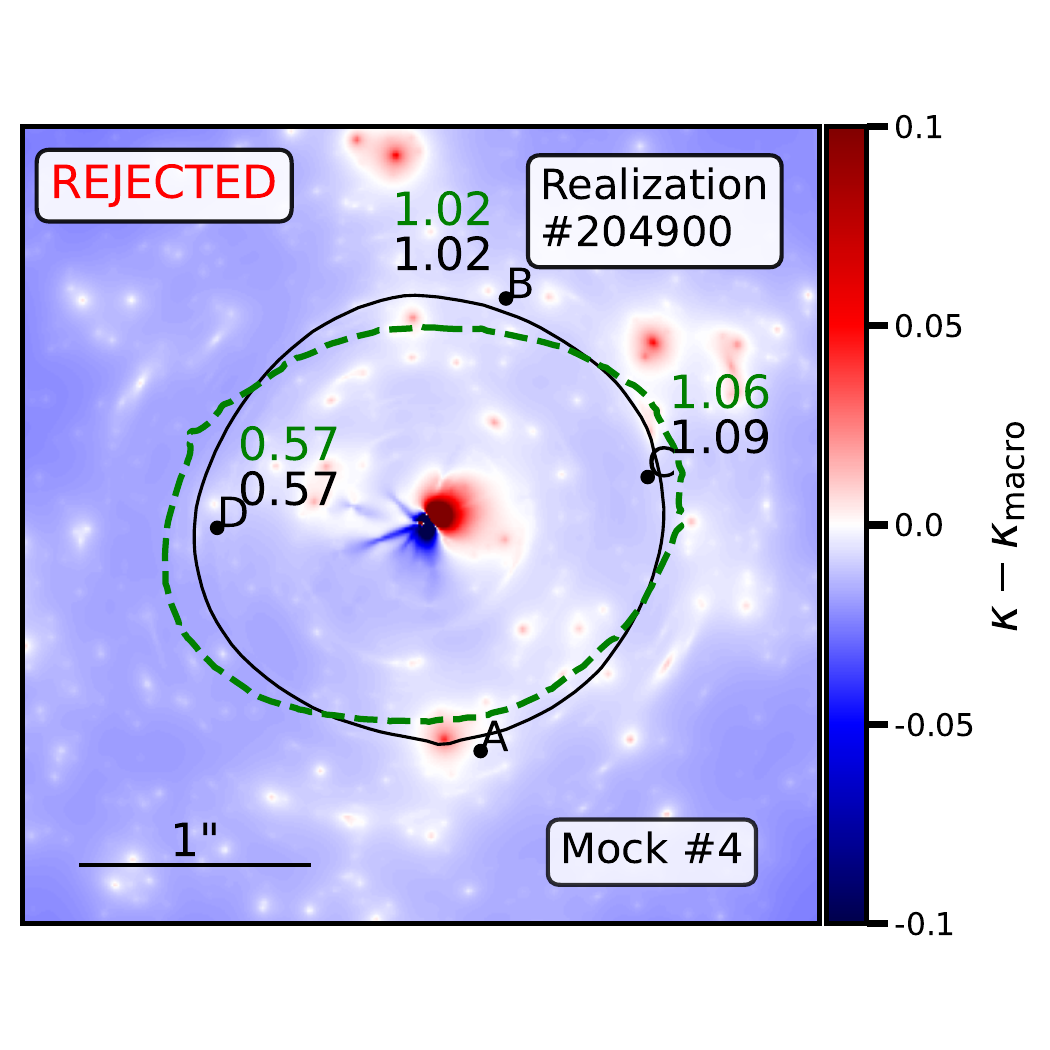}
			\includegraphics[trim=0.2cm 2.5cm 0.25cm
			1.5cm,width=0.33\textwidth]{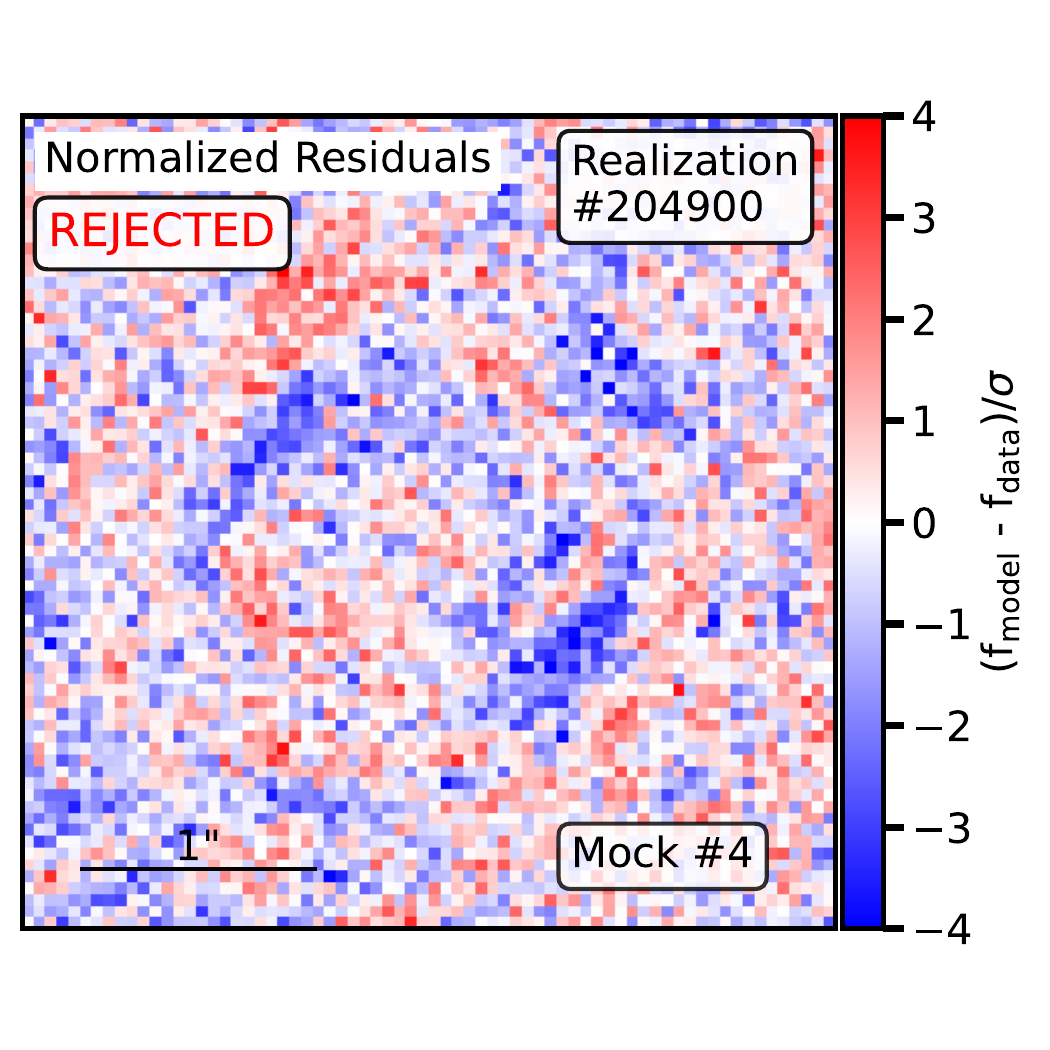}
			\caption{\label{fig:mock4cases} Reconstructed lensed image (left), substructure convergence (center), and normalized residual map from the reconstructed imaging data of Mock $4$ in the CDM ground truth sample. The top two rows depict realizations accepted based matching the image positions, flux ratios, and imaging data. The bottom rows show examples of systems that match the flux ratios, but which we reject due to a poor fit to the imaging data. The green (black) numbers and curves show the true (model-predicted) flux ratios and critical curves, respectively.}
		\end{figure*}
		\begin{figure*}
			\includegraphics[trim=0.2cm 0.2cm 0.25cm
			2.5cm,width=0.33\textwidth]{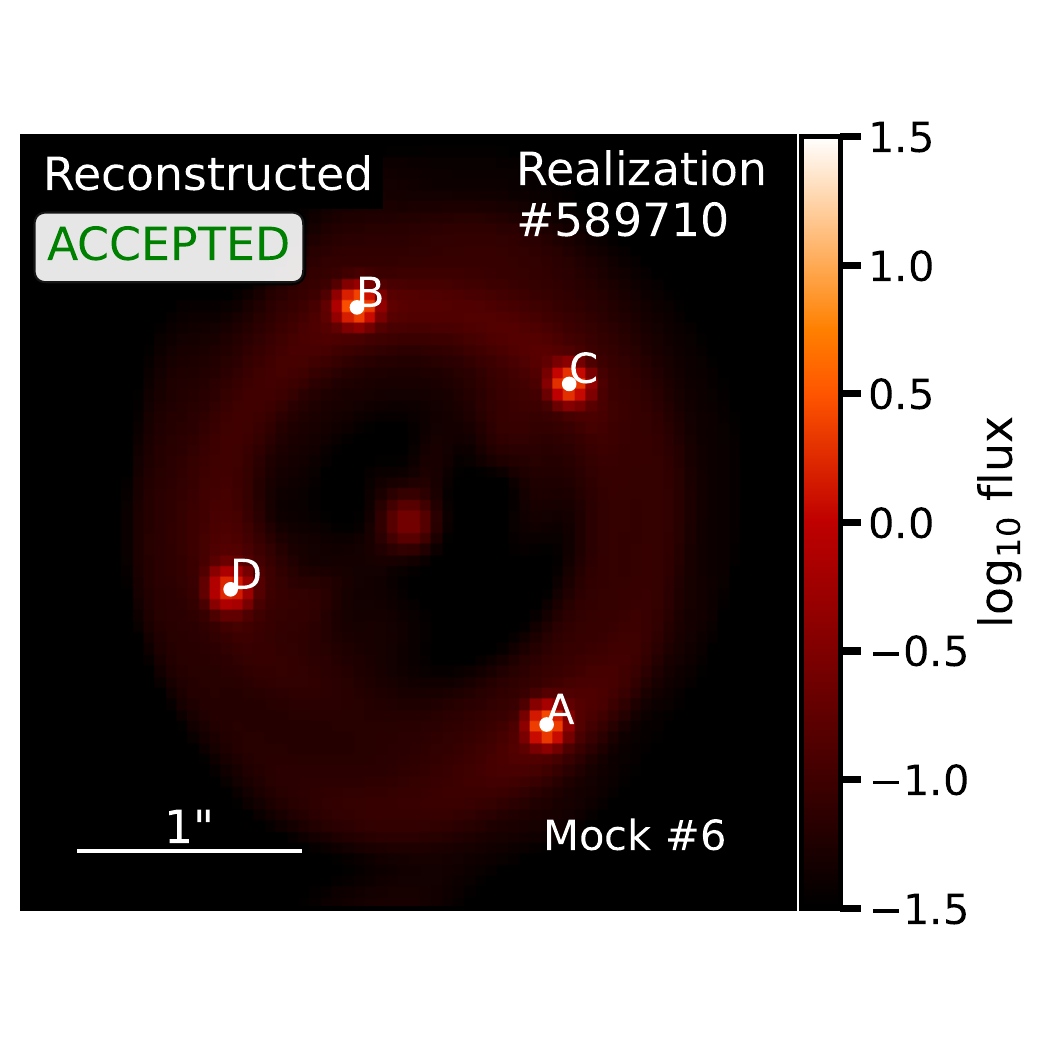}
			\includegraphics[trim=0.2cm 0.2cm 0.25cm
			2.5cm,width=0.33\textwidth]{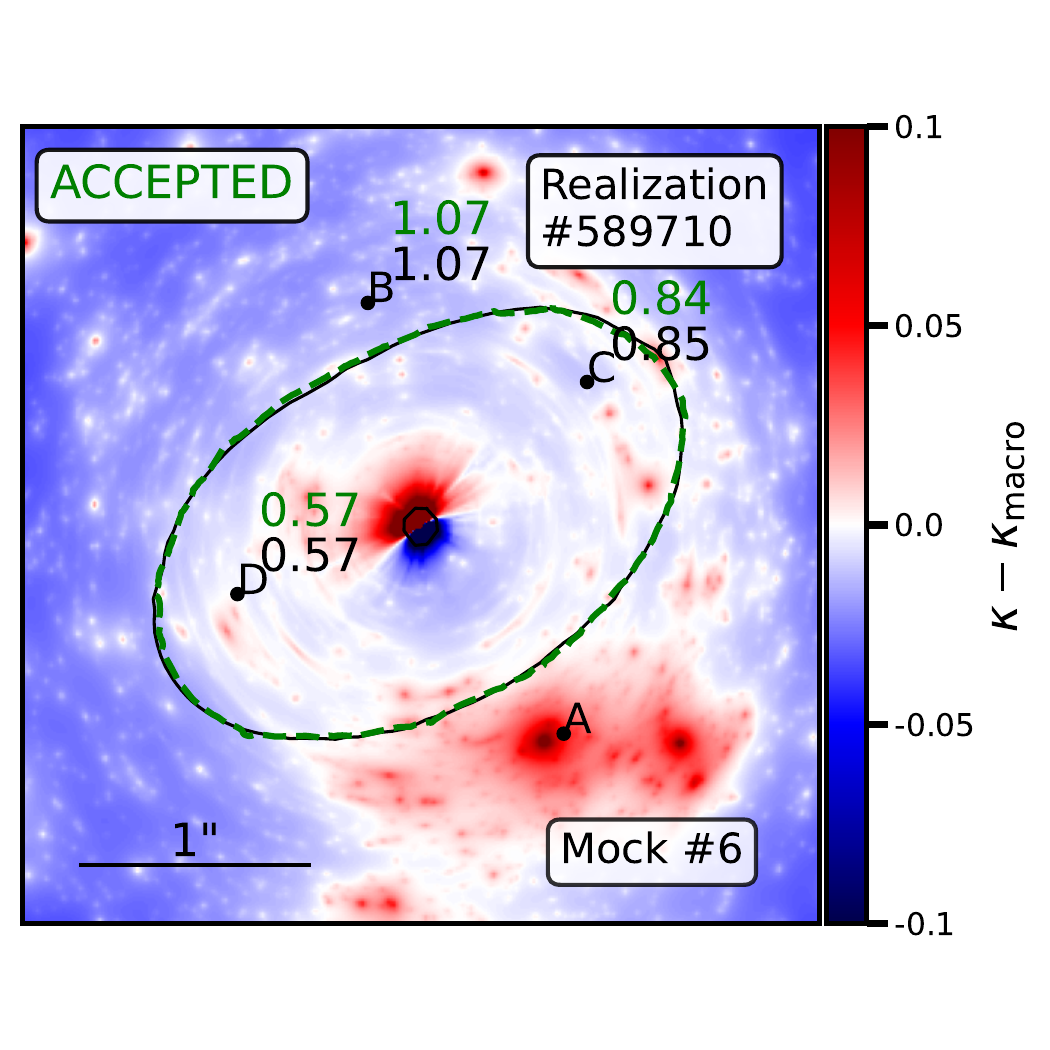}
			\includegraphics[trim=0.2cm 0.2cm 0.25cm
			2.5cm,width=0.33\textwidth]{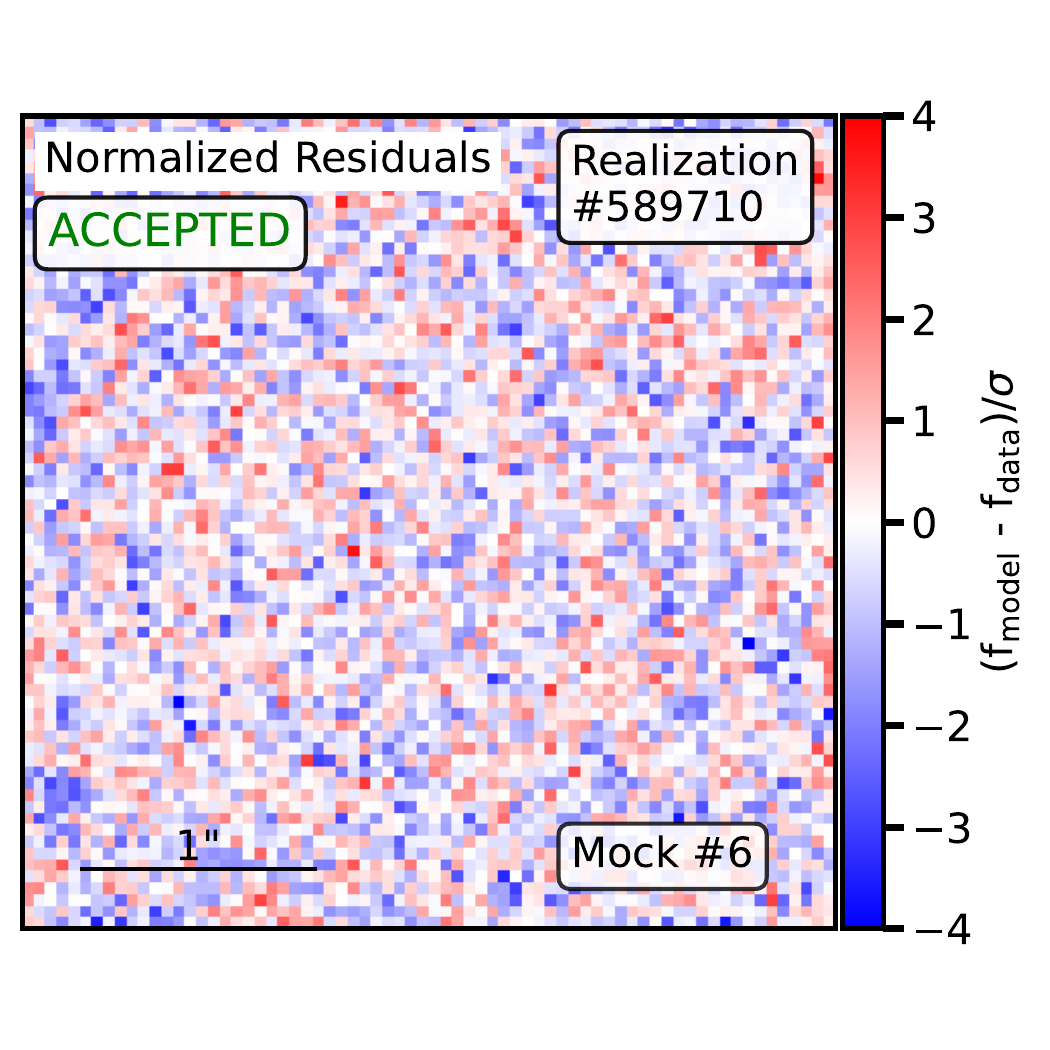}
			\includegraphics[trim=0.2cm 0.2cm 0.25cm
			1.5cm,width=0.33\textwidth]{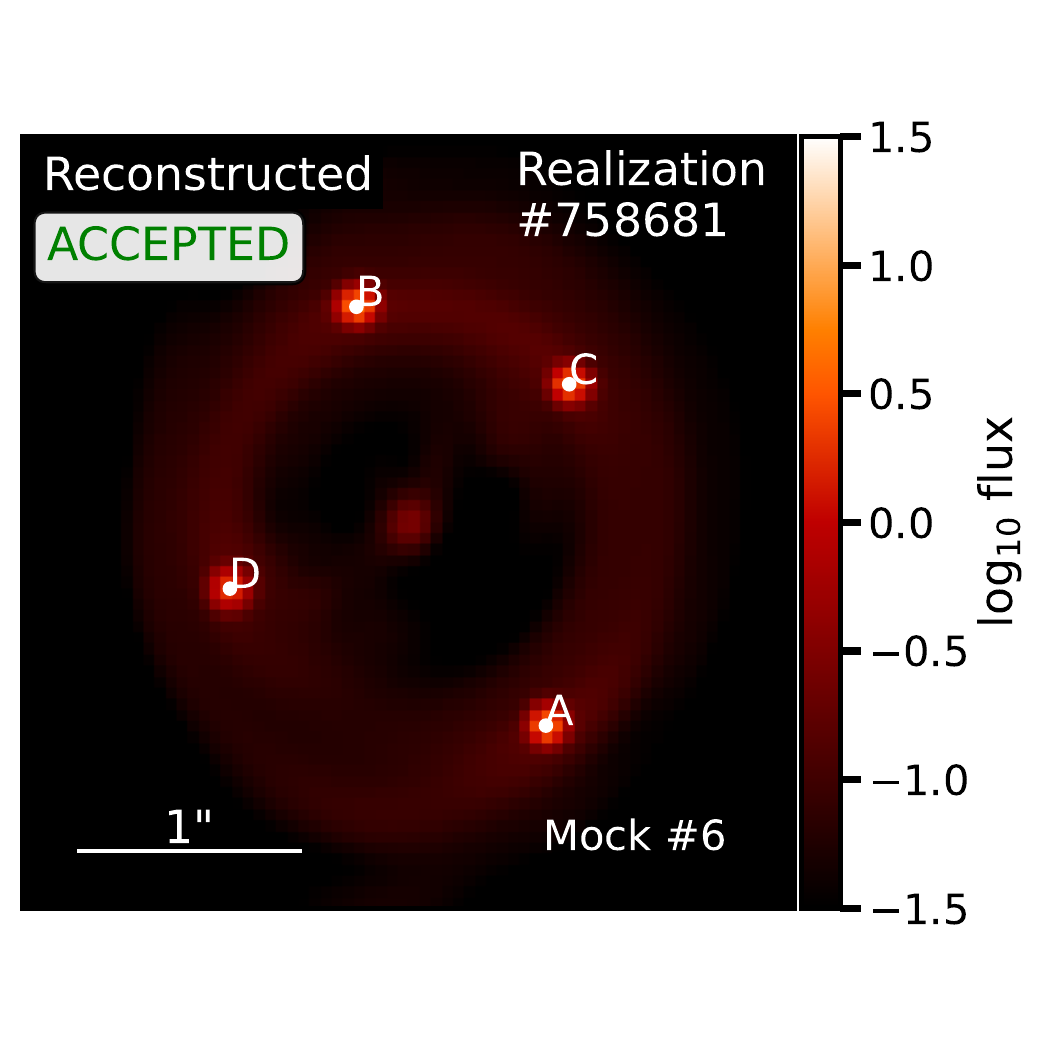}
			\includegraphics[trim=0.2cm 0.2cm 0.25cm
			1.5cm,width=0.33\textwidth]{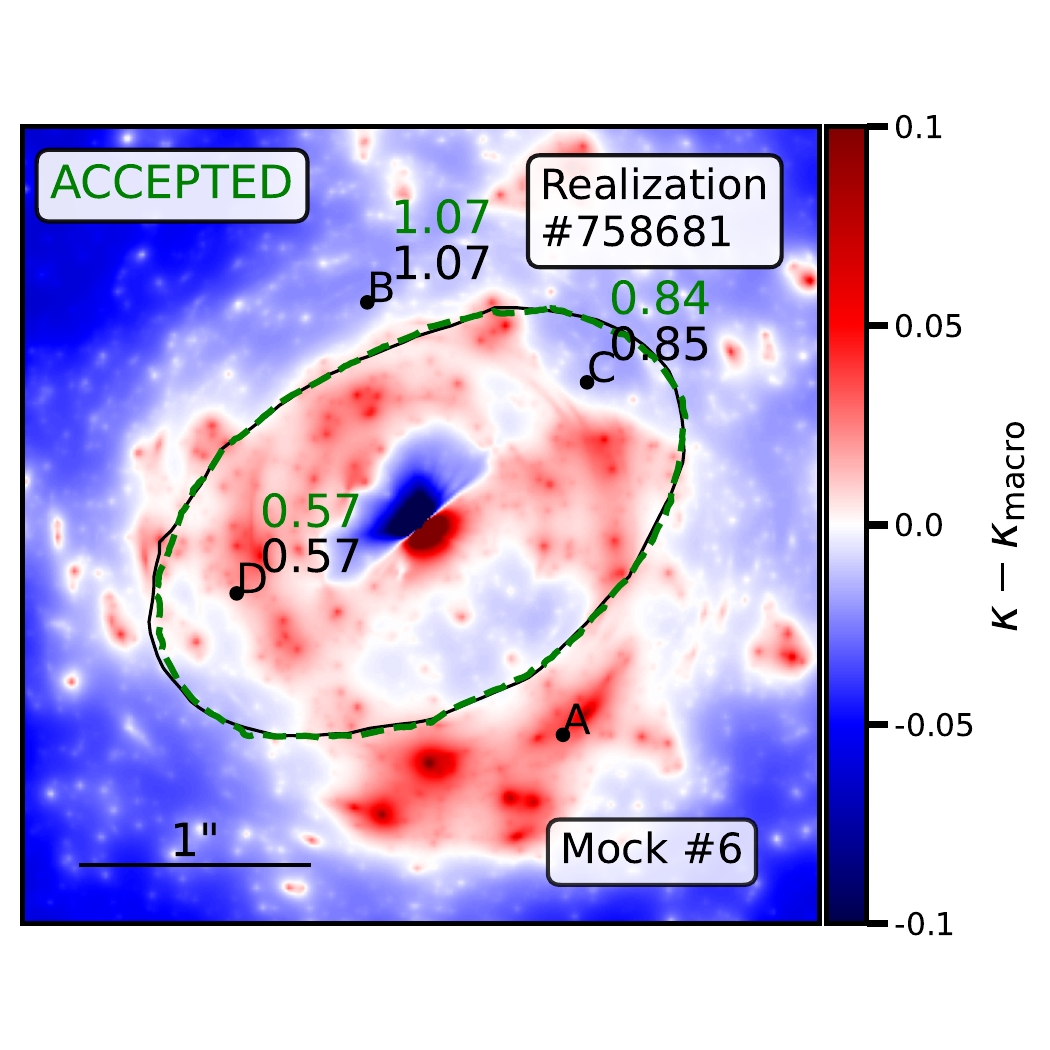}
			\includegraphics[trim=0.2cm 0.2cm 0.25cm
			1.5cm,width=0.33\textwidth]{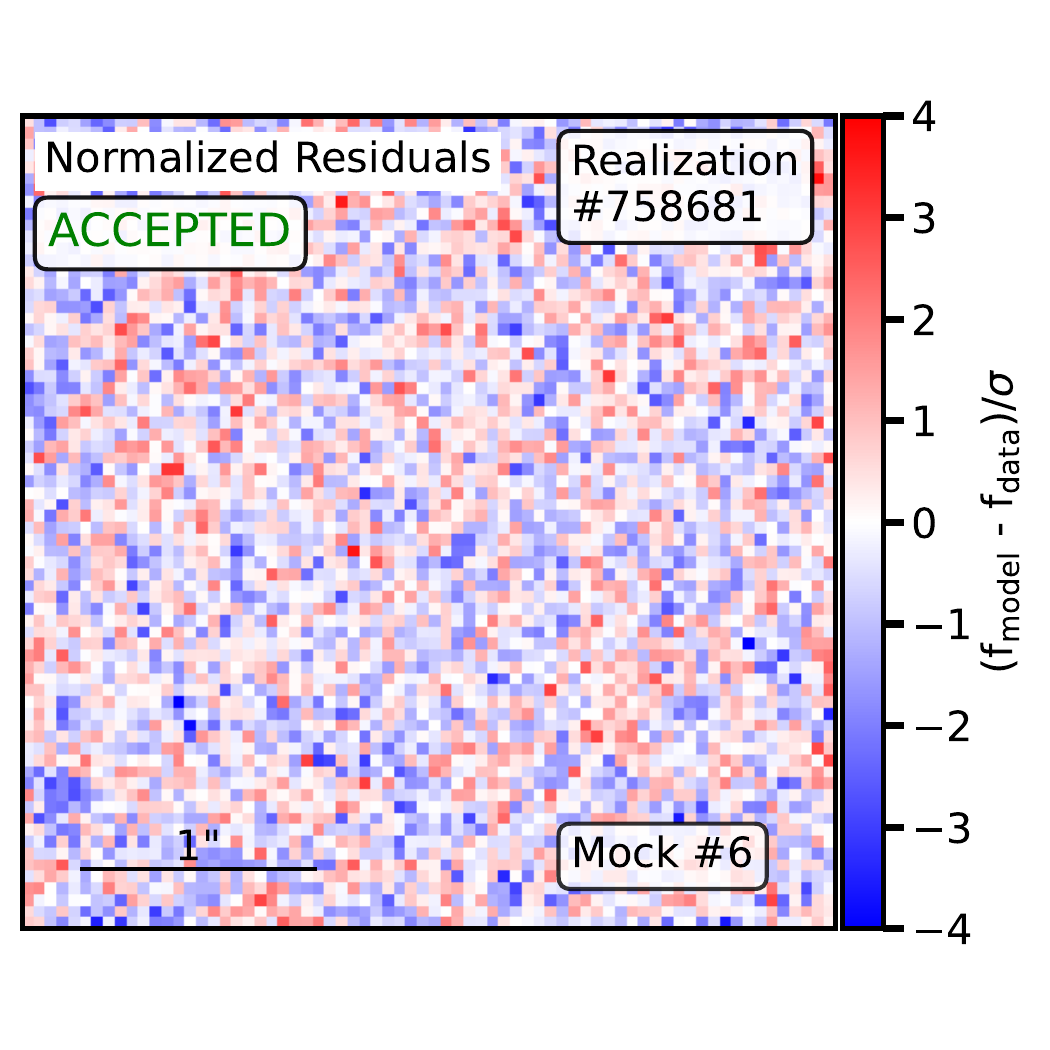}
			\includegraphics[trim=0.2cm 0.2cm 0.25cm
			1.5cm,width=0.33\textwidth]{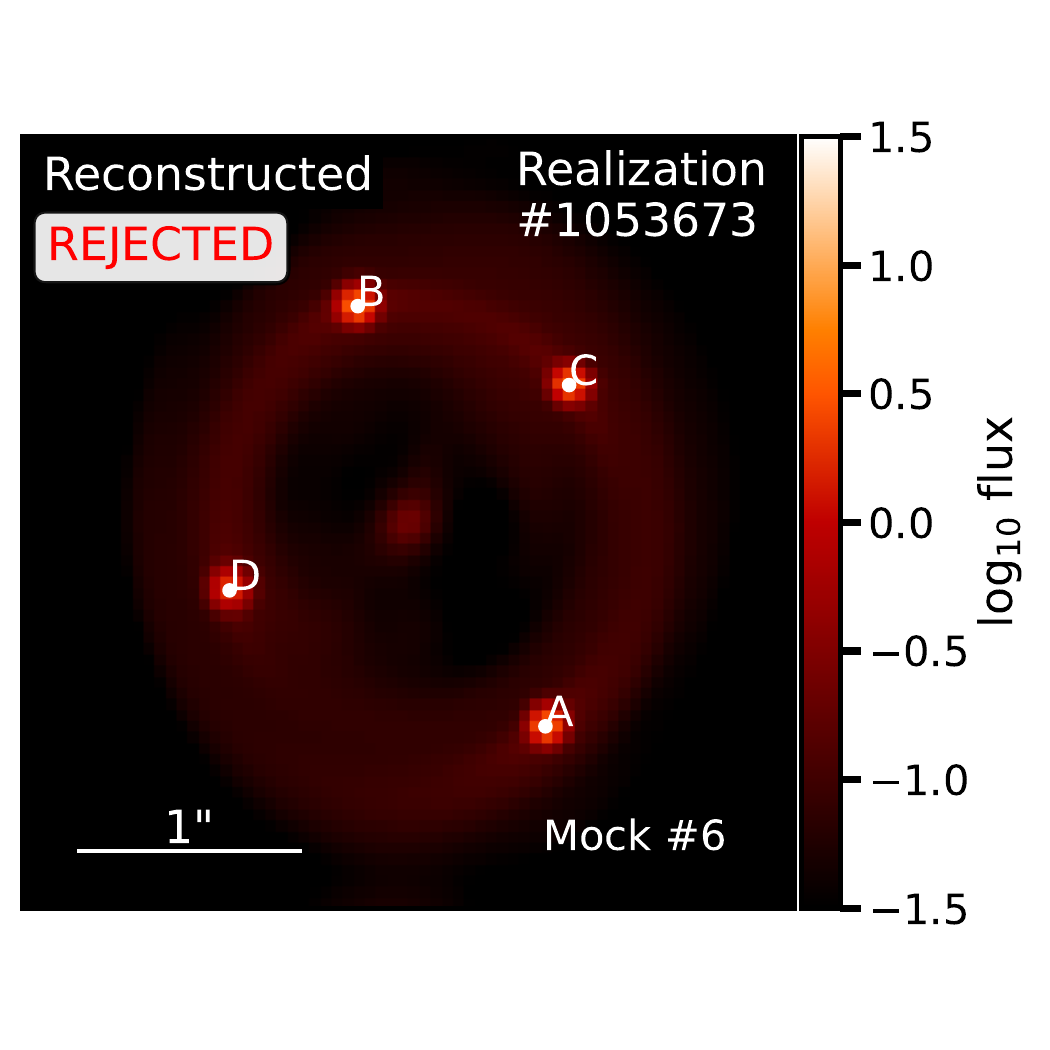}
			\includegraphics[trim=0.2cm 0.2cm 0.25cm
			1.5cm,width=0.33\textwidth]{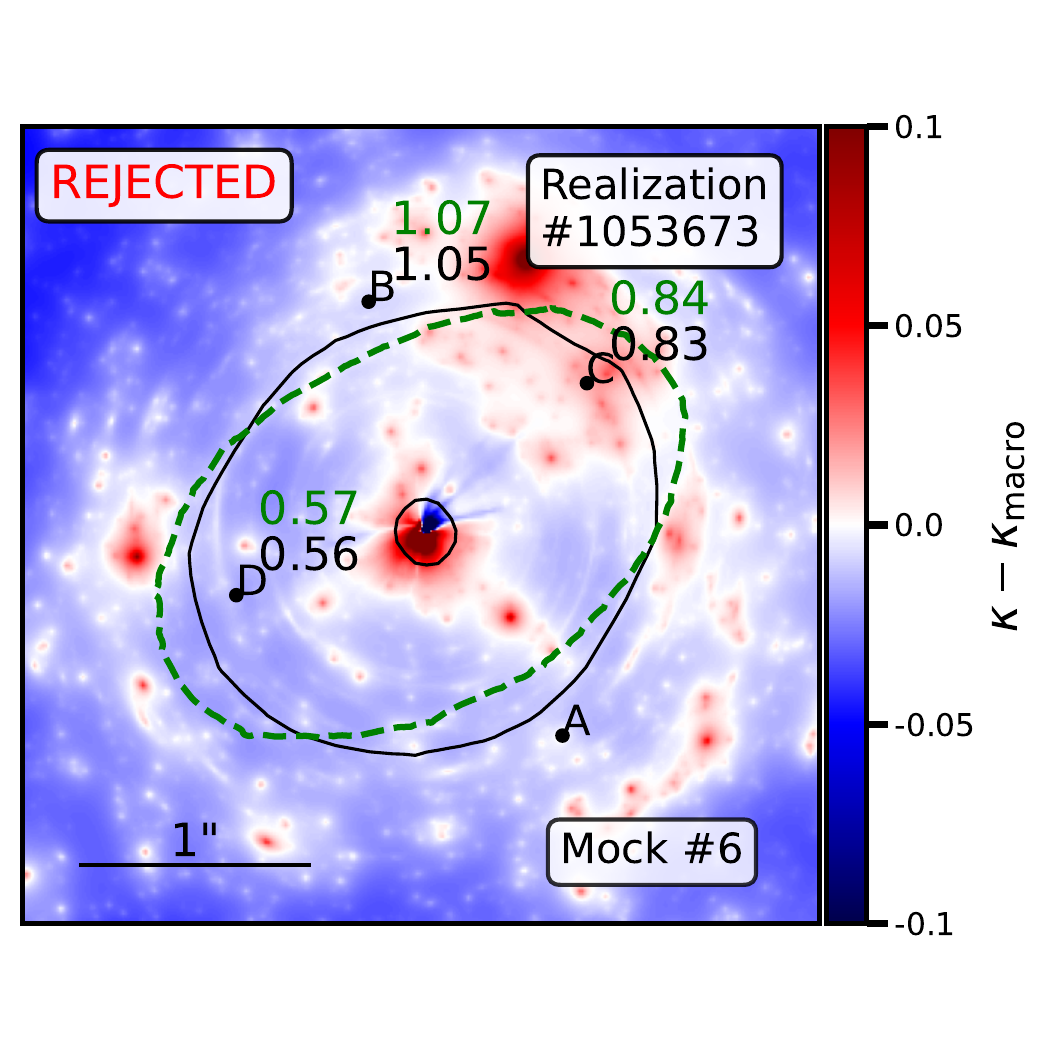}
			\includegraphics[trim=0.2cm 0.2cm 0.25cm
			1.5cm,width=0.33\textwidth]{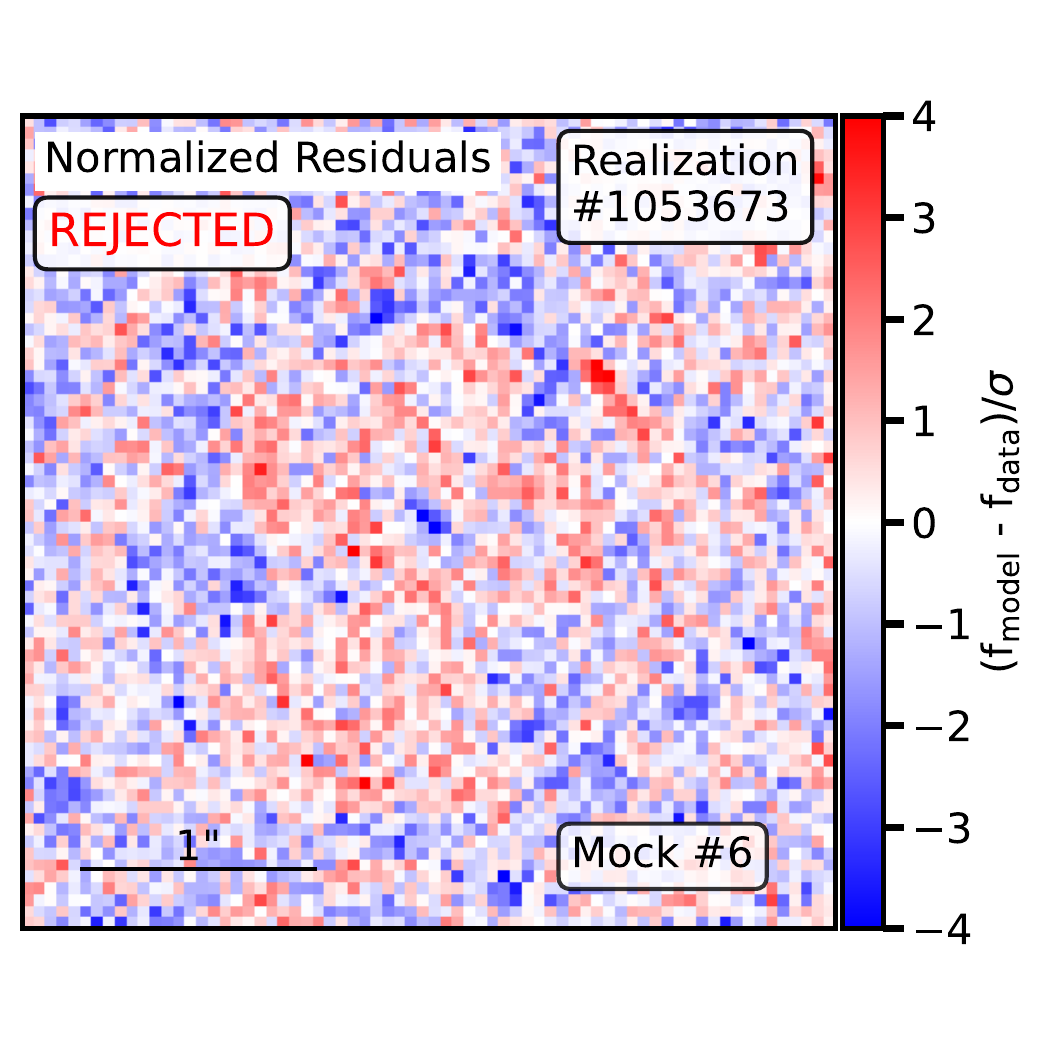}
			\includegraphics[trim=0.2cm 2.5cm 0.25cm
			1.5cm,width=0.33\textwidth]{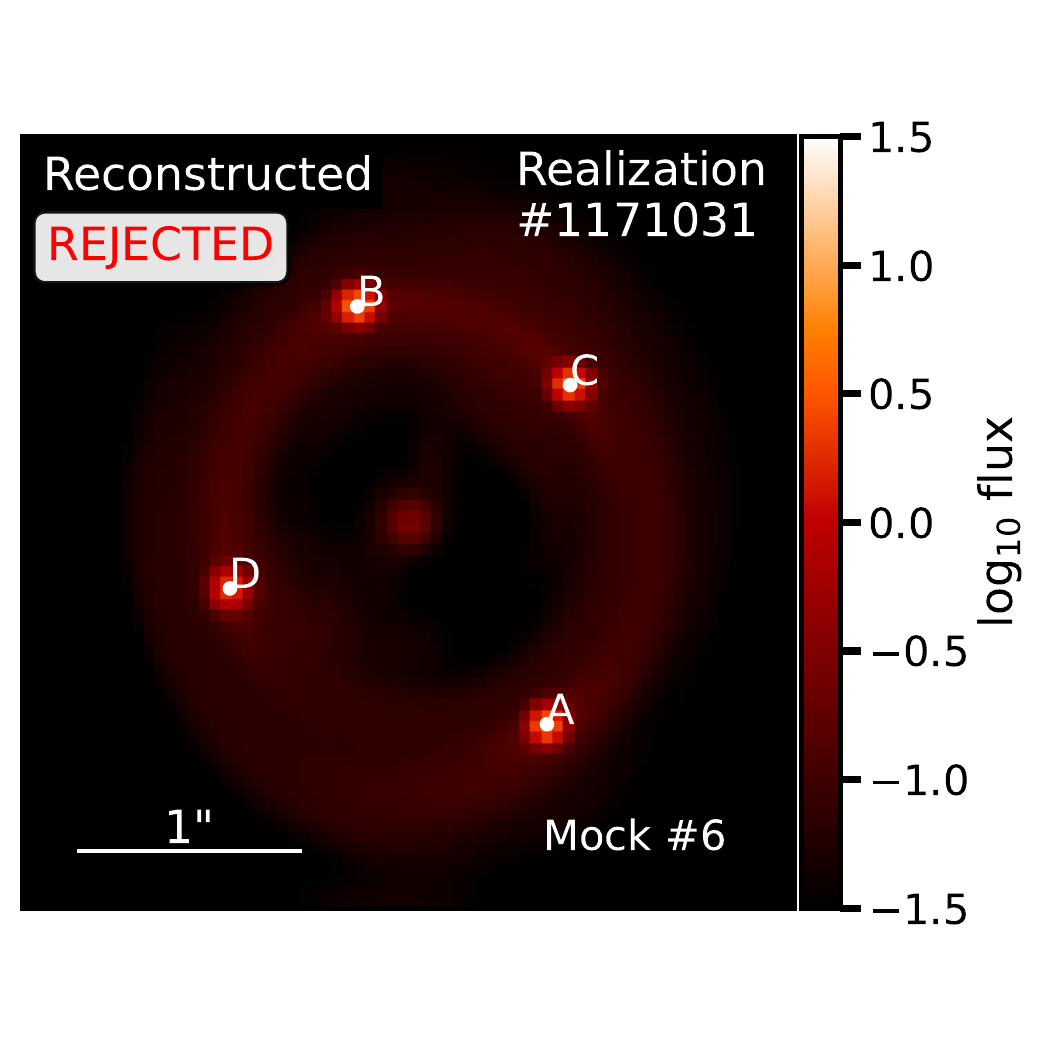}
			\includegraphics[trim=0.2cm 2.5cm 0.25cm
			1.5cm,width=0.33\textwidth]{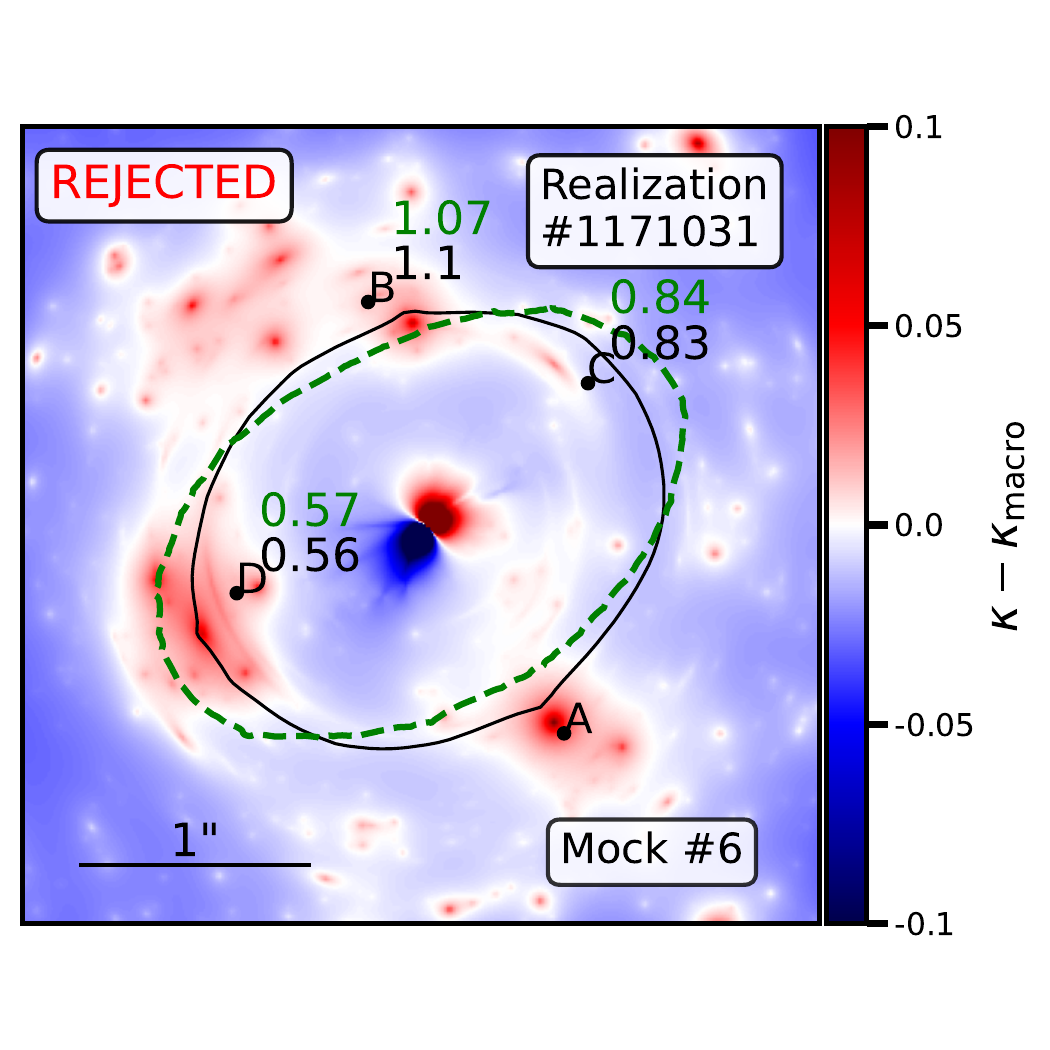}
			\includegraphics[trim=0.2cm 2.5cm 0.25cm
			1.5cm,width=0.33\textwidth]{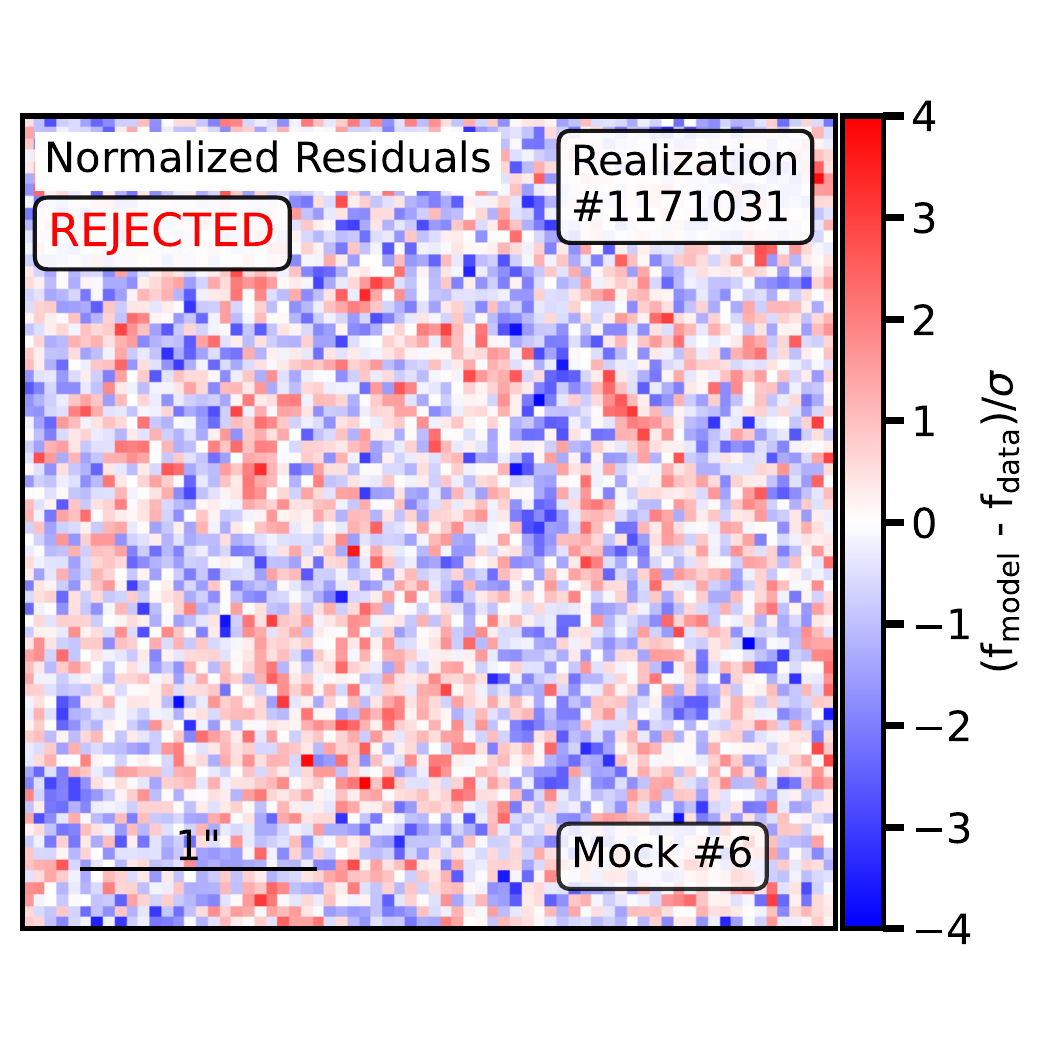}
			\caption{\label{fig:mock6cases} The same as Figure \ref{fig:mock4cases}, but with four example realizations generated for Mock $6$ in the CDM ground truth sample. Simultaneously reproducing the flux ratios and imaging data in this system requires dark matter substructure near image A. As shown by the rejected lens model configuration in the third row, incorporating imaging data allows us to rule out lens model configurations that match the flux ratios through large-scale deformation of the lens mass profile, isolating the effects of substructure on these data.}
		\end{figure*}
		\begin{figure*}
			\includegraphics[trim=0.2cm 0.2cm 0.25cm
			2.5cm,width=0.33\textwidth]{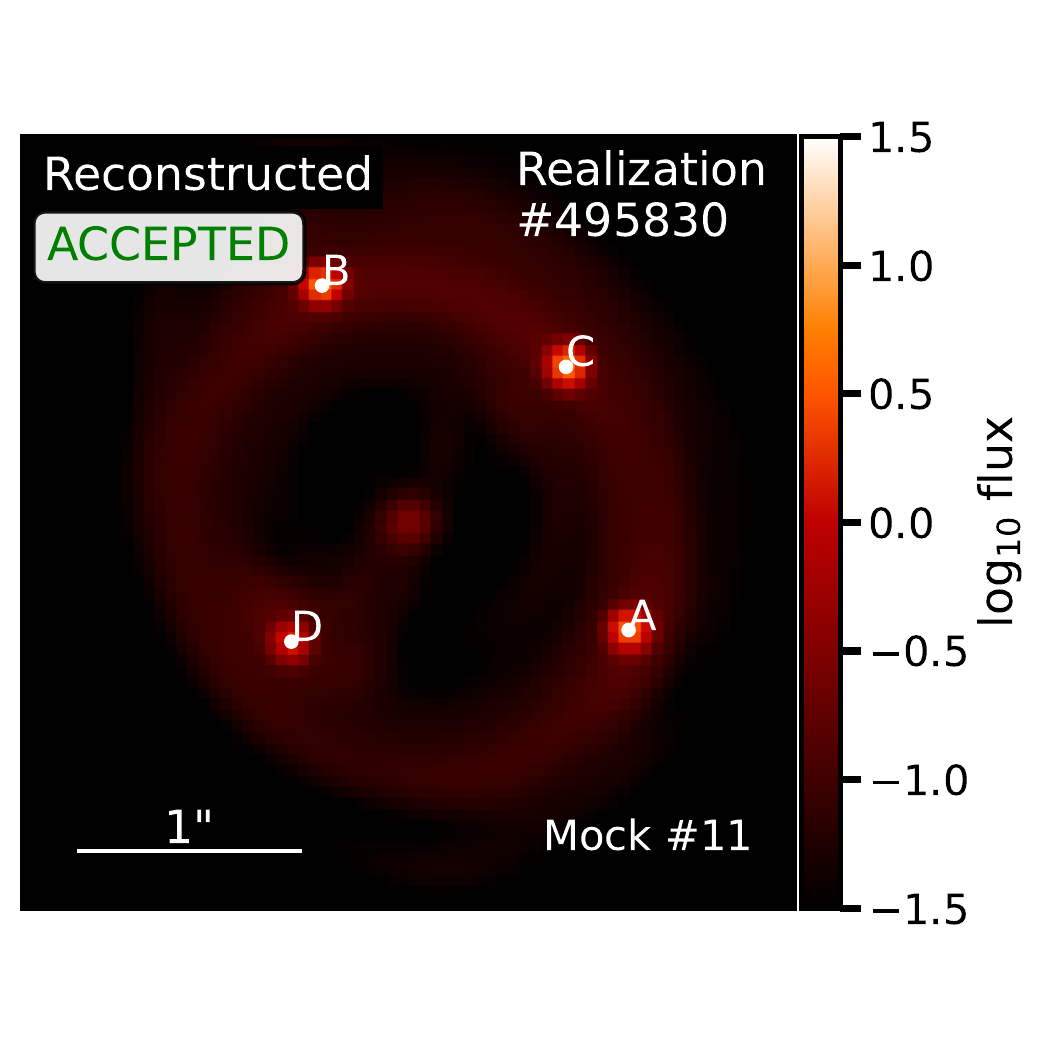}
			\includegraphics[trim=0.2cm 0.2cm 0.25cm
			2.5cm,width=0.33\textwidth]{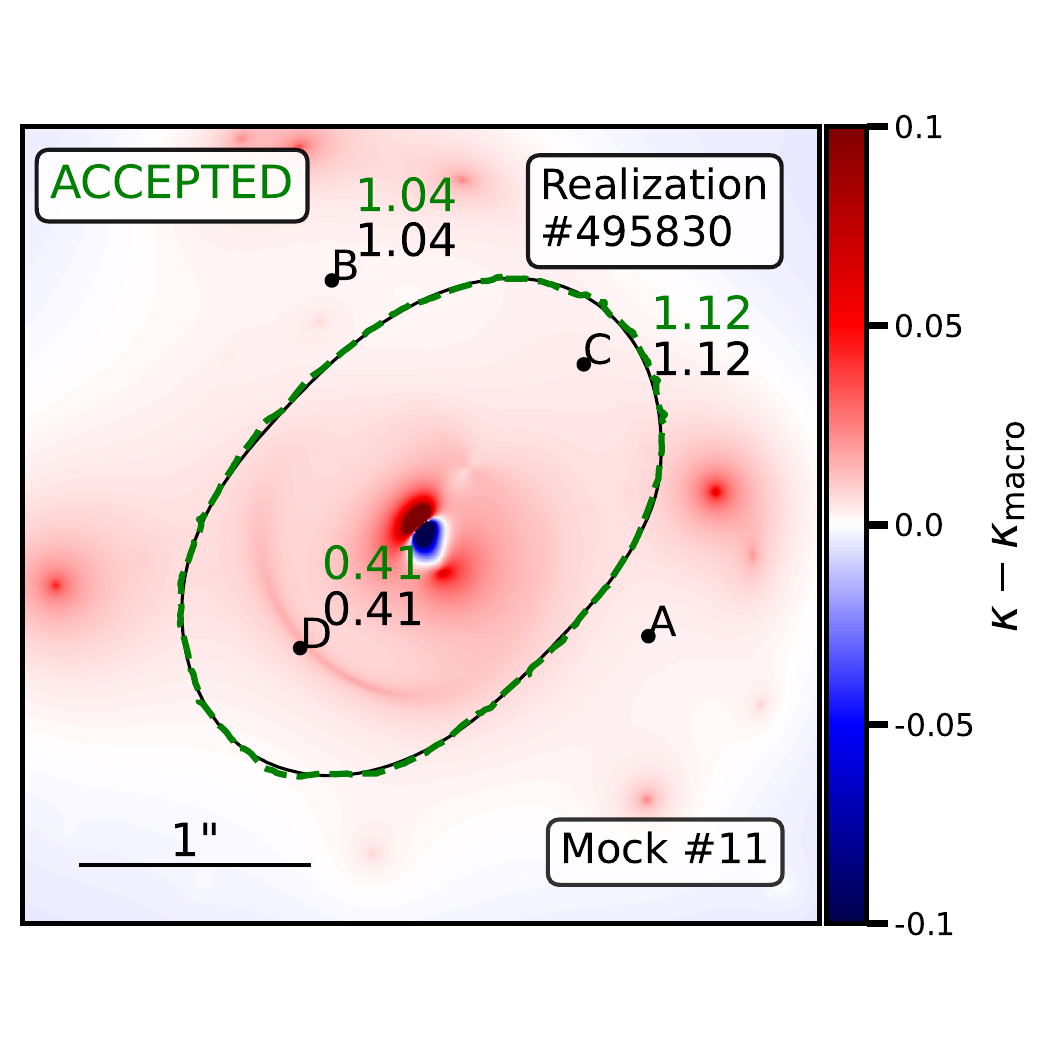}
			\includegraphics[trim=0.2cm 0.2cm 0.25cm
			2.5cm,width=0.33\textwidth]{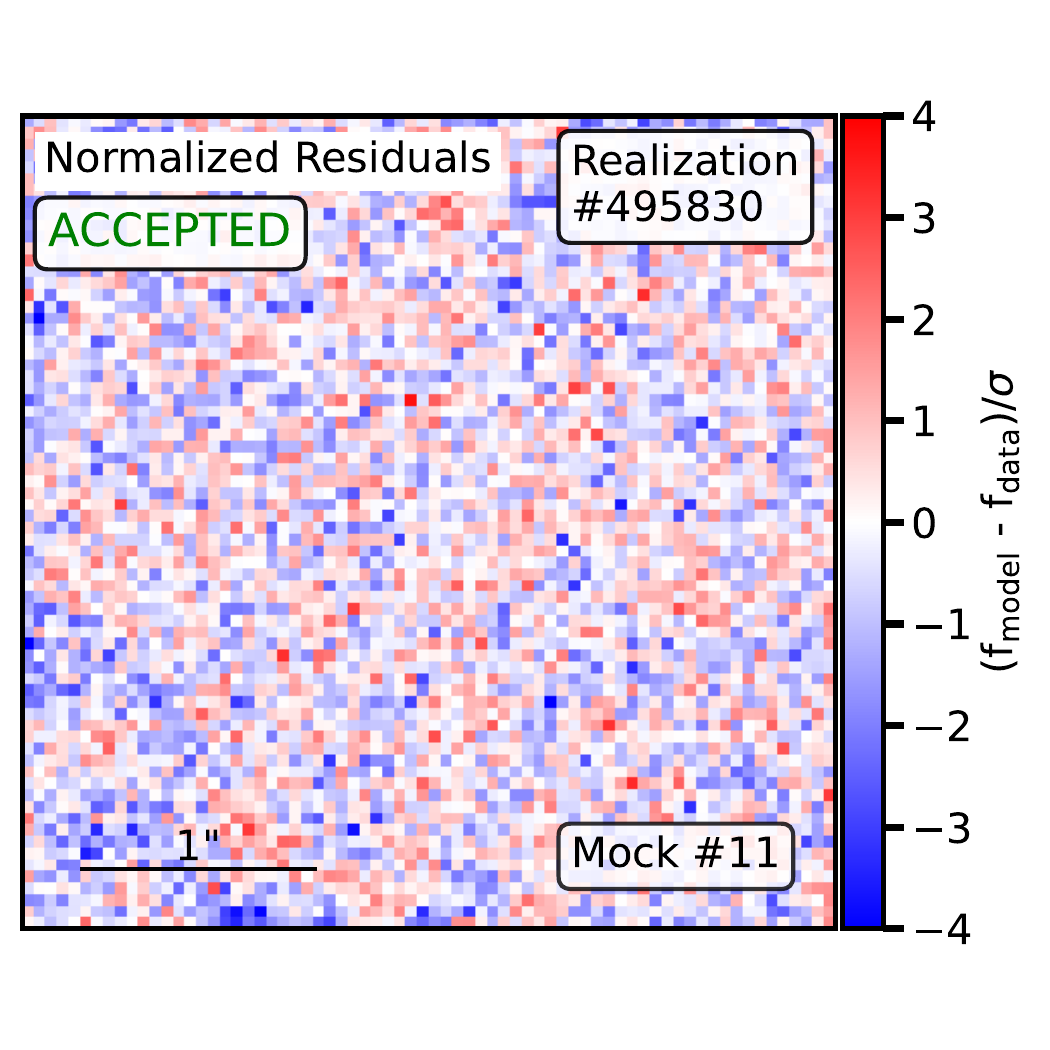}
			\includegraphics[trim=0.2cm 0.2cm 0.25cm
			1.5cm,width=0.33\textwidth]{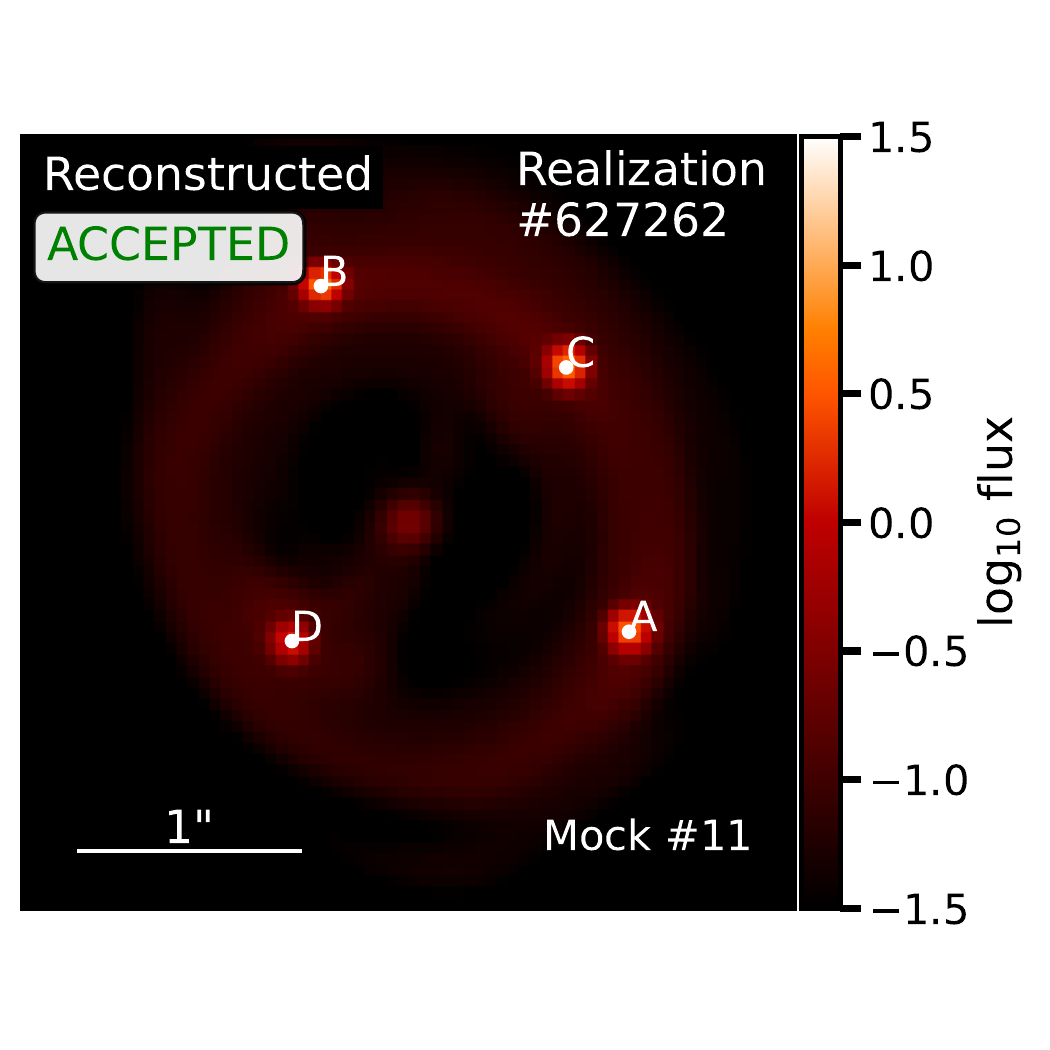}
			\includegraphics[trim=0.2cm 0.2cm 0.25cm
			1.5cm,width=0.33\textwidth]{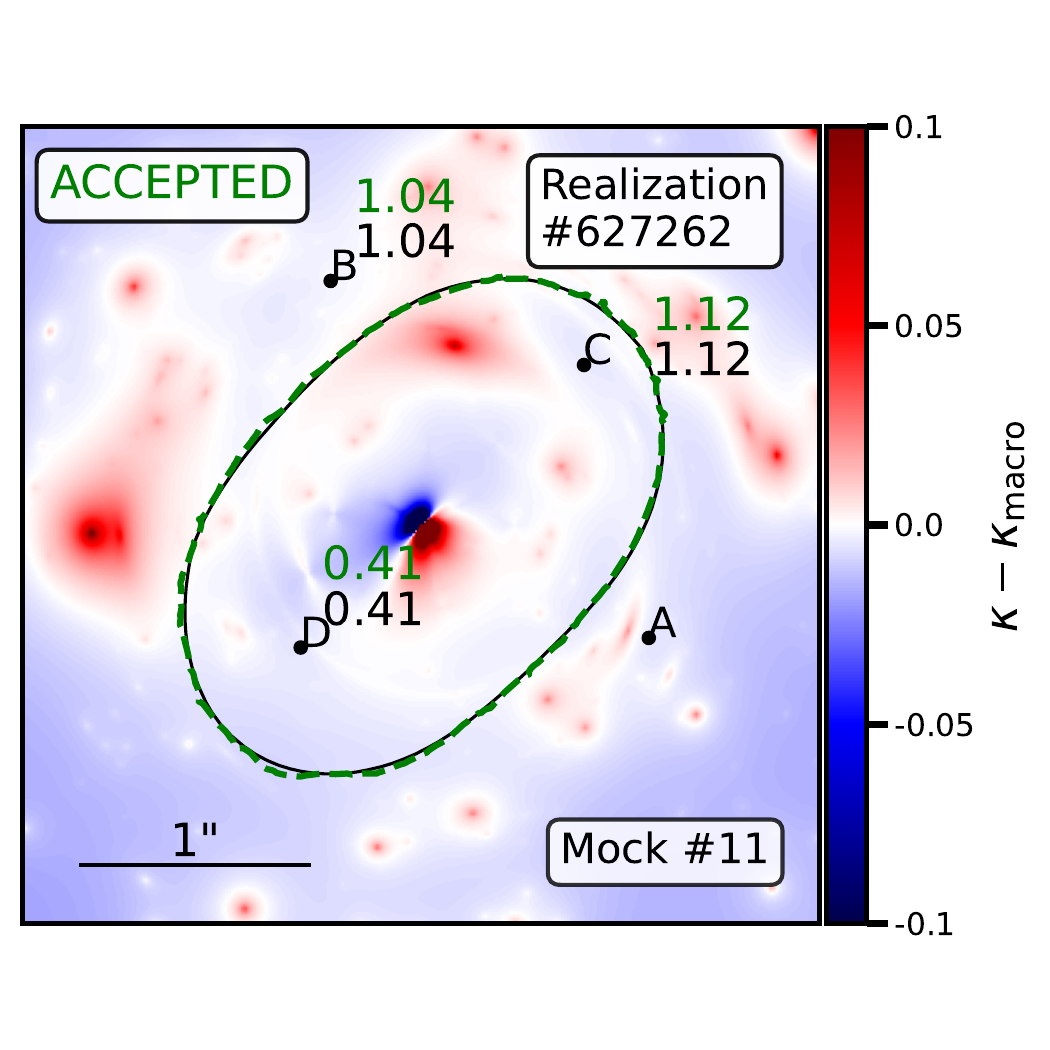}
			\includegraphics[trim=0.2cm 0.2cm 0.25cm
			1.5cm,width=0.33\textwidth]{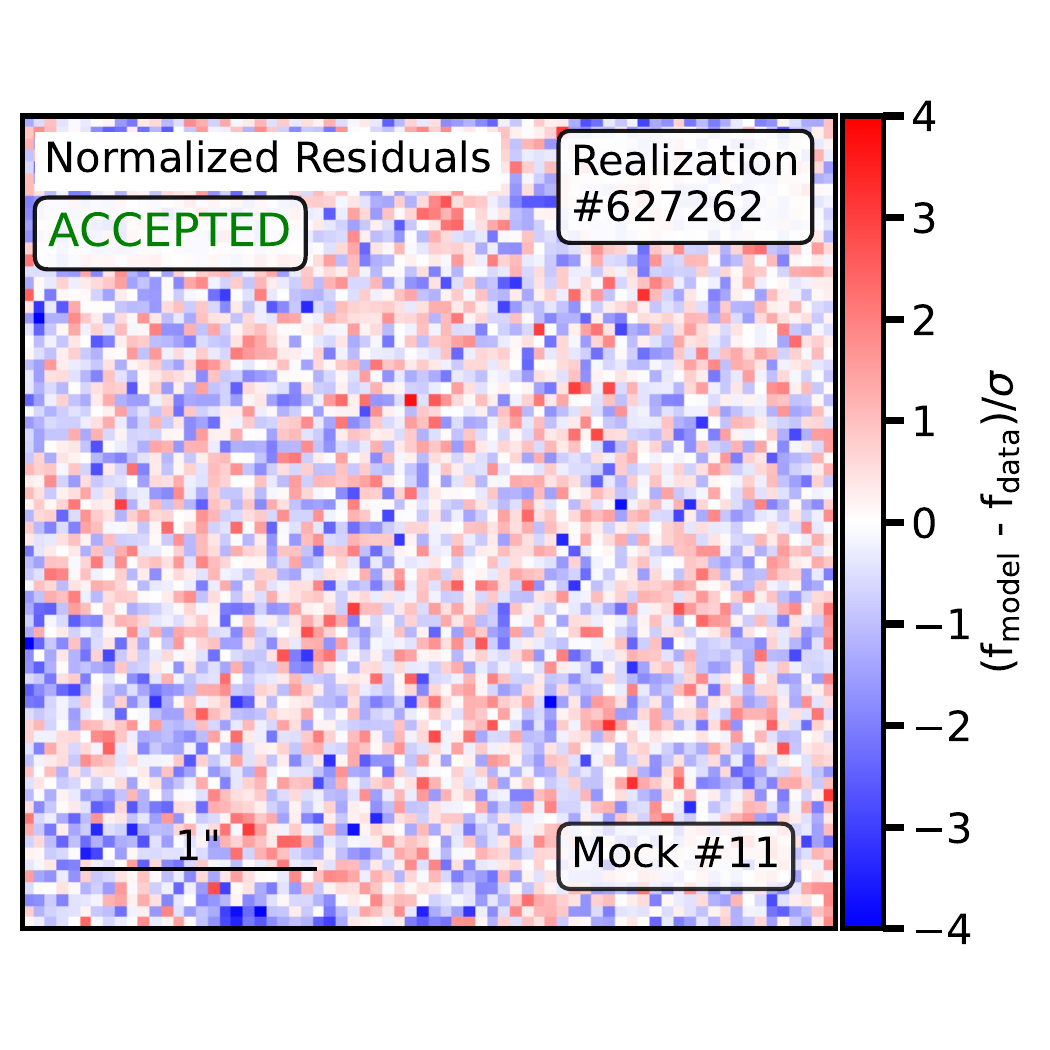}
			\includegraphics[trim=0.2cm 0.2cm 0.25cm
			1.5cm,width=0.33\textwidth]{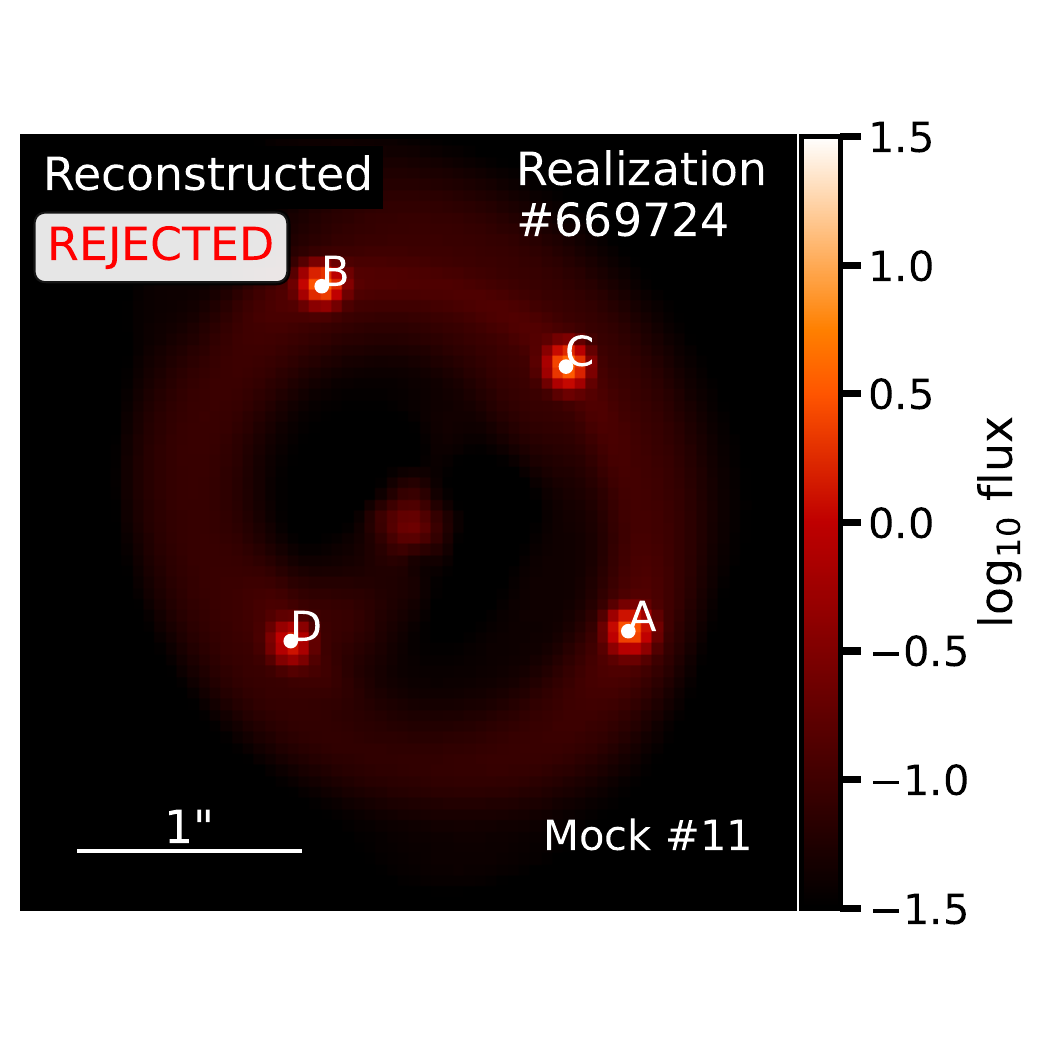}
			\includegraphics[trim=0.2cm 0.2cm 0.25cm
			1.5cm,width=0.33\textwidth]{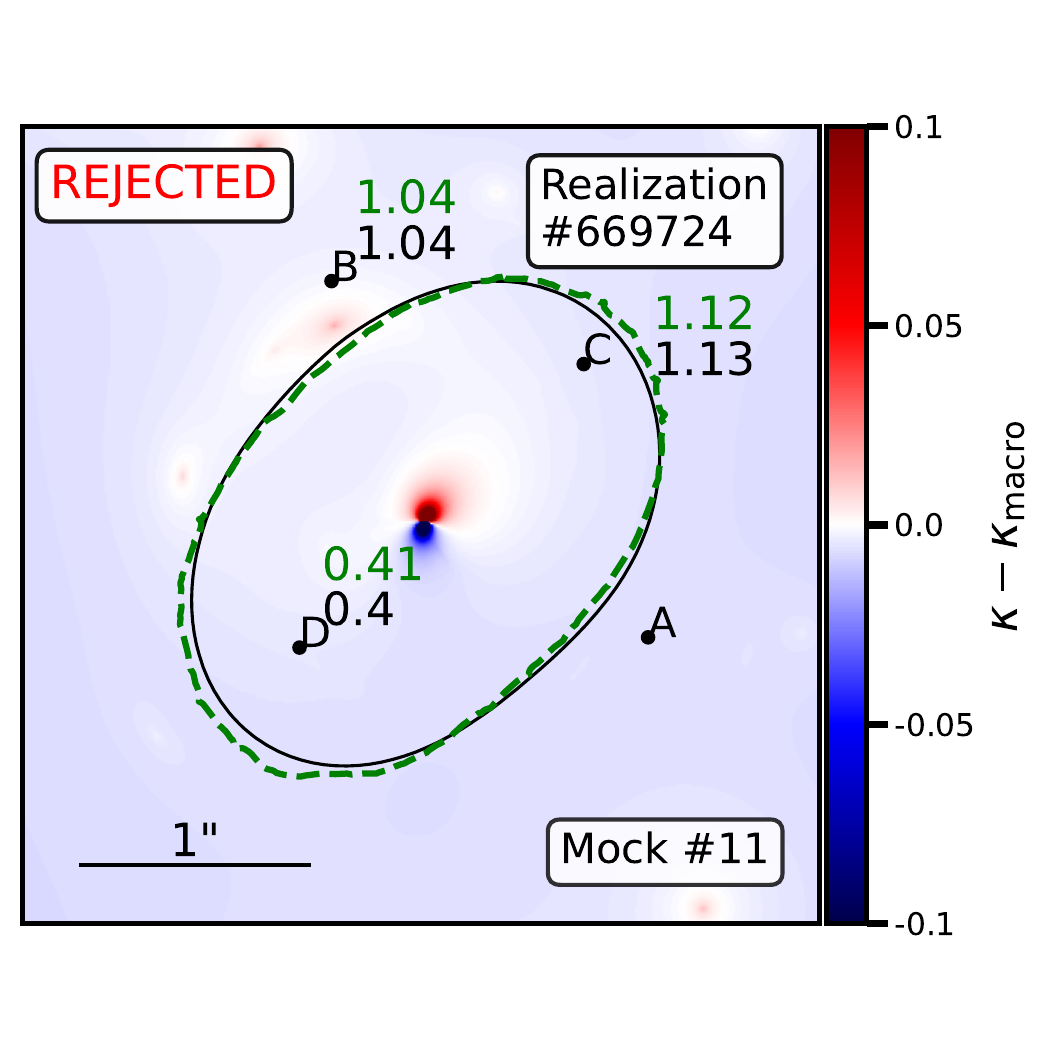}
			\includegraphics[trim=0.2cm 0.2cm 0.25cm
			1.5cm,width=0.33\textwidth]{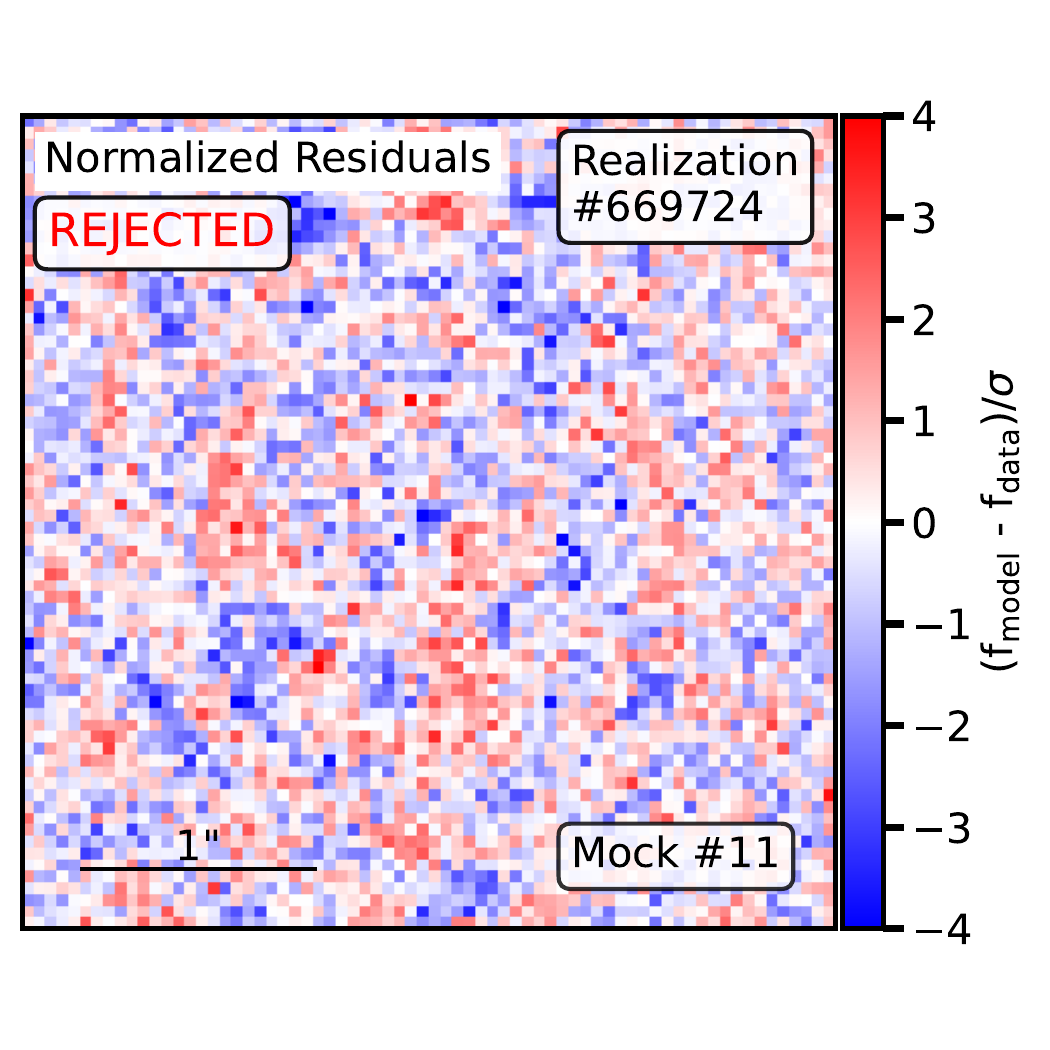}
			\includegraphics[trim=0.2cm 2.5cm 0.25cm
			1.5cm,width=0.33\textwidth]{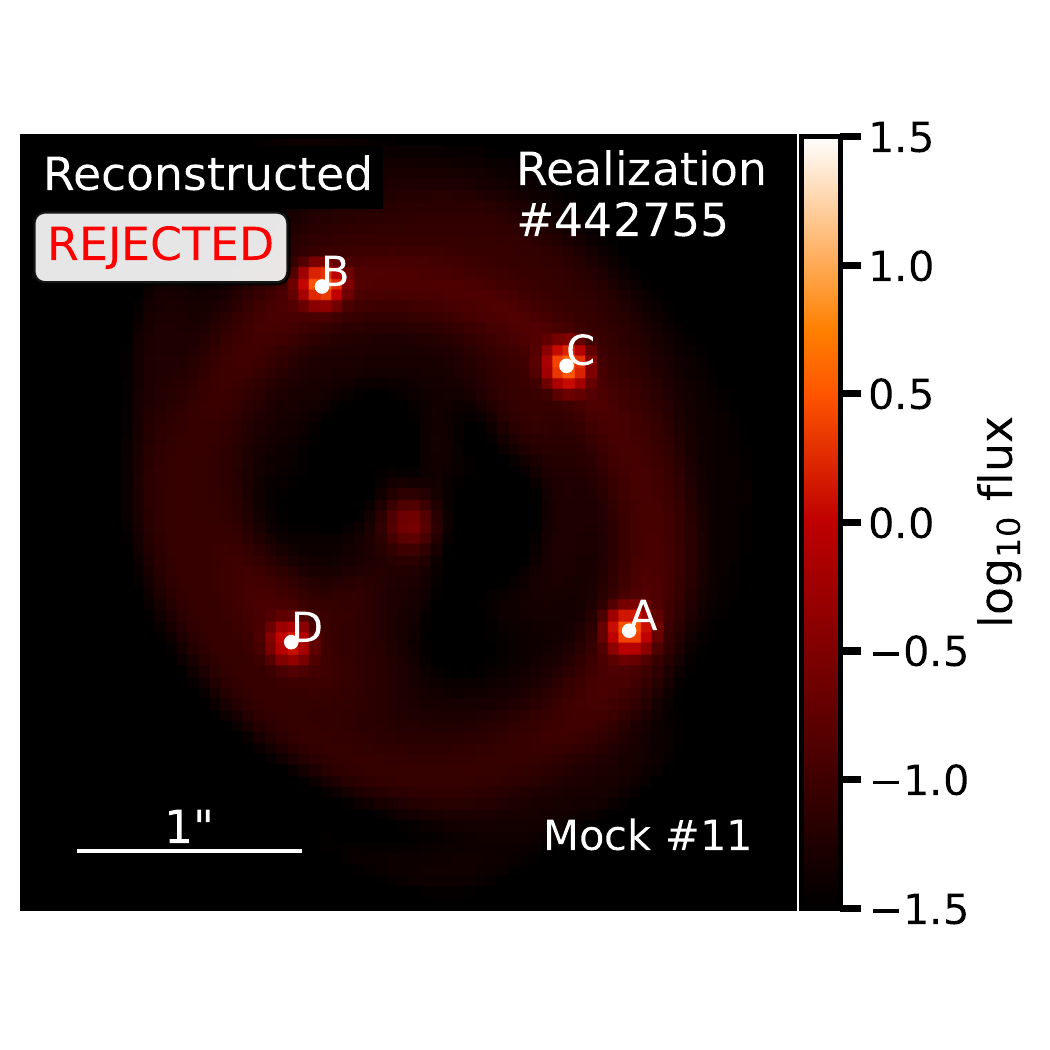}
			\includegraphics[trim=0.2cm 2.5cm 0.25cm
			1.5cm,width=0.33\textwidth]{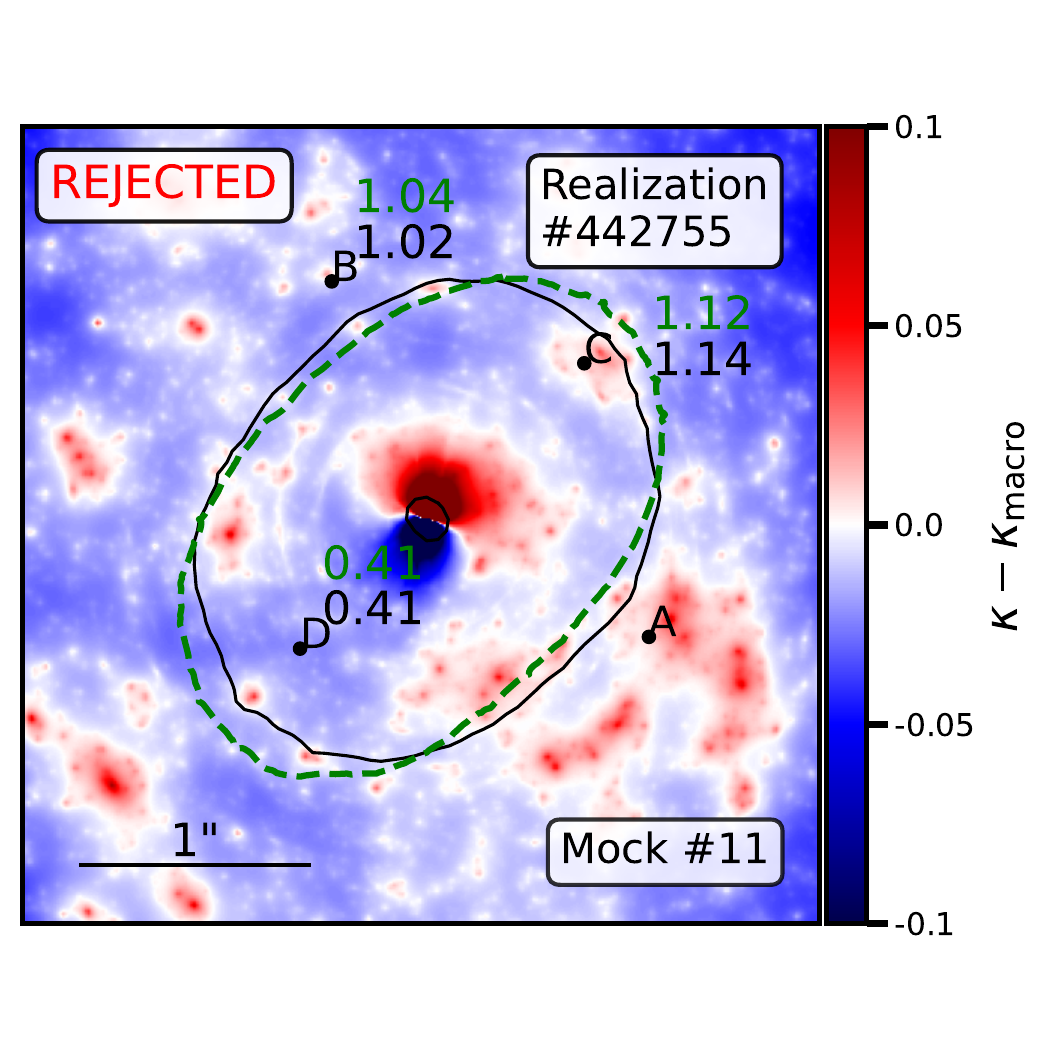}
			\includegraphics[trim=0.2cm 2.5cm 0.25cm
			1.5cm,width=0.33\textwidth]{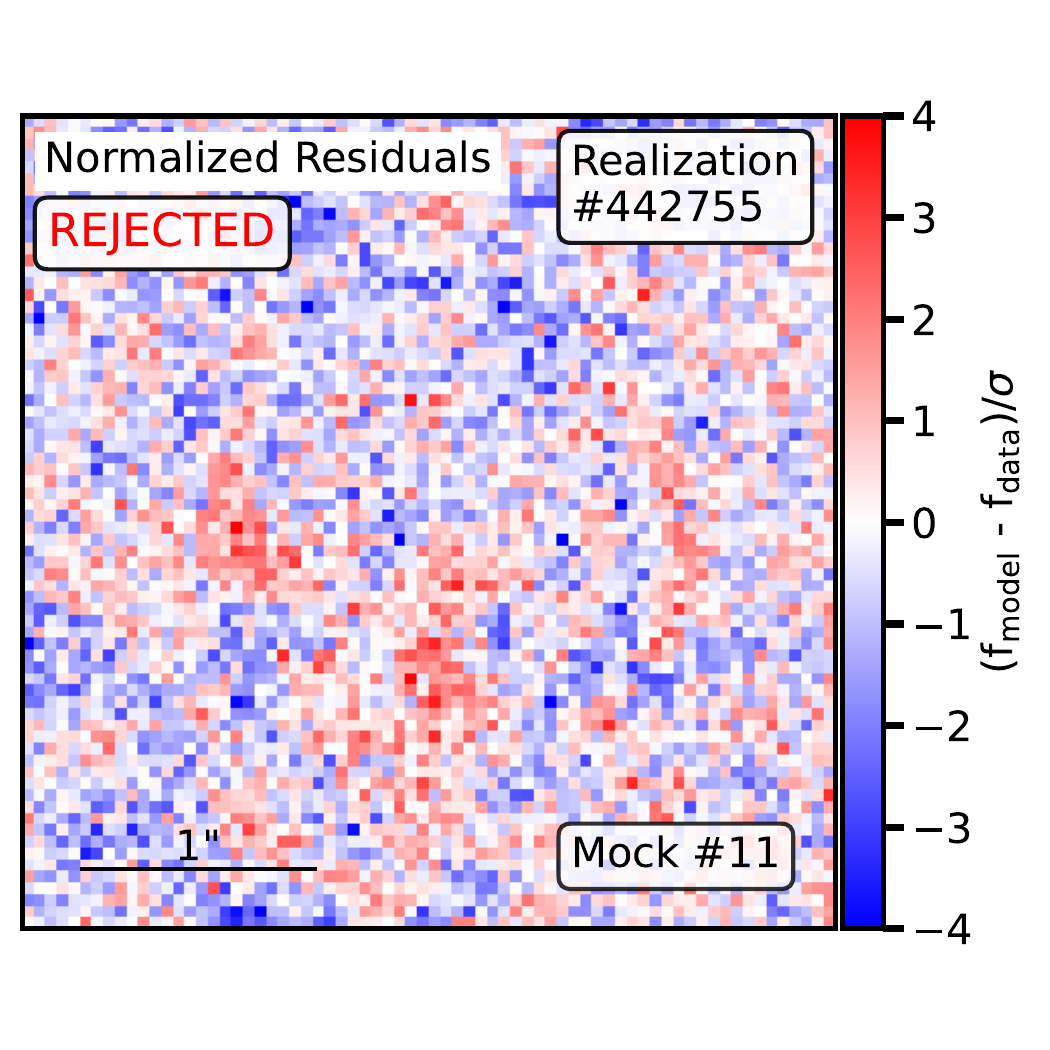}
			\caption{\label{fig:mock11cases} The same as Figure \ref{fig:mock4cases}, but with four example realizations generated for Mock $11$ in the CDM ground truth sample. This system has flux ratios consistent with those predicted by a smooth lens model, so accepted realizations are more likely to have fewer halos. As seen in the bottom row, the information encoded by the imaging data allows us to rule out lens model configurations that match the flux ratios through a combination of substructure and large-scale deformation of the lens mass profile.}
		\end{figure*}
		\begin{figure*}
			\includegraphics[trim=0.2cm 0.2cm 0.25cm
			2.5cm,width=0.33\textwidth]{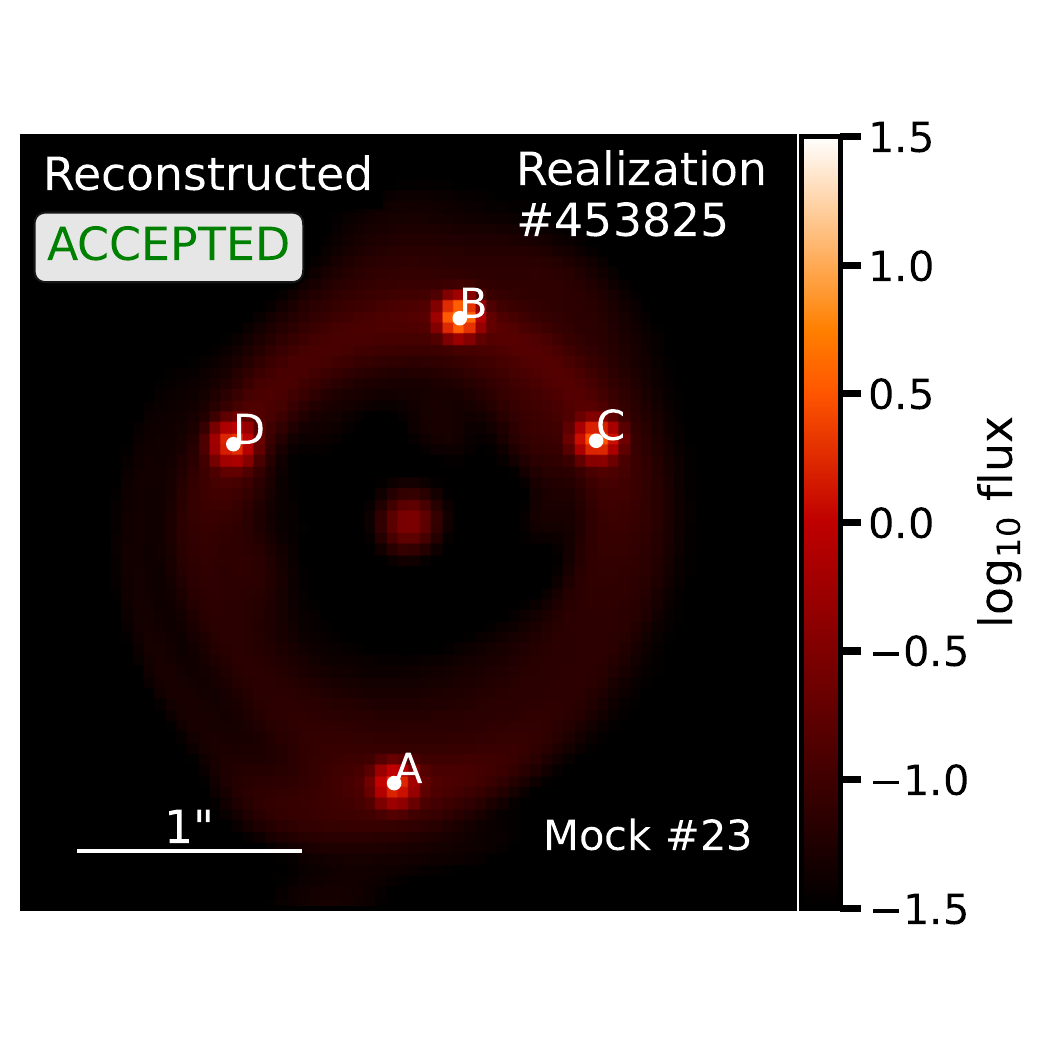}
			\includegraphics[trim=0.2cm 0.2cm 0.25cm
			2.5cm,width=0.33\textwidth]{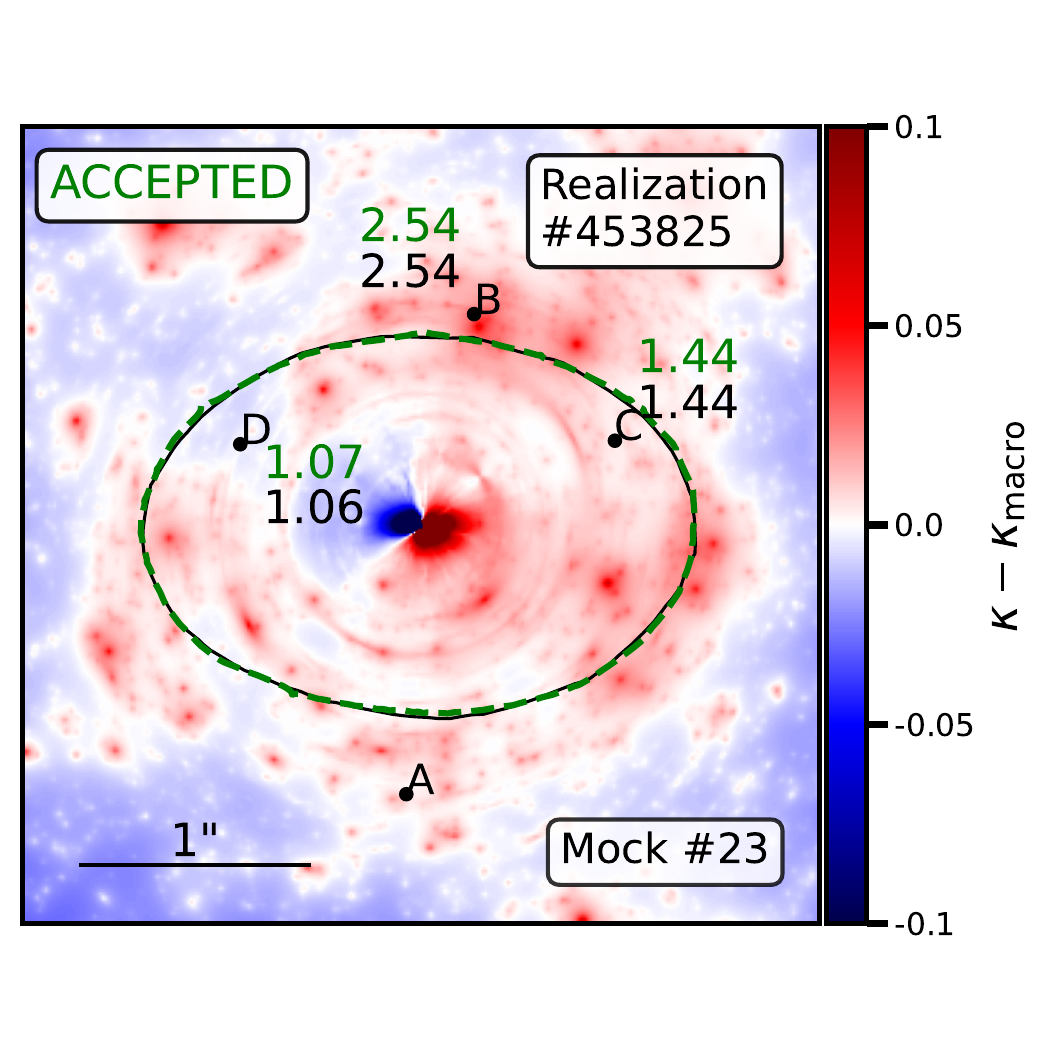}
			\includegraphics[trim=0.2cm 0.2cm 0.25cm
			2.5cm,width=0.33\textwidth]{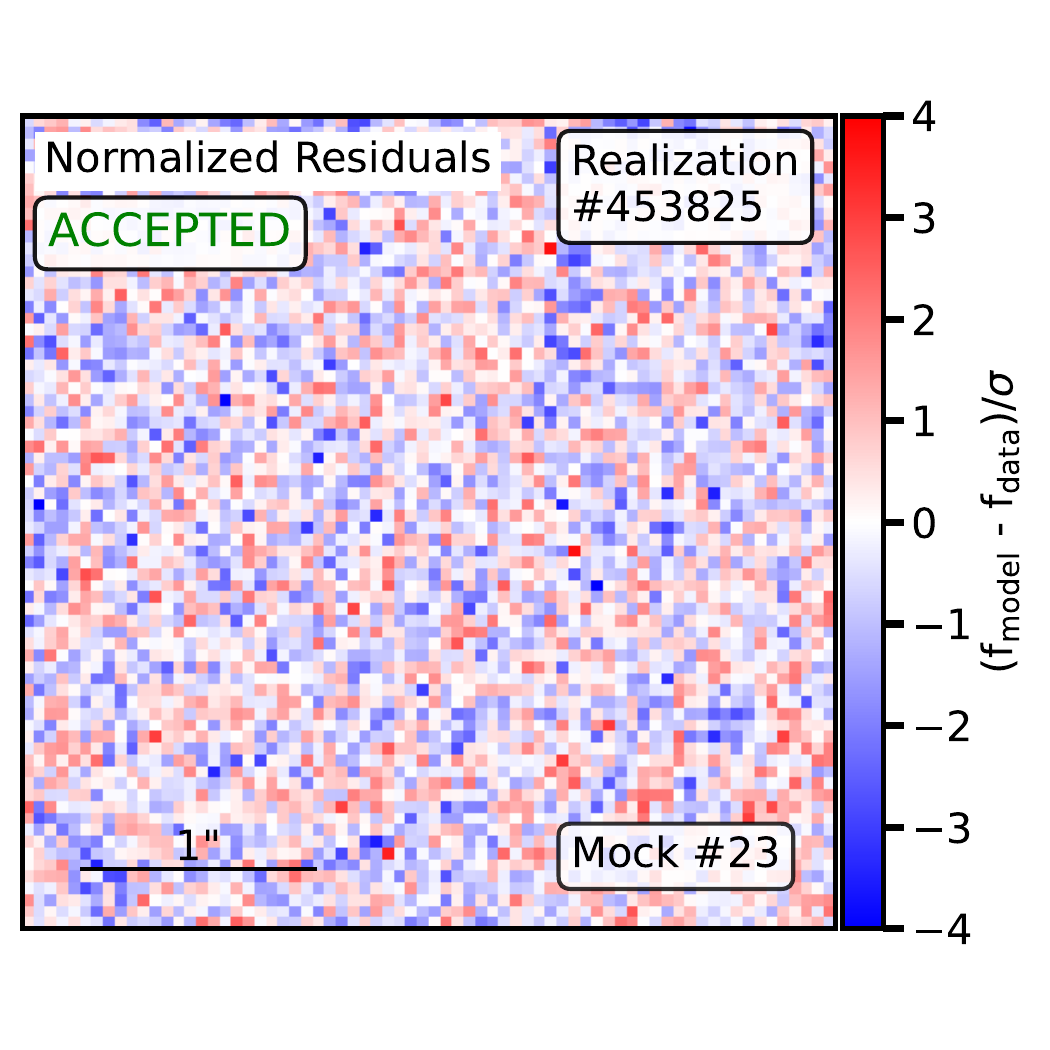}
			\includegraphics[trim=0.2cm 0.2cm 0.25cm
			1.5cm,width=0.33\textwidth]{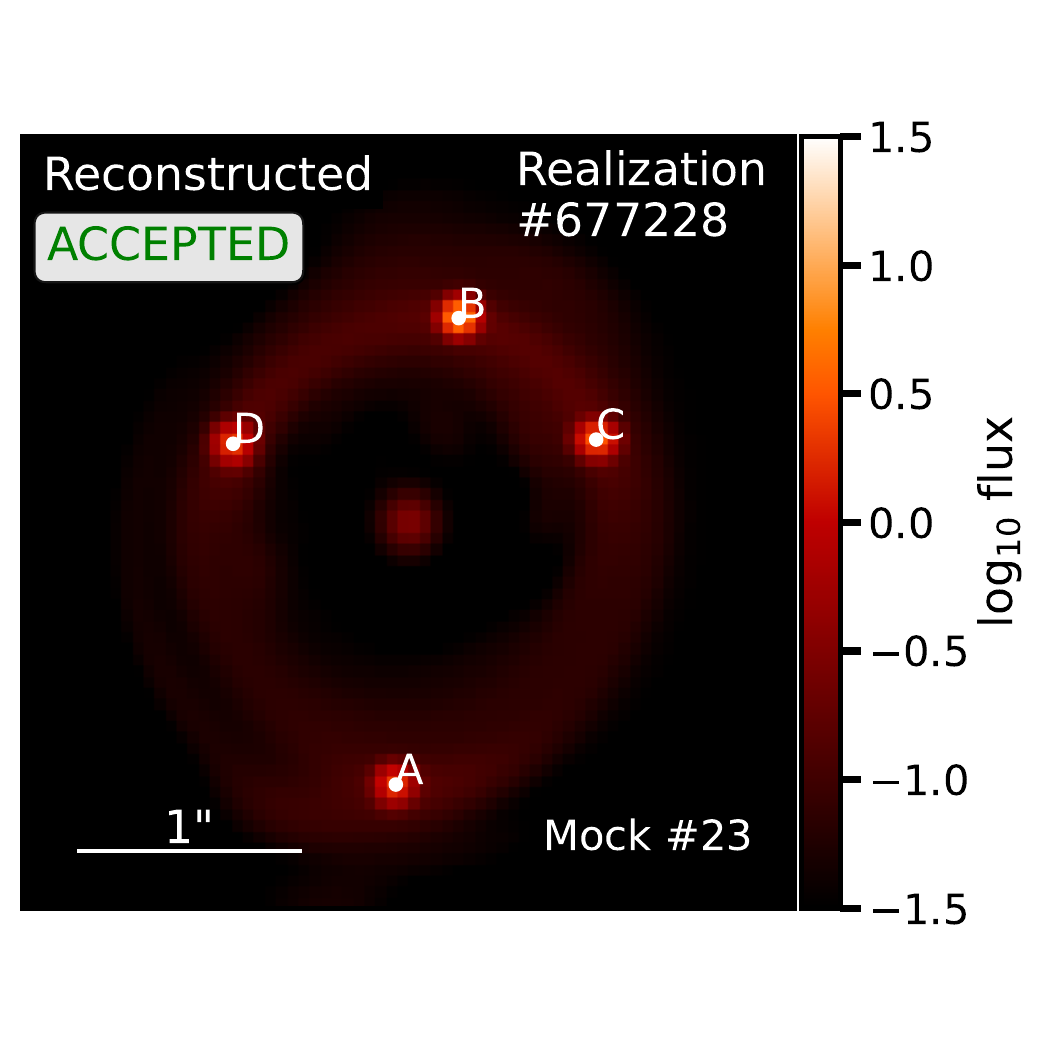}
			\includegraphics[trim=0.2cm 0.2cm 0.25cm
			1.5cm,width=0.33\textwidth]{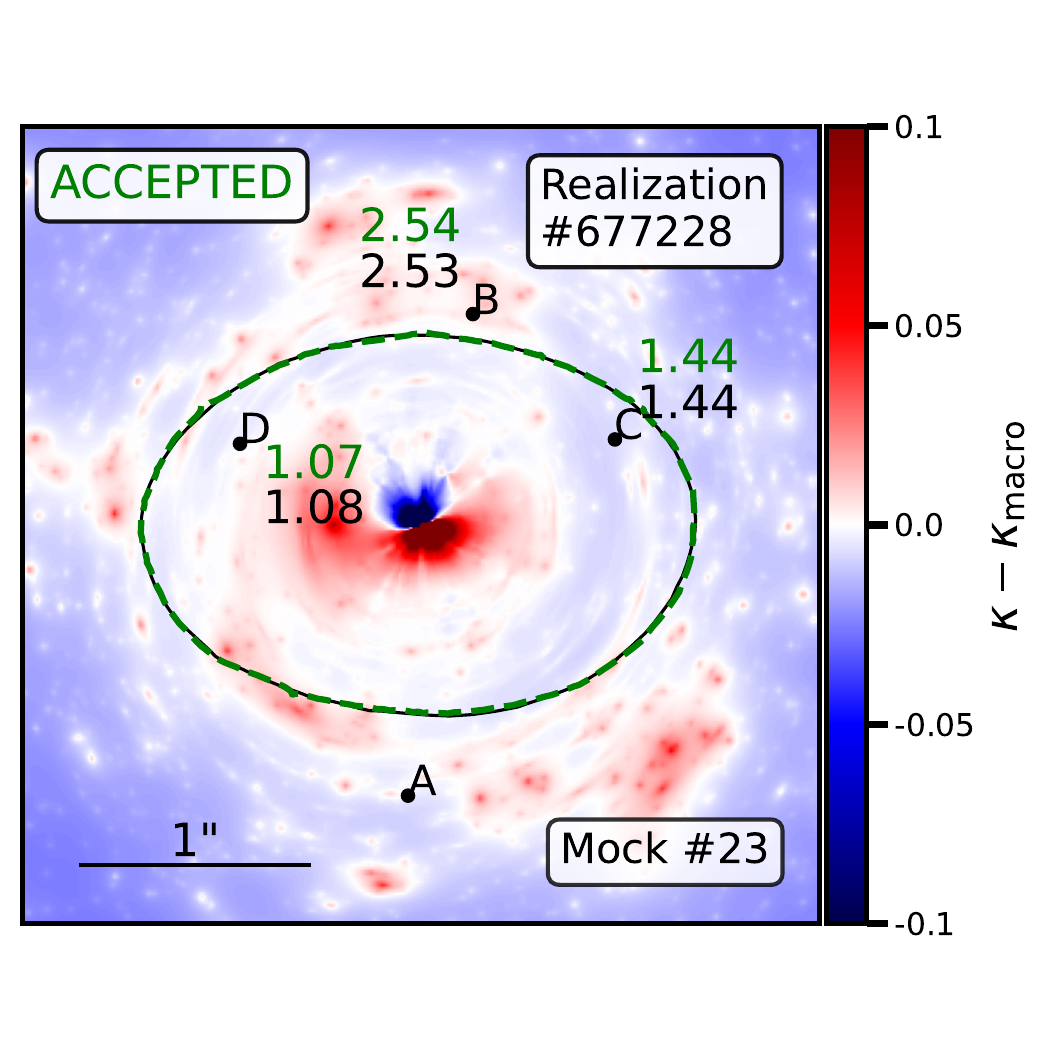}
			\includegraphics[trim=0.2cm 0.2cm 0.25cm
			1.5cm,width=0.33\textwidth]{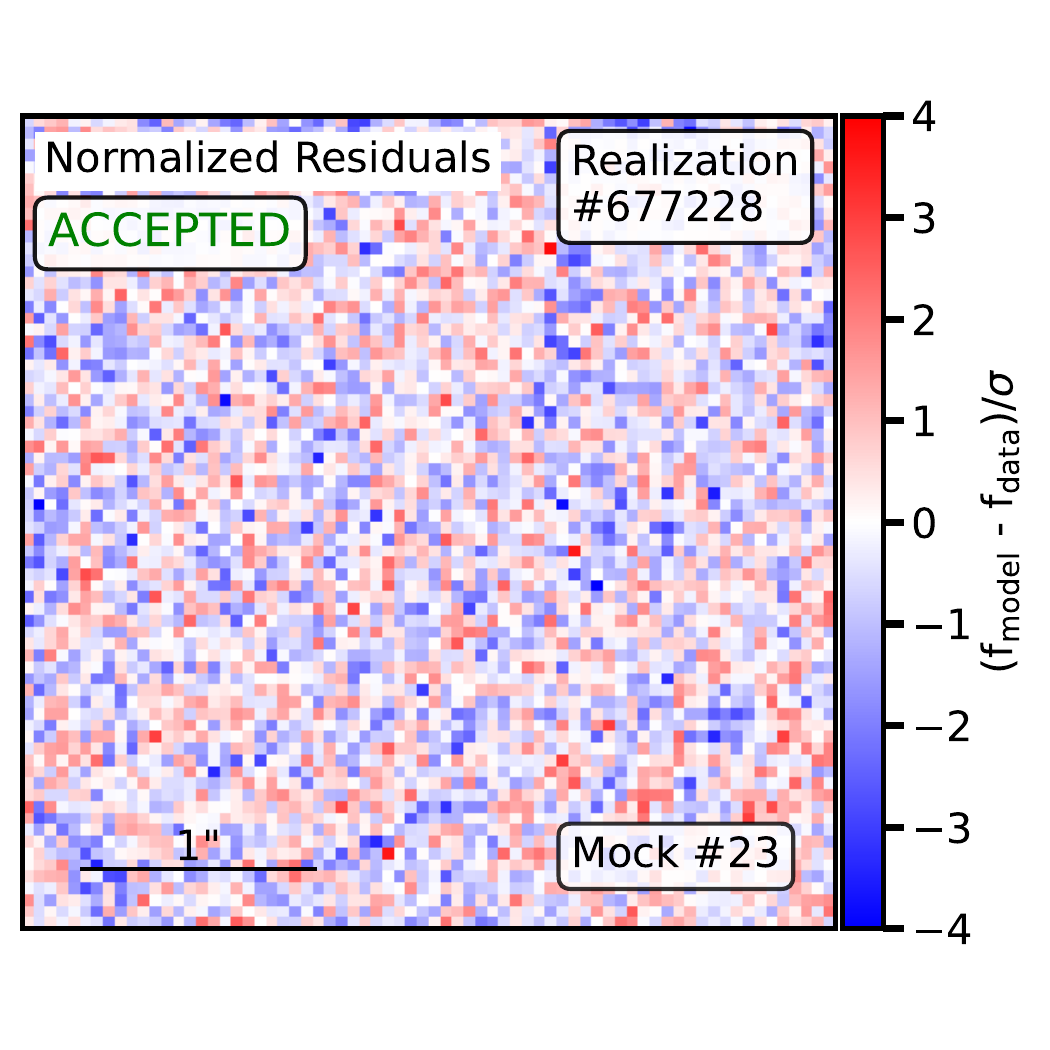}
			\includegraphics[trim=0.2cm 0.2cm 0.25cm
			1.5cm,width=0.33\textwidth]{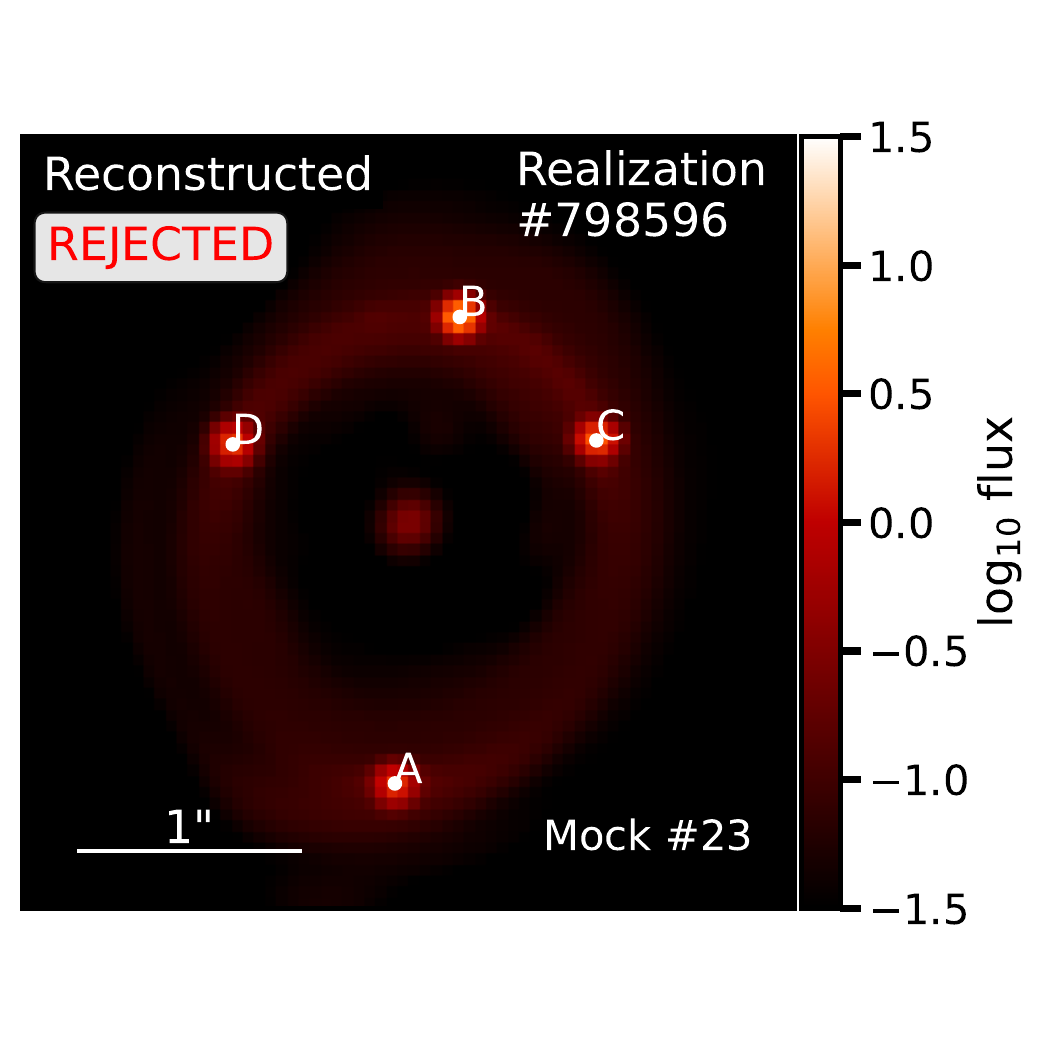}
			\includegraphics[trim=0.2cm 0.2cm 0.25cm
			1.5cm,width=0.33\textwidth]{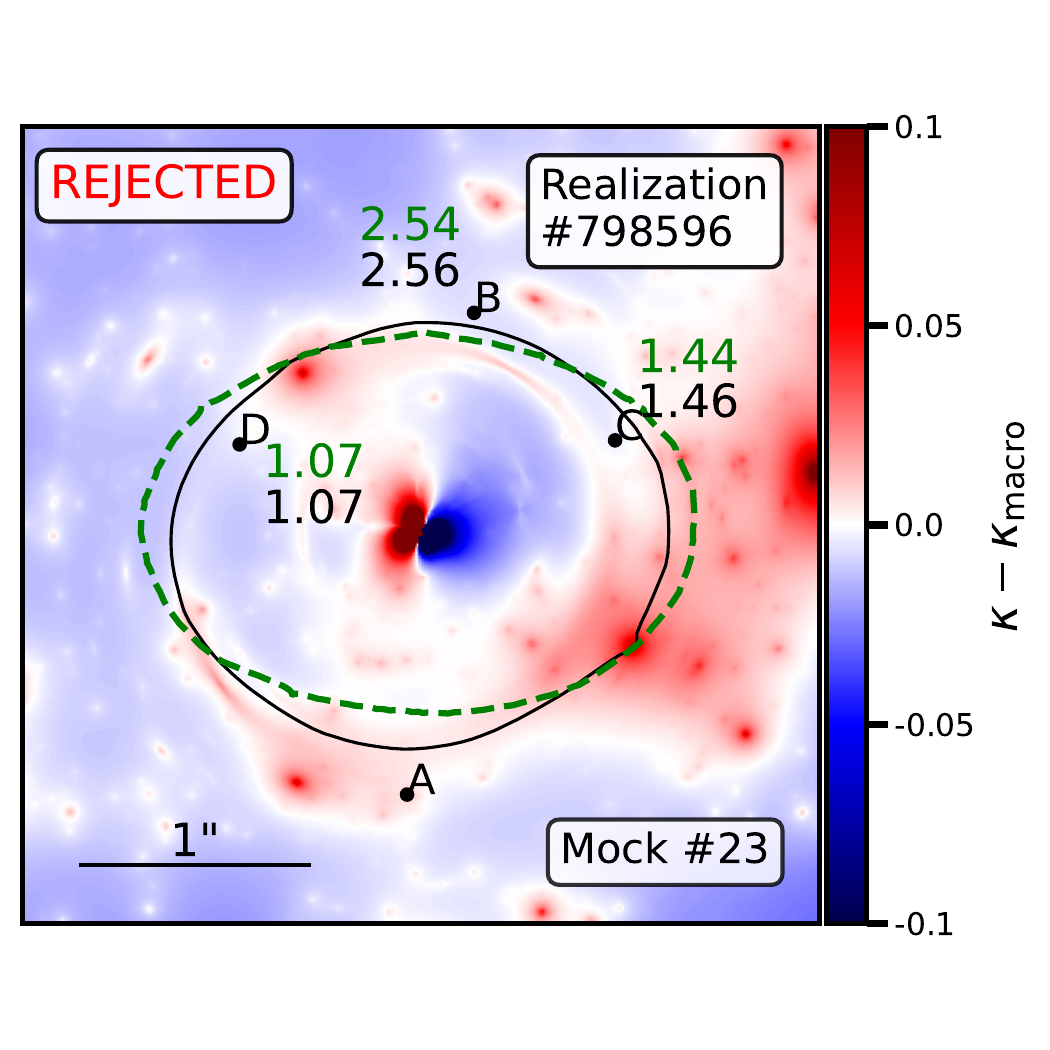}
			\includegraphics[trim=0.2cm 0.2cm 0.25cm
			1.5cm,width=0.33\textwidth]{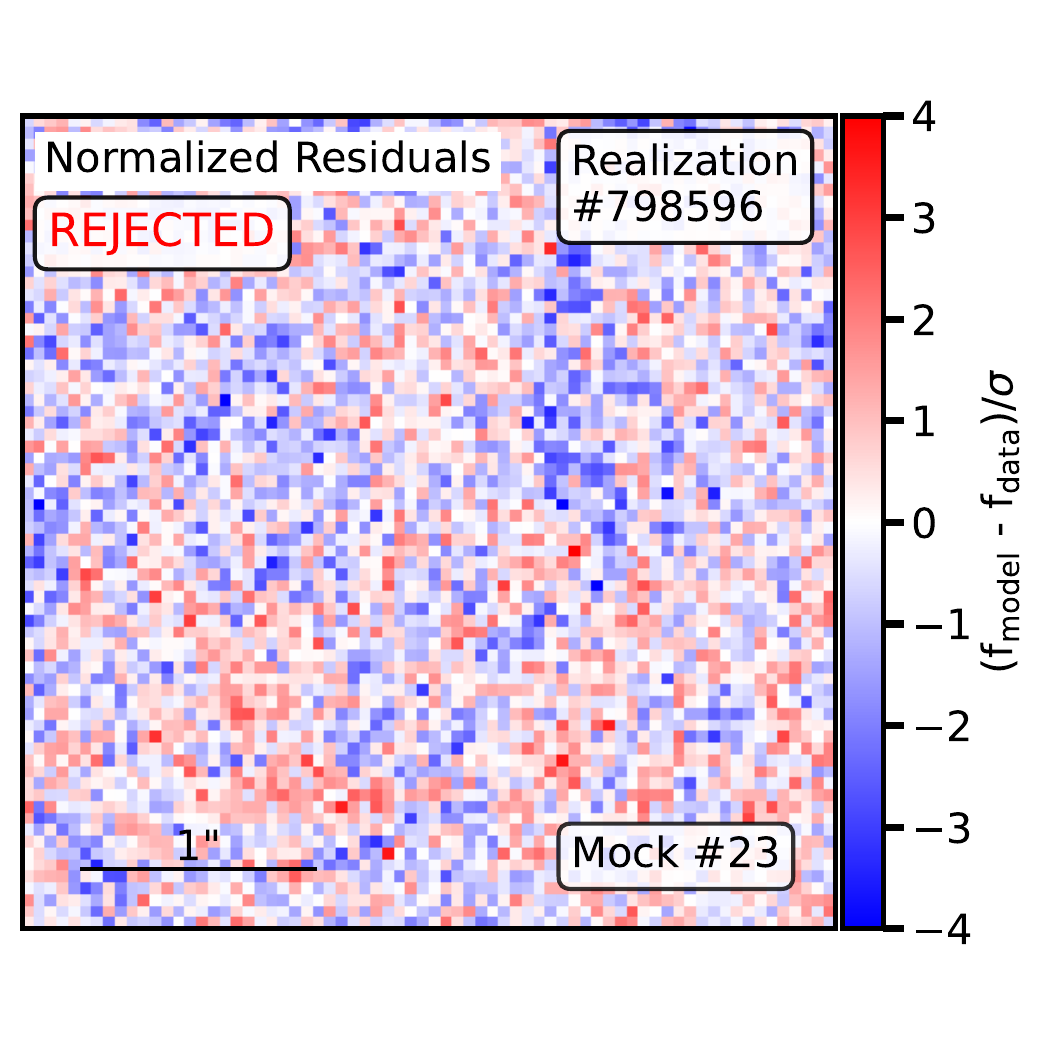}
			\includegraphics[trim=0.2cm 2.5cm 0.25cm
			1.5cm,width=0.33\textwidth]{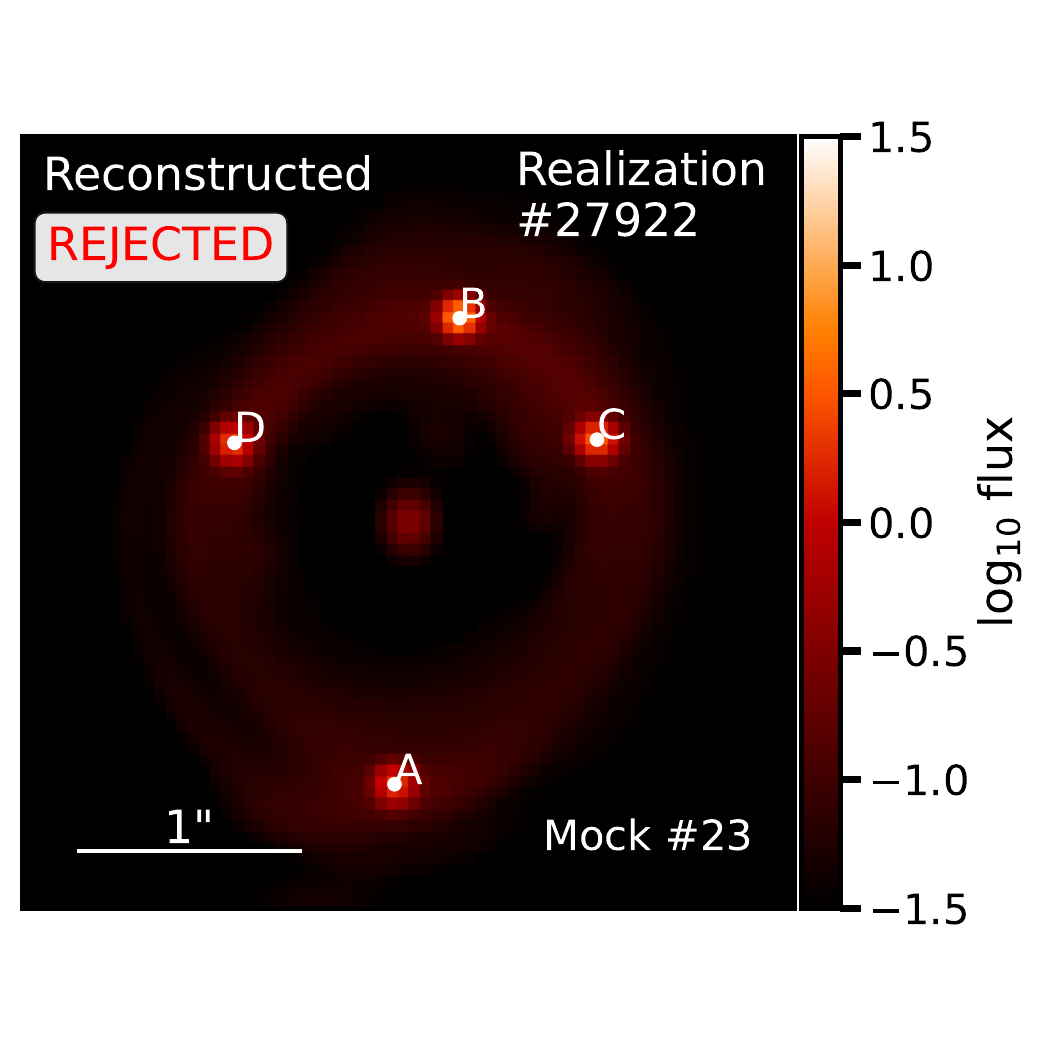}
			\includegraphics[trim=0.2cm 2.5cm 0.25cm
			1.5cm,width=0.33\textwidth]{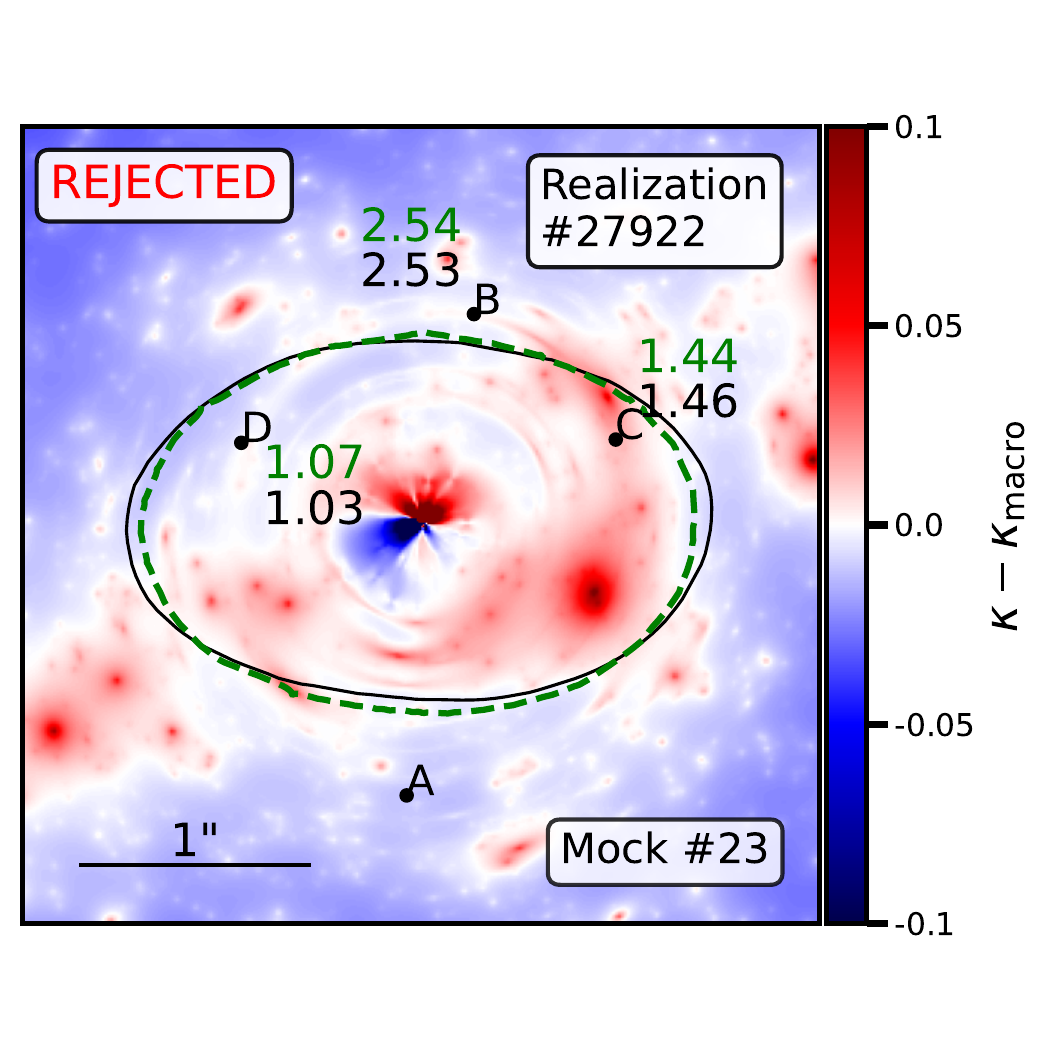}
			\includegraphics[trim=0.2cm 2.5cm 0.25cm
			1.5cm,width=0.33\textwidth]{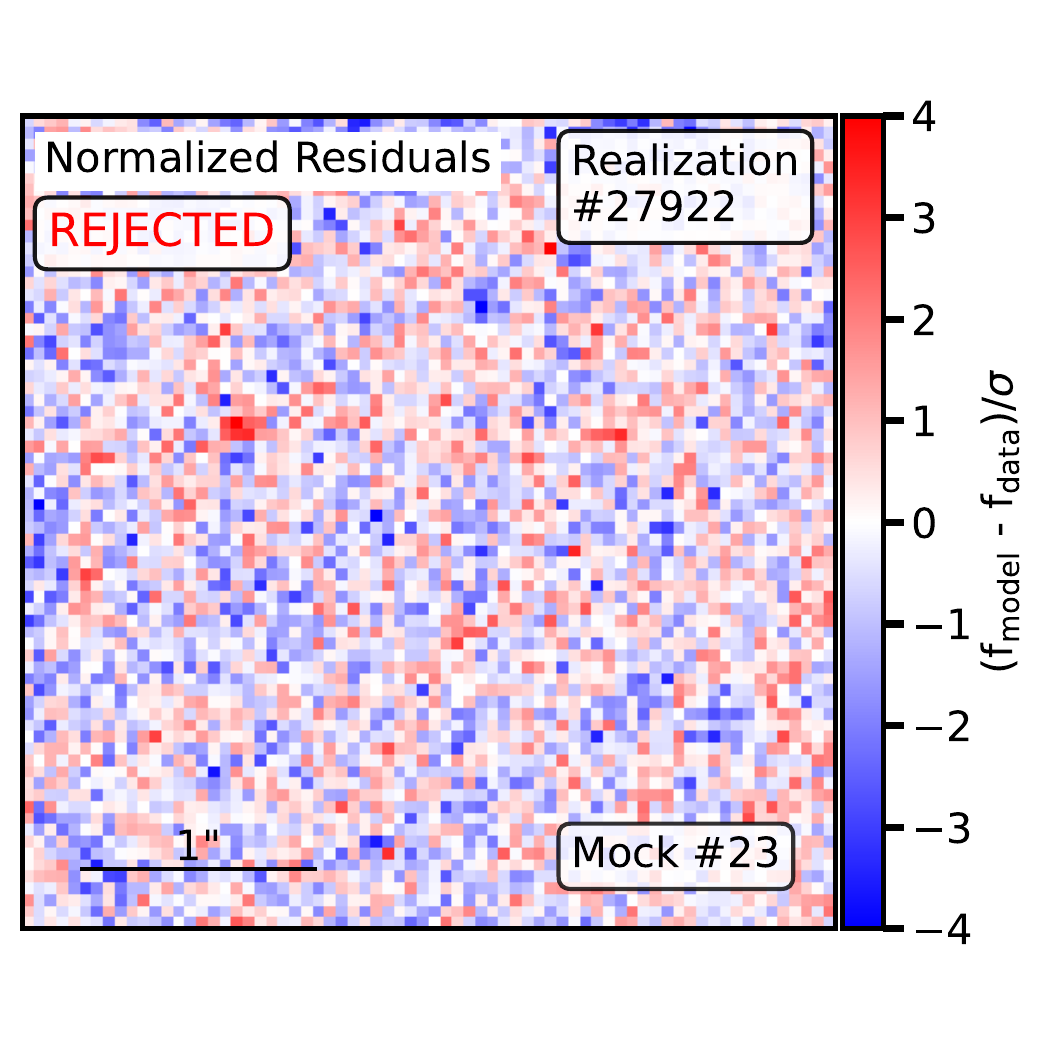}
			\caption{\label{fig:mock23cases} The same as Figure \ref{fig:mock4cases}, but with four example realizations generated for Mock $23$ in the CDM ground truth sample. Proposed lens models accepted for this system tend to have substructure perturbing image B.}
		\end{figure*}
		\subsection{Mass profile of the main deflectors}
		\label{ssec:massprofilemaindeflector}
		In this section we discuss how we create simulated main deflector mass profiles for the mock lenses (Section \ref{sssec:creatingmockmass}) and how we model the mock lenses during the inference performed on the mock datasets (Section \ref{sssec:modellingmassprofile}).
		\subsubsection{Creating mock lens mass profiles}
		\label{sssec:creatingmockmass}
		To test the methodology discussed in Section \ref{sec:lensmodeling} we create a sample of 25 mock lens systems with properties broadly comparable to the know population of such systems \citep{Auger++10,Oguri++10}. The mocks have lens (source) redshifts in the range 0.3 - 0.9 (0.9 - 3.0). We model the main deflector galaxy as a elliptical power-law (EPL) mass profile with an Einstein radius set to $1 \ \rm{arcsec}$ for each system, axis ratios in the range 0.50 - 0.95, and logarithmic mass profile slopes $\gamma$ drawn from a Gaussian prior $\mathcal{N}\left(2.0, 0.1\right)$. We apply external shear across the main lens plane with a random orientation and a strength $\gamma_{\rm{ext}}$ in the range 0.02 - 0.16. 
		
		The observed population of elliptical galaxies sometimes exhibit deviations from ellipticity quantified in terms of multipole perturbations on top of the elliptical mass profile \citep{Bender++89,Hao++06}. A multipole perturbation adds convergence\footnote{Convergence refers to a projected mass normalized by the critical surface mass density for lensing.}
		\begin{equation}
			\label{eqn:multipole}
			\kappa_{m}\left(r, \phi \right) = \frac{a_{\rm{m(phys)}}}{r} \cos \left(m \left(\phi - \phi_m\right) \right)
		\end{equation}
		where $r$ is the separation in arcseconds from the mass centroid, the angle $\phi_{m}$ determines the orientation, and $a_{\rm{m(phys)}}$ is the amplitude of the convergence associated with the multipole perturbation.  \citet{Oh++24} further explore the implications of these multipole perturbations in the context of flux ratio analyses in strong lenses.  
		
		We base our implementation of the multipole perturbations on the observed properties of 847 elliptical galaxies presented by \citet{Hao++06}. The shape of the iso-density contours inferred from the light (and assuming light traces mass) can be related to the physical amplitude of the perturbation by $a_{\rm{m(phys)}} = a_{\rm{m}} \times \frac{\theta_E}{\sqrt{q}}$, where $\theta_{\rm{E}}$ is the Einstein radius, $q$ is the axis ratio of the ellipse and $a_m$ represents the deviation from ellipticity of the light. The third ($m=3$) and fourth-order ($m=4$) moments make the dominant contribution to the multipole expansion of galaxy shapes \citep{Bender++89}. \citet{Hao++06} find $a_{\rm{3}} $ distributed as $\mathcal{N}\left(0.0, 0.005\right)$ with values of $\phi_3$ uncorrelated with the position angle of the underlying mass profile. \citet{Hao++06} also measure $a_4 \in \mathcal{N}\left(0.0, 0.01\right)$ with an orientation that tends to align with the orientation of the underlying mass profile. When the $a_4$ orientation aligns with the orientation of the underlying ellipse the profile appears boxy ($a_4<0$) or disky ($a_4>0$). The observed amplitudes $a_3$ and $a_4$ have no apparent correlation.
		
		Based on the findings by \citet{Hao++06}, when creating mock deflectors we include the dominant $m=3$ and $m=4$ terms and use priors for their observed amplitudes $a_3 \in \mathcal{N}\left(0.0, 0.005\right)$ and $a_4 \in \mathcal{N}\left(0.0, 0.01\right)$. Based on the measurements by \citet{Hao++06} and \citep{Oh++24}, we enforce alignment between the $a_4$ orientation and the underlying EPL profile (producing boxy or disky contours), and sample $\phi_3 \in \mathcal{U}\left(-\pi/6, \ \pi/6\right)$. For additional discussion regarding the role of multipole perturbations in dark matter inferences with quadruply-imaged quasars, we refer to Appendix \ref{app:B}. 
		
		\subsubsection{Modeling of the lens mass profile}
		\label{sssec:modellingmassprofile}
		When modeling the mock lenses, all of the parameters that describe the EPL profile are left free to vary while reconstructing the imaging data, as well as the strength and position angle of the external shear. We draw  $a_3$, $a_4$, and $\phi_3$ from the same priors as those used to create the simulated datasets. During the reconstruction of the imaging data for each realization, we keep the multipole terms fixed to the values drawn from their respective priors. We eventually constrain these parameters when down-selecting on the imaging data and flux ratio summary statistics, as discussed in Section \ref{ssec:inferencesummaries}. 
		
		We have experimented with allowing $a_3$ and $a_4$ to vary freely while reconstructing the imaging data, but find that this causes their inferred amplitudes to take on unphysical values. This unexpected behavior could arise from degeneracies in the imaging data likelihood between multipole perturbations to the lens model and full populations of dark matter halos, as well as limitations associated with the PSF and source reconstruction. During the lens modeling, we include a Gaussian prior on $\gamma$, the logarithmic slope of the main deflector mass profile, with a mean of -2.0 and standard deviation of 0.2. 
		
		\subsection{Source and lens light models}
		\label{ssec:lightmodels}
		In the next two sections, we discuss the lens and source light models used to create the simulated datasets (Section \ref{sssec:creatingmocklight}), and how we model the lens and source light profiles in the inference (Section \ref{sssec:modelingmocklight}). 
		\subsubsection{Creating mock lens and source light profiles}
		\label{sssec:creatingmocklight}
		We parameterize the main deflector light as a circular S{\'e}rsic profile \citep{Sersic63}. The source light model includes two components: the lensed quasar and its host galaxy. For the quasar, we assume flux ratios measured from the warm dust region now observable with JWST \citep{Nierenberg++23} with a physical size $\sim1 - 10 \ \rm{pc}$. To create the mocks, we model the quasar emission as a circular Gaussian with a full-width at half maximum sampled from a uniform prior $\mathcal{U}\left(1-10\right) \rm{pc}$. We assume a flux ratio measurement precision of $3 \%$, and an astrometric precision in the relative quasar image positions of 5 milli-arcseconds, based on the recent measurements \citep{Nierenberg++23}. 
		
		To create realistic lensed arcs, we place the spiral galaxy shown in Figure \ref{fig:source} at the source redshift for each of the 25 mock deflectors. We extract this source from the COSMOS survey catalog \citep{Koekemoer++07} using the software package {\tt{paltas}} \citep{Wagner-Carena++23}. This particular galaxy was selected because it exhibits enough morphological complexity (spiral arms) to warrant a small-scale basis set expansion of its light profile to reproduce a lensed image. Many real strong lens systems exhibit this level of complexity in their inferred source light profiles, and we include this feature in our simulations to determine whether our methodology yields unbiased inferences of substructure properties with a complex source morphology. 
		
		\subsubsection{Modeling of lens and source light profiles}
		\label{sssec:modelingmocklight}
		We model the main deflector light as an elliptical S{\'e}rsic profile. We model the quasar emission region as a circular Gaussian, and sample its size uniformly from a prior $\mathcal{U}\left(1, 10\right) \ \rm{pc}$ for each system. We model the quasar host galaxy with an ellpitical S{\'e}rsic profile with additional small-scale complexity implemented through shapelet basis functions \citep{Birrer++15}. An integer $n_{\rm{max}}$ determines the degree of complexity these functions can describe, with the complexity of the reconstructed image with increasing $n_{\rm{max}}$. 
		
		To choose an appropriate level of source complexity in the model we choose the value of $n_{\rm{max}}$ that minimizes the Bayesian information criterion (BIC) defined as 
		\begin{equation}
			\rm{BIC} = k \log \left(n\right) - \log \left(\mathcal{\max\left[L\right]}\right),
		\end{equation} 
		where $k$ is the number of model parameters, $n$ represents the number of data points, and $\max\left[\mathcal{L}\right]$ represents the maximum likelihood of the data given the model. The BIC statistic compensates between obtaining a better fit to data and over-fitting a model. To calculate the BIC, we fit a smooth lens model (i.e. a lens model without substructure) to each mock. For each mock system, we found the BIC reaches a minimum for a source model consisting of an elliptical S{\'e}rsic profile plus shapelets at $n_{\rm{max}} = 10$. With this source model we obtain a reduced $\chi^2$ per degree of freedom $\chi^2 / \rm{DOF} \approx 1.2$, with DOF=3694, for the lens models we accept based on the imaging data likelihood. Appendix \ref{app:A} provides additional discussion regarding the reconstruction of the source light in our simulations. 
		
		\subsection{Point spread function and observing conditions}
		\label{ssec:obs}
		Most of the known quadruply-imaged quasars, and many of those with flux ratios recently measured by JWST, have archival HST imaging \citep{Shajib++19,Schmidt++23}. Anticipating use of this data, we therefore assume HST-like observations with a pixel size of $0.05 \ \rm{arcsec}$, an r.m.s. background noise per pixel of 0.006 photons/sec, and an exposure time of $1600 \ \rm{sec}$. We use a Gaussian point spread function model with a width of $100$ m.a.s. in the creation and modeling of the mock lenses\footnote{In practice, one typically reconstructs the PSF simultaneously with the lensed image and source light. This methodology is possible within our analysis framework, but it was not included in our tests on simulated data.}. 
		
		\section{Results}
		\label{sec:results}
		This section presents the results of the inference of the lens model and substructure properties obtained from 25 simulated strong lens systems assuming a CDM ground truth, and 25 lens systems created assuming a WDM ground truth. Both sets of lens systems are analyzed using the inference methodology discussed in Section \ref{sec:lensmodeling}, the lens modeling approach discussed in Section \ref{sec:multiplanelensmodeling}, and the simulation details presented in Section \ref{sec:simulationsetup}. We begin in Section \ref{ssec:casestudy} with case studies of four mock lens systems created with a CDM ground truth. We discuss in detail how the image positions, flux ratios, and imaging data work in tandem to simultaneously constrain the mass profile of the main deflector and the properties of substructure in each system. In Section \ref{ssec:finalinference}, we present the joint constraints on the normalization of the subhalo mass function and the free-streaming length of dark matter obtained from the full sample, and quantify the degree to which including constraints from lensed arcs improves constraints on substructure properties relative to existing analysis methods that use only image positions and flux ratios. 
		
		\subsection{Case studies}
		\label{ssec:casestudy}
		We begin by analyzing the results of applying our inference methodology to four mock lenses created with a CDM ground truth to gain physical insight into how the joint modeling of lensed arcs, image positions and flux ratios constrains the lens model and the properties of substructure. Figures \ref{fig:mock4image}, \ref{fig:mock6image}, \ref{fig:mock11image}, and \ref{fig:mock23image} show the four lenses chosen for these case studies. We pick these four systems because they exhibit a variety of image configurations, lens and source redshifts, experience varying degrees of perturbation by halos, and have among the most informative likelihood functions. The left panels show the simulated lensed images, while the right-hand panels display the true convergence in dark matter substructure for each mock. As in previous work \citep{Gilman++19}, we define the convergence for a multi-plane lens system in terms of the divergence of the deflection field (see Equation \ref{eqn:lenseqn})
		\begin{equation}
			\label{eqn:multiplaneconv}
			\kappa \equiv \frac{1}{2}\boldsymbol{\nabla} \cdot \boldsymbol{\alpha_{\rm{eff}}}.
		\end{equation}
		To illustrate the distribution of dark matter substructure, we subtract the convergence from the lens macromodel, $\kappa_{\rm{macro}}$, from the total convergence given by Equation \ref{eqn:multiplaneconv}. In place of the deflection field from background halos, we take the divergence of $\boldsymbol{\alpha_{\beta}}$, the effective deflection field from halos behind the main deflector defined in Section \ref{sec:multiplanelensmodeling}. 
		
		\subsubsection{The complementary information conveyed by imaging data and flux ratios}
		As the imaging data subtends angular scales comparable to the image separation, we expect the reconstruction of the lensed arcs will impose the strongest constraints on the large-scale mass profile of the main deflector. Figures \ref{fig:mock4macro}, \ref{fig:mock6macro}, \ref{fig:mock11macro}, and \ref{fig:mock23macro} show the inference on a subset of the parameters that describe the main deflector mass profile. The parameters shown in each figure include the normalization of the main deflector mass profile $\theta_{\rm{E}}$, the axis ratio $q$, the ellipticity position angle $\phi_{\rm{q}}$, the external shear strength $\gamma_{\rm{ext}}$, the external shear position angle $\phi_{\rm{ext}}$, and the amplitude of the $m=3$ and $m=4$ multipole moments $a_3$ and $a_4$. Contours show the $68 \%$ and $95 \%$ credible intervals for the parameters after marginalizing over the properties of substructure in each lens. The black contours show constraints obtained from the ray tracing methods presented by \citet{Gilman++19} that use only image positions and flux ratios to constrain the lens model. Blue contours show constraints obtained from applying methods discussed in Sections \ref{ssec:inferencesummaries} and \ref{sec:multiplanelensmodeling} to compute the likelihood and model the image positions and imaging data. The magenta contours show constraints obtained from combining the flux ratio likelihood with the image position and imaging data likelihoods. 
		
		By comparing the volume enclosed by the black and blue distributions in Figures \ref{fig:mock4macro}-\ref{fig:mock23macro}, we conclude that the imaging data imposes significantly stronger constraints on the main deflector mass profile than those one obtains from only image positions and flux ratios. Using all available data (magenta), we obtain the tightest constraints on the macromodel parameters, but adding flux ratio information to lens models already constrained by imaging data leads to only marginal improvement. These trends persist among all the mock lenses we analyze in our simulations, and are consistent with a physical picture in which the large-scale mass distribution of the lens is primarily constrained by imaging data.  
		
		The flux ratios, however, convey vital information for constraining the properties of the deflection field on angular scales below those probed by imaging data\footnote{We recall that we have constructed the likelihood function in such a way that the posterior distribution of substructure properties given the data can only differ from the prior if one incorporates constraints from the flux ratios (see Section \ref{ssec:inferencesummaries} and Equation \ref{eqn:importanceweights2}).}. Figure \ref{fig:likelihoods} shows the joint likelihood function computed with the image positions, flux ratios, and imaging data for each of the four case study lenses shown in Figures \ref{fig:mock4image}-\ref{fig:mock23image}. The parameters shown in the figures are $\Sigma_{\rm{sub}}$ and $m_{\rm{hm}}$, the normalization of the subhalo mass function (Equation \ref{eqn:shmf}) and the cutoff scale of the halo mass function (Equations \ref{eqn:suppression} and \ref{eqn:suppressionc}). The color scale in Figure \ref{fig:likelihoods} corresponds to the relative likelihood between positions in the parameter space. The top-left region of parameter space includes dark matter models with a significant suppression of the halo mass function and relatively few subhalos, while the bottom right corner of parameter space includes a plethora of subhalos and a CDM-like halo mass function. 
		
		The likelihoods shown in Figure \ref{fig:likelihoods} exhibit clear preferences for models with a plethora of substructure in the case of Mocks 6 and 23, and for relatively little substructure for Mocks 4 and 11. Note that the improvement in the inference of the macromodel parameters shown in Figures \ref{fig:mock4macro}-\ref{fig:mock23macro} after adding flux ratio information does not significantly change between the four cases study lenses, and thus the degree to which the flux ratios constrain the large-scale mass distribution of the lens does not depend on the amount of substructure required to fit the flux ratios. However, as we will discuss in the next section, the imaging data strengthens inferences of substructure properties by breaking degeneracies between large-scale deformation of the main deflector mass profile and small-scale perturbations caused by halos. 
		
		\subsubsection{Visualizing accepted and rejected realizations}
		A useful feature of the open-source code we present with our analysis, $\tt{samana}$\footnote{https://github.com/dangilman/samana}, is the ability to recreate a lens model from a random seed assigned to each realization that we record as a model parameter. Thus, after down-selecting on lens models and substructure realizations following the methodology discussed in Section \ref{ssec:inferencesummaries}, we can regenerate lens models from the random seeds to gain physical insight as to why we reject or accept certain realizations when computing the likelihood.  
		
		Figures \ref{fig:mock4cases}, \ref{fig:mock6cases}, \ref{fig:mock11cases}, and \ref{fig:mock23cases} each show two examples of lens models accepted based on fitting the imaging data and flux ratios (top two rows), and two examples of lens models that we would accept based on the image positions and flux ratios, but which we reject based on a poor fit to the imaging data (bottom two rows). From left, the columns show the reconstructed lensed image, the projected multiplane convergence in substructure obtained from substracting the macromodel convergence from the total convergence, i.e $ \frac{1}{2}\boldsymbol{\nabla} \cdot \boldsymbol{\alpha_{\rm{eff}}}-\kappa_{\rm{macro}}$, and the normalized residuals from the fit to the imaging data. In the center panels, we label the observed (model-predicted) flux ratios in green (black), and the true (model-predicted) critical curves in green (black). We have included the critical curves in these figures to serve as a proxy for the shape of the main deflector. Alignment of the critical curves indicates an accurate reconstruction of the main deflector mass profile. 
		
		As shown in the top two rows of each figure, the realizations that we accept simultaneously match the small-scale properties of the deflection field constrained by the flux ratios and the large-scale properties of the deflection field primarily constrained by the imaging data. The bottom two rows of Figures \ref{fig:mock4cases}-\ref{fig:mock23cases} show examples of realizations that match the observed image positions and flux ratios, but which we reject based on a poor fit to the imaging data. In the rejected lens models, a particular configuration of dark matter halos conspires with a large-scale deformation of the deflection field to produce the correct flux ratios, but these configurations of the lens model cannot reproduce the observed lensed arcs. These figures provide a visual illustration of how incorporating imaging data breaks degeneracies between large and small-scale properties of the lens model, isolating flux ratio perturbations caused by halos from large-scale deformations of the deflection field constrained by the arcs. 
		
		Figures \ref{fig:mock4cases}-\ref{fig:mock23cases} also serve to aid in the interpretation of the  likelihood functions shown in Figure \ref{fig:likelihoods}, with Mock $6$ providing a particularly clear illustration of how perturbations by dark matter halos manifest in the likelihood. Closely examining the true multiplane convergence map for Mock $6$ in right panel of Figure \ref{fig:mock6image}, a collection of dark matter subhalos and line-of-sight halos appear in close proximity to image A (the bottom-right image). The collective impact of these structures imparts a $\sim 12 \%$ perturbation to the magnification of this image. To match this feature in the data, substructure models that match the flux ratios in Mock 6 tend to have a single massive halo, or a collection of low-mass halos, perturbing image A. These halos appear prominently in the convergence maps shown in the central panels of Figure \ref{fig:mock6cases}. 
		
		Substructure realizations with halos perturbing the images in Mock 6 and 23 occur more frequently in CDM than in WDM, and thus the likelihood functions shown in Figure \ref{fig:likelihoods} for these systems punish models with fewer halos. On the other hand, Mocks $4$ and $11$ have flux ratios consistent with those predicted by a smooth lens model to within the measurement uncertainties. As such, the substructure realizations that match the flux ratios for Mocks 4 and 11 tend to have fewer halos than the dark matter models that match the flux ratios in Mocks 6 and 23, as reflected in the likelihoods. The likelihood function from the full sample of mock lenses results from a product of the individual likelihoods, and is discussed in the next section. 
		\begin{figure*}
			\includegraphics[trim=0cm 0cm 0cm
			0cm,width=0.9\textwidth]{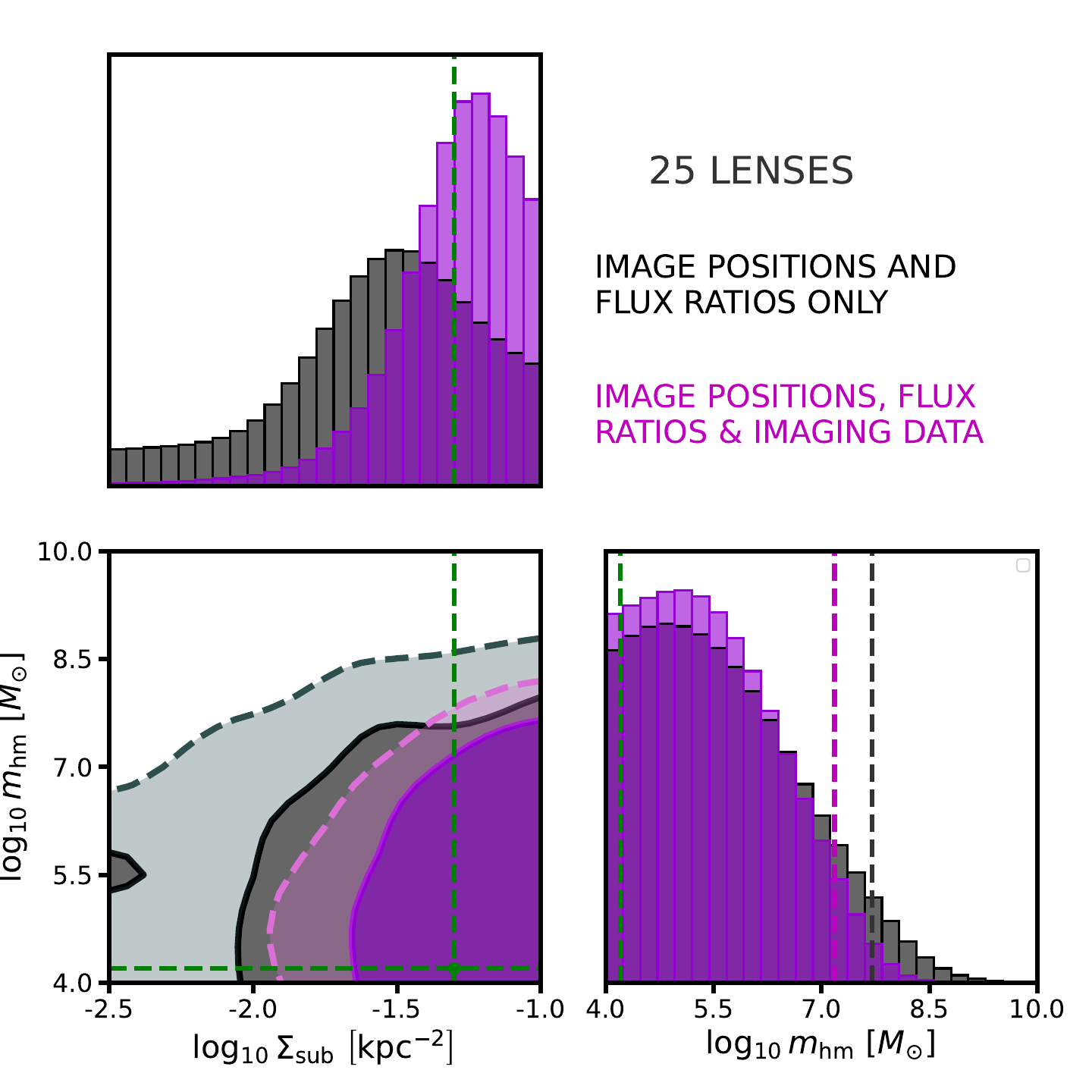}
			\caption{\label{fig:inference} The posterior distribution for $\Sigma_{\rm{sub}}$ and $m_{\rm{hm}}$ obtained from modeling only image positions and flux ratios (black), and image positions, flux ratios, and imaging data (magenta). These posteriors are marginalized over nuisance parameters that include the mass profile of the main deflector, including multipole moments $a_3$ and $a_4$, the finite-size of the quasar emission region, and the surface brightness of the lensed quasar host galaxy. Black and magenta vertical bars in the marginal likelihood for $m_{\rm{hm}}$ correspond to $95\%$ exclusion limits. The red crosshairs indicates the ground-truth used to create the simulated data, and the vertical lines in the marginal likelihood for $m_{\rm{hm}}$ correspond to $95 \%$ exclusion limits. For visualization we have marked the CDM ground truth $m_{\rm{hm}} = 0$ as $m_{\rm{hm}} = 10^{4.2} M_{\odot}$. The inference assumes flux ratio measurement precision of 3 percent and astrometric precision of 0.005 milli-arcseconds.}
		\end{figure*}
		\begin{table*}
			\label{tab:likelihoods}
			\begin{tabular}{lccccr} 
				\hline
				dataset used & likelihood ratio CDM:WDM & likelihood ratio CDM:WDM  & likelihood ratio CDM:WDM  & likelihood ratio CDM:WDM  & \\
				& $7.0 < \log_{10} m_{\rm{hm}} / M_{\odot}<7.5$ & $7.5 < \log_{10} m_{\rm{hm}} / M_{\odot}<8.0$  &$8.0 < \log_{10} m_{\rm{hm}} / M_{\odot}<8.5$  &$8.5 < \log_{10} m_{\rm{hm}} / M_{\odot}<9.0$  & \\
				\hline
				image positions $\&$ & 3:1 & 4:1 & 9:1 & 23:1 & \\  \vspace{0.025in}
				and flux ratios &  &  & & & \\ 
				image positions, & 4:1 & 10:1 & 50:1 & 301:1 & \\
				flux ratios,  $\&$ &  &  & & & \\
				imaging data &  &  & & & \\
				\hline		
			\end{tabular}
		\end{table*}
		\begin{figure}
			\includegraphics[trim=0cm 0cm 0cm
			0cm,width=0.45\textwidth]{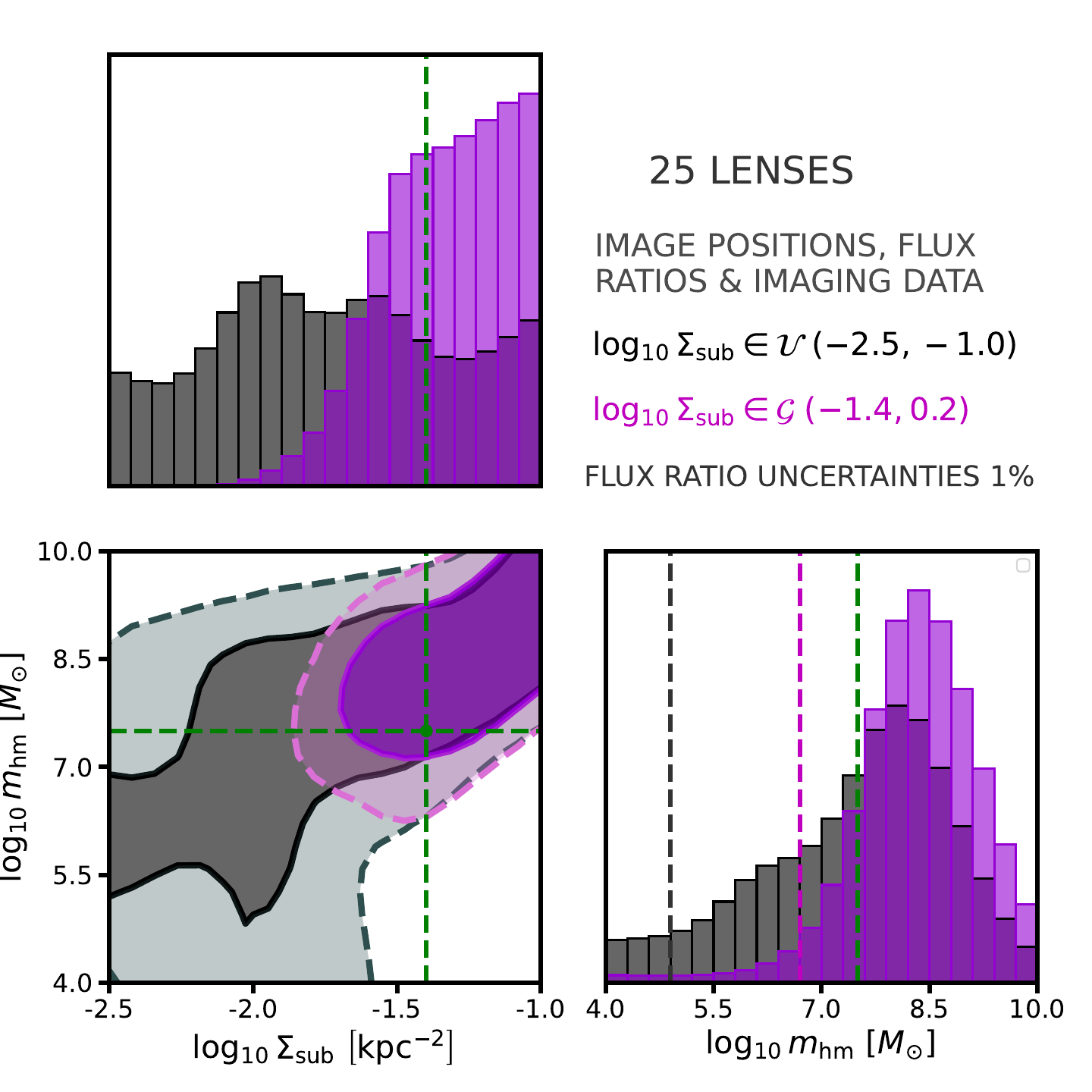}
			\caption{\label{fig:inferencewdm1} The joint posterior on the normalization of the subhalo mass function $\Sigma_{\rm{sub}}$ and the half-mode mass $m_{\rm{hm}}$ for a simulated dataset with a WDM ground truth $m_{\rm{hm}} = 10^{7.5} M_{\odot}$. The posterior results from flux ratio measurement uncertainties of 1 percent. Black contours show the posterior with a log-uniform prior on $\Sigma_{\rm{sub}}$, and magenta contours show the result of incorporating a Gaussian prior on $\Sigma_{\rm{sub}}$ centered on the ground truth value of $\log_{10} \Sigma_{\rm{sub}} = -1.4$ with a width 0.2 dex. As in Figure \ref{fig:inference}, contours correspond to $68 \%$ and $95\%$ credible intervals, the red crosshairs mark the input ground truth, and the vertical bars in the $m_{\rm{hm}}$ marginal likelihood represent $95 \%$ exclusion limits.}
		\end{figure}
		\begin{figure}
			\includegraphics[trim=0cm 0cm 0cm
			0cm,width=0.45\textwidth]{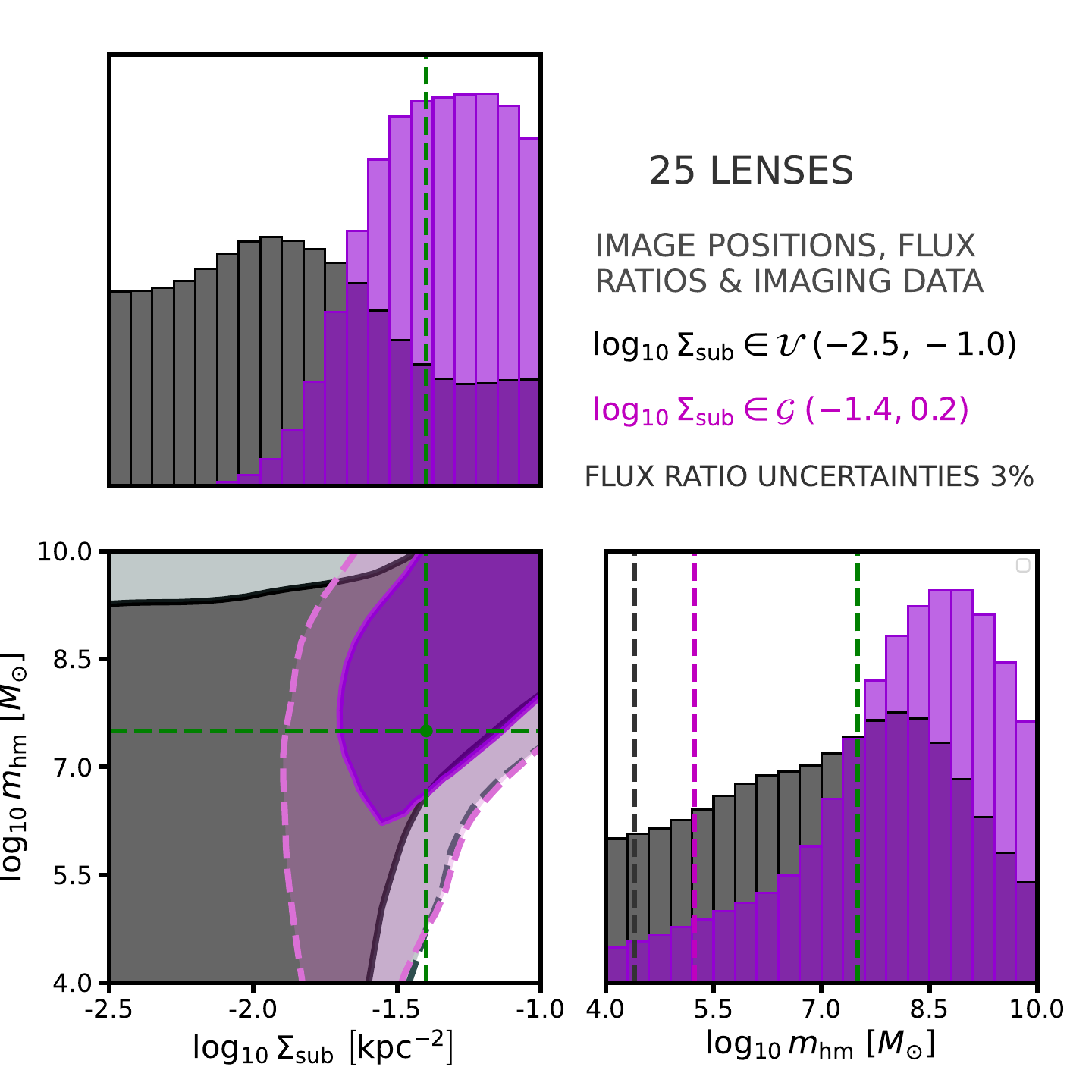}
			\caption{\label{fig:inferencewdm3} The same as Figure \ref{fig:inferencewdm1}, but assuming flux ratio measurement uncertainties of 3 percent.}
		\end{figure}
		\subsection{Inference on substructure properties from the full sample}
		\label{ssec:finalinference}
		Figure \ref{fig:inference} shows the joint inference on the normalization of the subhalo mass function, $\Sigma_{\rm{sub}}$, and the half-mode mass, $m_{\rm{hm}}$, obtained from 25 mock lenses.  The black contours show the inference resulting from applying the ray tracing methods presented by \citet{Gilman++19}, which use only the image positions and flux ratios to constrain the lens model and substructure properties. The magenta contour in Figure \ref{fig:inference} shows the constraints resulting from applying the methodology presented in Section \ref{sec:lensmodeling}, in which we use the lensed arcs, image positions, and flux ratios to constrain the lens model. Both distributions assume a flux ratio measurement precision of 3$\%$. To make a direct comparison between these two approaches and evaluate the relative improvement, we compute the likelihood obtained from only the image positions and flux ratios (black contours) using the same tolerance threshold $\epsilon$ to down-select on the flux ratio summary statistics when computing the magenta constraints, which make use of all available data. 
		
		In terms of credible intervals, incorporating imaging data improves the $95 \%$ exclusion level by 0.5 dex; with only image positions and flux ratios, we find $\log_{10} m_{\rm{hm}} < 10^{7.7} M_{\odot}$, and $\log_{10} m_{\rm{hm}} < 10^{7.2} M_{\odot}$ when incorporating imaging data with a log-uniform prior on $m_{\rm{hm}}\in \mathcal{U}\left(4, 10\right)$. We can also quantify the improvement gained by incorporating imaging data in terms of relative likelihoods, which do not depend on the prior. First, we define a region of parameter space associated with CDM as having $4.0 < \log_{10} m_{\rm{hm}} / M_{\odot}<4.5$, and regions of parameter space associated with WDM as bins in $\log_{10} m_{\rm{hm}}$ with a width of 0.5 dex between $10^7 M_{\odot}$ and $10^9 {M_\odot}$. We then compute the relative likelihood between CDM and WDM as the volume of the posterior with $\log_{10} m_{\rm{hm}} < {4.5}$ to the volume of the posterior with $\log_{10} m_{\rm{hm}}$ in each bin. 
		
		Table \ref{tab:likelihoods} summarizes the inferred likelihood ratios. In the ``coldest'' WDM bin with $m_{\rm{hm}}$ in the range $10^7 - 10^{7.5} M_{\odot}$, incorporating constraints from the lensed arcs improves the likelihood ratios punishing WDM models by a factor of $1.3$. At scales $m_{\rm{hm}} \in 10^{7.5} - 10^{8} M_{\odot}$, adding imaging data strengthens the likelihood ratios by a factor of 2.5, eventually reaching a factor 13.1 for $m_{\rm{hm}} \in 10^{8.5} - 10^{9} M_{\odot}$. The likelihood function shrinks the volume of the posterior distribution relative to the prior volume by a factor of 4 using only image positions and flux ratios, and by a factor of 7 using the image positions, flux ratios, and imaging data. 
		
		Figures \ref{fig:inferencewdm1} and \ref{fig:inferencewdm3} show the inference on 25 lenses created with a WDM ground truth $m_{\rm{hm}} = 10^{7.5} M_{\odot}$ using the lensed arcs and flux ratios. Figure \ref{fig:inferencewdm1} assumes a flux ratio measurement precision of 1$\%$, and Figure \ref{fig:inferencewdm3} assumes a flux ratio measurement uncertainty of 3$\%$. The black distribution corresponds to a log-uniform prior on the amplitude of the subhalo mass function, and the magenta results from assuming a prior on the amplitude of the subhalo mass function $\log_{10} \Sigma_{\rm{sub}} = -1.4 \pm 0.2$.
		
		Our inference method recovers the input values for these parameters, even after marginalizing over the main deflector mass profile including third and fourth order multipole perturbations, the size of the warm dust region surrounding the lensed quasar, and the lensed quasar host galaxy light. The covariance between $\Sigma_{\rm{sub}}$ and $m_{\rm{hm}}$ manifests more prominently for these inferences than for the CDM ground truth (Figure \ref{fig:inference}), but incorporating an informative prior for the amplitude of the subhalo mass function can aid in breaking this covariance. In practice, such a prior could come from N-body simulations or semi-analytic models \citep[e.g.][]{Benson++12,Fiacconi++16,Jiang++21,Nadler++23b,Mansfield++23,Du++24}, which make increasingly robust predictions for the number of main deflector subhalos that appear in projection near the Einstein radius in typical host halos $\sim 10^{13} M_{\odot}$. Alternatively, a prior on the amplitude of the subhalo mass function could come from measurements of the halo mass function in lenses from gravitational imaging \citep[e.g.][]{Vegetti++14,Hezaveh++16,He++22,Wagner-Carena++23}. The marginal likelihood of $m_{\rm{hm}}$ excludes CDM for the posteriors that assume a prior on $\Sigma_{\rm{sub}}$, with $\log_{10} m_{\rm{hm}}  / M_{\odot} > 6.7$ and $\log_{10} m_{\rm{hm}} / M_{\odot} > 5.2$ for flux ratio measurement uncertainties of $1\%$ and $3\%$, respectively. 
		
		Comparing Figures \ref{fig:inferencewdm1} and \ref{fig:inferencewdm3}, the inference on WDM parameters is particularly sensitive to the measurement uncertainty in the flux ratios. The systems observed by JWST \citep{Nierenberg++23} have typical uncertainties of $3\%$, but we have included the inference assuming $1\%$ uncertainties to demonstrate that our inference methodology can recover input ground truth with sufficient measurement precision. The constraints obtained assuming $1 \%$ precision suggest that a larger sample of lenses with $3 \%$ uncertainties could yield a statistically significant constrain on WDM models with a turnover $m_{\rm{hm}} \sim 3 \times 10^7 M_{\odot}$. 
		
		\section{Conclusions}
		\label{sec:conclusions}
		We present a self-consistent formalism to jointly model lensed image positions, flux ratios, and extended lensed arcs in strong lens systems comprised of a multiply-imaged quasar and extended emssion from a background galaxy. To address the computational challenge that has precluded the joint reconstruction of flux ratios and lensed arcs to date, we develop an approximation scheme for full multi-plane ray tracing. To validate the methodology, we test it on 25 simulated strong lens systems to constrain a fiduciary warm dark matter (WDM) model with simulated datasets prepared assuming CDM. We assume observations of lensed arcs come from the HST, and flux ratios from the warm dust region observed by JWST with measurement uncertainties of $3 \%$. Our simulations account for finite source size effects in the calculation of image magnifications, and we marginalize over the unknown source size for each system. The simulated lenses include a complex source morphology in the form of a spiral galaxy, and deviations from an elliptical symmetry in the main lens mass profile parameterized by $m=3$ and $m=4$ multipole terms. As a point of comparison for the new lens modeling techniques we present, we analyze the same set of twenty-five mock lenses using the lens modeling methods presented by \citet{Gilman++19}, which use only image positions and flux ratios to constrain the lens mass profile and substructure properties. Our main results are summarized as follows: 
		
		\begin{itemize}
			\item Incorporating lensed arcs leads to stronger constraints on the free-streaming length of WDM than analyses that use only image positions and flux ratios to constrain the lens model. The $95 \%$ exclusion limit on the half-model mass $m_{\rm{hm}}$ improves by 0.5 dex. The relative likelihood of CDM to WDM improves by factors of $1.3$, $2.5$, $5.6$, and $13.1$ for WDM models with mass function turnovers $m_{\rm{hm}} \in \left[10^7 - 10^{7.5} M_{\odot}\right]$,  $\left[10^{7.5} - 10^{8.0} M_{\odot}\right]$, $\left[10^{8.0} - 10^{8.5} M_{\odot}\right]$, and $\left[10^{8.5} - 10^{9.0} M_{\odot}\right]$, respectively, and the posterior volume shrinks by a factor of $1.8$. In Section \ref{ssec:casestudy}, we show that this additional constraining power comes from breaking degeneracies between large-scale deformation of the deflection field, which we constrain with the lensed arcs, and small-scale perturbation to image magnifications by dark matter halos. 
			\item Our method can recover the free-streaming cutoff in a WDM mass function, although the ability to detect a WDM cutoff near $10^7 M_{\odot}$ requires measurement precision of $1 \%$, or a larger sample of lenses with both flux ratio measurements and lensed arcs than the 25 considered here. Theoretically-motivated priors for the amplitude of the subhalo mass function can aid in breaking covariance between the number of subhalos and the free-streaming cutoff. 
			\item The presence of multipole perturbations in the lens mass profile, provided we also include these terms in the lens models used to analyze data, does not incur a detectable source of systematic bias in inferred substructure properties when only image positions and flux ratios are used to constrain the lens model. Thus, we can apply the methods presented by \citet{Gilman++19} to model quadruply-imaged quasars without prominent lensed arcs.
		\end{itemize}
		
		The methodology we present can be applied to any dark matter model that predicts the form of the halo mass function and halo density profiles. For example, in self-interacting dark matter theories halos can undergo core collapse, a process that significantly increases their central density and therefore their lensing efficiency \citep{Gilman++21,Minor++21,Yang++21,Gilman++23,Nadler++23}. In fuzzy dark matter models, wave interference effects produce density fluctuations in the lens mass profile that impact both flux ratios and lensed arcs \citep{Laroche++22,Powell++23}. In terms of a direct test of a key prediction of cold dark matter, incorporating constraints from lensed arcs increases sensitivity to perturbation by low-mass substructure, as indicated by the stronger constraints on mass function turnovers on scales below $10^{7.5} M_{\odot}$. This increased sensitivity will aid in pushing constraints from strong lensing to scales below the threshold of galaxy formation, to scales $\sim 10^{7} M_{\odot}$ and below. 
		
		To test the methods described in this paper we have made various simplifying assumptions when creating the simulated datasets and in the modeling of dark matter substructure. First, we have assumed perfect knowledge of the PSF when reconstructing the imaging data. In practice, one must simultaneously reconstruct the PSF with the lensed image and source. We can easily incorporate the PSF reconstruction in our analysis when analyzing real data. Second, we have made several simplifying assumptions regarding the dark matter substructure model, including perfect knowledge of the amplitude of the line-of-sight halo mass function and logarithmic slope of the subhalo mass function. Improved treatments of the substructure models based on the semi-analytic model {\tt{galacticus}} are being developed for use in forthcoming analyses, but these models were not developed at the time we created the simulated data used in this work\footnote{Recent updates to {\tt{pyHalo}} (versions 1.2.1 and later) includes an improvement treatment of the tidal evolution of dark matter subhalos calibrated against the most up-to-date version of {\tt{galacticus}} (Gannon et al., in prep).}. We expect an improved treatment of the tidal evolution of subhalos will lead to stronger constraints on dark matter models, but it does not affect our conclusions regarding the relative improvement from incorporating constraints from lensed arcs. 
		
		The number of mock lenses we have analyzed in this work is partially motivated by the number of strong lens systems for which we currently have archival HST imaging data of lensed arcs and flux ratio measurements suitable for a milli-lensing analysis of dark substructure. Suitable flux ratios for milli-lensing must come from a region surrounding the background quasar spatially extended by $\gtrsim 1 \rm{pc}$ such that it becomes immune to contamination from stellar micro-lensing. Such measurements can come from observations of nuclear narrow-line emission from Keck and the HST \citep{Nierenberg++14,Nierenberg++17,Nierenberg++20}, radio measurements from VLBI \citep{Koopmans++04,McKean++07,Hsueh++20}, or emission from the warm dust region measured with JWST \citep{Nierenberg++23}. The synthesis of strong lensing flux ratios and extended lensed arcs, in combination with these various datasets from ground and space-based observatories, advances the observational frontier of cosmic probes of dark matter physics to uncharted territory. 
		
		\section*{Acknowledgments}
		DG thanks Jo Bovy for his continuous support throughout the course of this project. We also thank Tommaso Treu for useful discussions. We also thank the anonymous referee for a very constructive report that improved the presentation of the methodology. 
		
		DG acknowledges support for this work provided by The Brinson Foundation through a Brinson Prize Fellowship grant. SB is supported by NASA Roman WFS Award No: 80NSSC24K0095 and the Department of Physics $\&$ Astronomy, Stony Brook University. AMN and MO acknowledge support by the NSF through grant 2206315 "Collaborative Research: Measuring the physical properties of dark matter with strong gravitational lensing", and from GO-2046 which was provided by NASA through a grant from the Space Telescope Science Institute, which is operated by the Association of Universities for Research in Astronomy, Inc., under NASA contract NAS 5-03127.
		
		Computations performed for this analysis were performed on the Midway2 and Midway3 clusters hosted by the University of Chicago’s Research Computing Center, the Niagara supercomputer at the SciNet HPC Consortium, and the Hoffman2 Shared Cluster provided by the UCLA Institute for Digital Research and Education’s Research Technology Group. SciNet is funded by the Canada Foundation for Innovation; the Government of Ontario; Ontario Research Fund - Research Excellence; and the University of Toronto.
		
		\section*{Data Availability}
		The simulated data we use in this article can be reproduced using the notebooks in the main {\tt{samana}} code repository at https://github.com/dangilman/samana. The code version used for this article was {\tt{samana}} version 1.0.0, or github commit 1708597. The simulated data used in this article were created with {\tt{pyHalo}} commit 4dc87c8.  
		
		\bibliographystyle{mnras}
		\bibliography{bibliography} 

\begin{thebibliography}{}
\makeatletter
\relax
\def\mn@urlcharsother{\let\do\@makeother \do\$\do\&\do\#\do\^\do\_\do\%\do\~}
\def\mn@doi{\begingroup\mn@urlcharsother \@ifnextchar [ {\mn@doi@}
  {\mn@doi@[]}}
\def\mn@doi@[#1]#2{\def\@tempa{#1}\ifx\@tempa\@empty \href
  {http://dx.doi.org/#2} {doi:#2}\else \href {http://dx.doi.org/#2} {#1}\fi
  \endgroup}
\def\mn@eprint#1#2{\mn@eprint@#1:#2::\@nil}
\def\mn@eprint@arXiv#1{\href {http://arxiv.org/abs/#1} {{\tt arXiv:#1}}}
\def\mn@eprint@dblp#1{\href {http://dblp.uni-trier.de/rec/bibtex/#1.xml}
  {dblp:#1}}
\def\mn@eprint@#1:#2:#3:#4\@nil{\def\@tempa {#1}\def\@tempb {#2}\def\@tempc
  {#3}\ifx \@tempc \@empty \let \@tempc \@tempb \let \@tempb \@tempa \fi \ifx
  \@tempb \@empty \def\@tempb {arXiv}\fi \@ifundefined
  {mn@eprint@\@tempb}{\@tempb:\@tempc}{\expandafter \expandafter \csname
  mn@eprint@\@tempb\endcsname \expandafter{\@tempc}}}

\bibitem[\protect\citeauthoryear{{Abazajian} \& {Kusenko}}{{Abazajian} \&
  {Kusenko}}{2019}]{AbazajianKusenko19}
{Abazajian} K.~N.,  {Kusenko} A.,  2019, \mn@doi [\prd]
  {10.1103/PhysRevD.100.103513}, \href
  {https://ui.adsabs.harvard.edu/abs/2019PhRvD.100j3513A} {100, 103513}

\bibitem[\protect\citeauthoryear{{Adhikari}, {Dalal}  \&
  {Chamberlain}}{{Adhikari} et~al.}{2014}]{Adhikari++14}
{Adhikari} S.,  {Dalal} N.,   {Chamberlain} R.~T.,  2014, \mn@doi [\jcap]
  {10.1088/1475-7516/2014/11/019}, \href
  {https://ui.adsabs.harvard.edu/abs/2014JCAP...11..019A} {2014, 019}

\bibitem[\protect\citeauthoryear{{Akita} \& {Ando}}{{Akita} \&
  {Ando}}{2023}]{AkitaAndo23}
{Akita} K.,  {Ando} S.,  2023, \mn@doi [\jcap] {10.1088/1475-7516/2023/11/037},
  \href {https://ui.adsabs.harvard.edu/abs/2023JCAP...11..037A} {2023, 037}

\bibitem[\protect\citeauthoryear{{Amorisco} et~al.,}{{Amorisco}
  et~al.}{2022}]{Amorisco++22}
{Amorisco} N.~C.,  et~al., 2022, \mn@doi [\mnras] {10.1093/mnras/stab3527},
  \href {https://ui.adsabs.harvard.edu/abs/2022MNRAS.510.2464A} {510, 2464}

\bibitem[\protect\citeauthoryear{{Ando}, {Hiroshima}  \& {Ishiwata}}{{Ando}
  et~al.}{2022}]{Ando++22}
{Ando} S.,  {Hiroshima} N.,   {Ishiwata} K.,  2022, \mn@doi [\prd]
  {10.1103/PhysRevD.106.103014}, \href
  {https://ui.adsabs.harvard.edu/abs/2022PhRvD.106j3014A} {106, 103014}

\bibitem[\protect\citeauthoryear{{Auger}, {Treu}, {Bolton}, {Gavazzi},
  {Koopmans}, {Marshall}, {Moustakas}  \& {Burles}}{{Auger}
  et~al.}{2010}]{Auger++10}
{Auger} M.~W.,  {Treu} T.,  {Bolton} A.~S.,  {Gavazzi} R.,  {Koopmans}
  L.~V.~E.,  {Marshall} P.~J.,  {Moustakas} L.~A.,   {Burles} S.,  2010,
  \mn@doi [\apj] {10.1088/0004-637X/724/1/511}, \href
  {https://ui.adsabs.harvard.edu/abs/2010ApJ...724..511A} {724, 511}

\bibitem[\protect\citeauthoryear{{Balberg}, {Shapiro}  \& {Inagaki}}{{Balberg}
  et~al.}{2002}]{Balberg++02}
{Balberg} S.,  {Shapiro} S.~L.,   {Inagaki} S.,  2002, \mn@doi [\apj]
  {10.1086/339038}, \href
  {https://ui.adsabs.harvard.edu/abs/2002ApJ...568..475B} {568, 475}

\bibitem[\protect\citeauthoryear{{Ballard}, {Enzi}, {Collett}, {Turner}  \&
  {Smith}}{{Ballard} et~al.}{2024}]{Ballard++23}
{Ballard} D.~J.,  {Enzi} W. J.~R.,  {Collett} T.~E.,  {Turner} H.~C.,   {Smith}
  R.~J.,  2024, \mn@doi [\mnras] {10.1093/mnras/stae514}, \href
  {https://ui.adsabs.harvard.edu/abs/2024MNRAS.528.7564B} {528, 7564}

\bibitem[\protect\citeauthoryear{{Baltz}, {Marshall}  \& {Oguri}}{{Baltz}
  et~al.}{2009}]{Baltz++09}
{Baltz} E.~A.,  {Marshall} P.,   {Oguri} M.,  2009, \mn@doi [\jcap]
  {10.1088/1475-7516/2009/01/015}, \href
  {https://ui.adsabs.harvard.edu/abs/2009JCAP...01..015B} {2009, 015}

\bibitem[\protect\citeauthoryear{{Banik}, {Bertone}, {Bovy}  \&
  {Bozorgnia}}{{Banik} et~al.}{2018}]{Banik++18}
{Banik} N.,  {Bertone} G.,  {Bovy} J.,   {Bozorgnia} N.,  2018, \mn@doi [\jcap]
  {10.1088/1475-7516/2018/07/061}, \href
  {https://ui.adsabs.harvard.edu/abs/2018JCAP...07..061B} {2018, 061}

\bibitem[\protect\citeauthoryear{{Banik}, {Bovy}, {Bertone}, {Erkal}  \& {de
  Boer}}{{Banik} et~al.}{2021}]{Banik++21}
{Banik} N.,  {Bovy} J.,  {Bertone} G.,  {Erkal} D.,   {de Boer} T.~J.~L.,
  2021, \mn@doi [\jcap] {10.1088/1475-7516/2021/10/043}, \href
  {https://ui.adsabs.harvard.edu/abs/2021JCAP...10..043B} {2021, 043}

\bibitem[\protect\citeauthoryear{{Bechtol} et~al.,}{{Bechtol}
  et~al.}{2022}]{Bechtol++22}
{Bechtol} K.,  et~al., 2022, \mn@doi [arXiv e-prints]
  {10.48550/arXiv.2203.07354}, \href
  {https://ui.adsabs.harvard.edu/abs/2022arXiv220307354B} {p. arXiv:2203.07354}

\bibitem[\protect\citeauthoryear{{Bender}, {Surma}, {Doebereiner},
  {Moellenhoff}  \& {Madejsky}}{{Bender} et~al.}{1989}]{Bender++89}
{Bender} R.,  {Surma} P.,  {Doebereiner} S.,  {Moellenhoff} C.,   {Madejsky}
  R.,  1989, \aap, \href
  {https://ui.adsabs.harvard.edu/abs/1989A&A...217...35B} {217, 35}

\bibitem[\protect\citeauthoryear{{Benson}}{{Benson}}{2012}]{Benson++12}
{Benson} A.~J.,  2012, \mn@doi [\na] {10.1016/j.newast.2011.07.004}, \href
  {https://ui.adsabs.harvard.edu/abs/2012NewA...17..175B} {17, 175}

\bibitem[\protect\citeauthoryear{{Birrer} \& {Amara}}{{Birrer} \&
  {Amara}}{2018}]{BirrerAmara18}
{Birrer} S.,  {Amara} A.,  2018, \mn@doi [Physics of the Dark Universe]
  {10.1016/j.dark.2018.11.002}, \href
  {https://ui.adsabs.harvard.edu/abs/2018PDU....22..189B} {22, 189}

\bibitem[\protect\citeauthoryear{{Birrer}, {Amara}  \& {Refregier}}{{Birrer}
  et~al.}{2015}]{Birrer++15}
{Birrer} S.,  {Amara} A.,   {Refregier} A.,  2015, \mn@doi [\apj]
  {10.1088/0004-637X/813/2/102}, \href
  {https://ui.adsabs.harvard.edu/abs/2015ApJ...813..102B} {813, 102}

\bibitem[\protect\citeauthoryear{{Birrer}, {Amara}  \& {Refregier}}{{Birrer}
  et~al.}{2017}]{Birrer++17}
{Birrer} S.,  {Amara} A.,   {Refregier} A.,  2017, \mn@doi [\jcap]
  {10.1088/1475-7516/2017/05/037}, \href
  {https://ui.adsabs.harvard.edu/abs/2017JCAP...05..037B} {2017, 037}

\bibitem[\protect\citeauthoryear{{Birrer} et~al.,}{{Birrer}
  et~al.}{2021}]{Birrer++21}
{Birrer} S.,  et~al., 2021, \mn@doi [The Journal of Open Source Software]
  {10.21105/joss.03283}, \href
  {https://ui.adsabs.harvard.edu/abs/2021JOSS....6.3283B} {6, 3283}

\bibitem[\protect\citeauthoryear{{Blandford} \& {Narayan}}{{Blandford} \&
  {Narayan}}{1986}]{Blandford++86}
{Blandford} R.,  {Narayan} R.,  1986, \mn@doi [\apj] {10.1086/164709}, \href
  {https://ui.adsabs.harvard.edu/abs/1986ApJ...310..568B} {310, 568}

\bibitem[\protect\citeauthoryear{{Bode}, {Ostriker}  \& {Turok}}{{Bode}
  et~al.}{2001}]{Bode++01}
{Bode} P.,  {Ostriker} J.~P.,   {Turok} N.,  2001, \mn@doi [\apj]
  {10.1086/321541}, \href
  {https://ui.adsabs.harvard.edu/abs/2001ApJ...556...93B} {556, 93}

\bibitem[\protect\citeauthoryear{{Bonaca}, {Hogg}, {Price-Whelan}  \&
  {Conroy}}{{Bonaca} et~al.}{2019}]{Bonaca++19}
{Bonaca} A.,  {Hogg} D.~W.,  {Price-Whelan} A.~M.,   {Conroy} C.,  2019,
  \mn@doi [\apj] {10.3847/1538-4357/ab2873}, \href
  {https://ui.adsabs.harvard.edu/abs/2019ApJ...880...38B} {880, 38}

\bibitem[\protect\citeauthoryear{{Bond} \& {Szalay}}{{Bond} \&
  {Szalay}}{1983}]{Bond++83}
{Bond} J.~R.,  {Szalay} A.~S.,  1983, \mn@doi [\apj] {10.1086/161460}, \href
  {https://ui.adsabs.harvard.edu/abs/1983ApJ...274..443B} {274, 443}

\bibitem[\protect\citeauthoryear{{Bose}, {Hellwing}, {Frenk}, {Jenkins},
  {Lovell}, {Helly}  \& {Li}}{{Bose} et~al.}{2016}]{Bose++16}
{Bose} S.,  {Hellwing} W.~A.,  {Frenk} C.~S.,  {Jenkins} A.,  {Lovell} M.~R.,
  {Helly} J.~C.,   {Li} B.,  2016, \mn@doi [\mnras] {10.1093/mnras/stv2294},
  \href {https://ui.adsabs.harvard.edu/abs/2016MNRAS.455..318B} {455, 318}

\bibitem[\protect\citeauthoryear{{Bovy}, {Erkal}  \& {Sanders}}{{Bovy}
  et~al.}{2017}]{Bovy++17}
{Bovy} J.,  {Erkal} D.,   {Sanders} J.~L.,  2017, \mn@doi [\mnras]
  {10.1093/mnras/stw3067}, \href
  {https://ui.adsabs.harvard.edu/abs/2017MNRAS.466..628B} {466, 628}

\bibitem[\protect\citeauthoryear{{Bringmann}, {Scott}  \& {Akrami}}{{Bringmann}
  et~al.}{2012}]{Bringmann++12}
{Bringmann} T.,  {Scott} P.,   {Akrami} Y.,  2012, \mn@doi [\prd]
  {10.1103/PhysRevD.85.125027}, \href
  {https://ui.adsabs.harvard.edu/abs/2012PhRvD..85l5027B} {85, 125027}

\bibitem[\protect\citeauthoryear{{Buckley} \& {Peter}}{{Buckley} \&
  {Peter}}{2018}]{Buckley++18}
{Buckley} M.~R.,  {Peter} A. H.~G.,  2018, \mn@doi [\physrep]
  {10.1016/j.physrep.2018.07.003}, \href
  {https://ui.adsabs.harvard.edu/abs/2018PhR...761....1B} {761, 1}

\bibitem[\protect\citeauthoryear{{Cao} et~al.,}{{Cao} et~al.}{2022}]{Cao++22}
{Cao} X.,  et~al., 2022, \mn@doi [Research in Astronomy and Astrophysics]
  {10.1088/1674-4527/ac3f2b}, \href
  {https://ui.adsabs.harvard.edu/abs/2022RAA....22b5014C} {22, 025014}

\bibitem[\protect\citeauthoryear{{Cohen}, {Fassnacht}, {O'Riordan}  \&
  {Vegetti}}{{Cohen} et~al.}{2024}]{Cohen++24}
{Cohen} J.~S.,  {Fassnacht} C.~D.,  {O'Riordan} C.~M.,   {Vegetti} S.,  2024,
  \mn@doi [\mnras] {10.1093/mnras/stae1228}, \href
  {https://ui.adsabs.harvard.edu/abs/2024MNRAS.531.3431C} {531, 3431}

\bibitem[\protect\citeauthoryear{{Correa}}{{Correa}}{2021}]{Correa++21}
{Correa} C.~A.,  2021, \mn@doi [\mnras] {10.1093/mnras/stab506}, \href
  {https://ui.adsabs.harvard.edu/abs/2021MNRAS.503..920C} {503, 920}

\bibitem[\protect\citeauthoryear{{Dalal} \& {Kochanek}}{{Dalal} \&
  {Kochanek}}{2002}]{DalalKochanek02}
{Dalal} N.,  {Kochanek} C.~S.,  2002, \mn@doi [\apj] {10.1086/340303}, \href
  {https://ui.adsabs.harvard.edu/abs/2002ApJ...572...25D} {572, 25}

\bibitem[\protect\citeauthoryear{{Dekker}, {Ando}, {Correa}  \& {Ng}}{{Dekker}
  et~al.}{2022}]{Dekker++22}
{Dekker} A.,  {Ando} S.,  {Correa} C.~A.,   {Ng} K. C.~Y.,  2022, \mn@doi
  [\prd] {10.1103/PhysRevD.106.123026}, \href
  {https://ui.adsabs.harvard.edu/abs/2022PhRvD.106l3026D} {106, 123026}

\bibitem[\protect\citeauthoryear{{Despali}, {Vegetti}, {White}, {Powell},
  {Stacey}, {Fassnacht}, {Rizzo}  \& {Enzi}}{{Despali}
  et~al.}{2022}]{Despali++22}
{Despali} G.,  {Vegetti} S.,  {White} S. D.~M.,  {Powell} D.~M.,  {Stacey}
  H.~R.,  {Fassnacht} C.~D.,  {Rizzo} F.,   {Enzi} W.,  2022, \mn@doi [\mnras]
  {10.1093/mnras/stab3537}, \href
  {https://ui.adsabs.harvard.edu/abs/2022MNRAS.510.2480D} {510, 2480}

\bibitem[\protect\citeauthoryear{{Dhanasingham}, {Cyr-Racine}, {Peter},
  {Benson}  \& {Gilman}}{{Dhanasingham} et~al.}{2023}]{Dhanasingham++23}
{Dhanasingham} B.,  {Cyr-Racine} F.-Y.,  {Peter} A. H.~G.,  {Benson} A.,
  {Gilman} D.,  2023, \mn@doi [\mnras] {10.1093/mnras/stac2993}, \href
  {https://ui.adsabs.harvard.edu/abs/2023MNRAS.518.5843D} {518, 5843}

\bibitem[\protect\citeauthoryear{{Diemer}}{{Diemer}}{2018}]{Diemer18}
{Diemer} B.,  2018, \mn@doi [\apjs] {10.3847/1538-4365/aaee8c}, \href
  {https://ui.adsabs.harvard.edu/abs/2018ApJS..239...35D} {239, 35}

\bibitem[\protect\citeauthoryear{{Diemer} \& {Joyce}}{{Diemer} \&
  {Joyce}}{2019}]{DiemerJoyce19}
{Diemer} B.,  {Joyce} M.,  2019, \mn@doi [\apj] {10.3847/1538-4357/aafad6},
  \href {https://ui.adsabs.harvard.edu/abs/2019ApJ...871..168D} {871, 168}

\bibitem[\protect\citeauthoryear{{Diemer} \& {Kravtsov}}{{Diemer} \&
  {Kravtsov}}{2014}]{DiemerKravtsov14}
{Diemer} B.,  {Kravtsov} A.~V.,  2014, \mn@doi [\apj]
  {10.1088/0004-637X/789/1/1}, \href
  {https://ui.adsabs.harvard.edu/abs/2014ApJ...789....1D} {789, 1}

\bibitem[\protect\citeauthoryear{{Dike}, {Gilman}  \& {Treu}}{{Dike}
  et~al.}{2023}]{Dike++23}
{Dike} V.,  {Gilman} D.,   {Treu} T.,  2023, \mn@doi [\mnras]
  {10.1093/mnras/stad1313}, \href
  {https://ui.adsabs.harvard.edu/abs/2023MNRAS.522.5434D} {522, 5434}

\bibitem[\protect\citeauthoryear{{Dobler} \& {Keeton}}{{Dobler} \&
  {Keeton}}{2006}]{DoblerKeeton06}
{Dobler} G.,  {Keeton} C.~R.,  2006, \mn@doi [\mnras]
  {10.1111/j.1365-2966.2005.09809.x}, \href
  {https://ui.adsabs.harvard.edu/abs/2006MNRAS.365.1243D} {365, 1243}

\bibitem[\protect\citeauthoryear{{Drlica-Wagner} et~al.,}{{Drlica-Wagner}
  et~al.}{2022}]{DrlicaWagner++22}
{Drlica-Wagner} A.,  et~al., 2022, \mn@doi [arXiv e-prints]
  {10.48550/arXiv.2209.08215}, \href
  {https://ui.adsabs.harvard.edu/abs/2022arXiv220908215D} {p. arXiv:2209.08215}

\bibitem[\protect\citeauthoryear{{Du} et~al.,}{{Du} et~al.}{2024}]{Du++24}
{Du} X.,  et~al., 2024, \mn@doi [arXiv e-prints] {10.48550/arXiv.2403.09597},
  \href {https://ui.adsabs.harvard.edu/abs/2024arXiv240309597D} {p.
  arXiv:2403.09597}

\bibitem[\protect\citeauthoryear{{Esteban}, {Peter}  \& {Kim}}{{Esteban}
  et~al.}{2023}]{Esteban++23}
{Esteban} I.,  {Peter} A. H.~G.,   {Kim} S.~Y.,  2023, \mn@doi [arXiv e-prints]
  {10.48550/arXiv.2306.04674}, \href
  {https://ui.adsabs.harvard.edu/abs/2023arXiv230604674E} {p. arXiv:2306.04674}

\bibitem[\protect\citeauthoryear{{Fiacconi}, {Madau}, {Potter}  \&
  {Stadel}}{{Fiacconi} et~al.}{2016}]{Fiacconi++16}
{Fiacconi} D.,  {Madau} P.,  {Potter} D.,   {Stadel} J.,  2016, \mn@doi [\apj]
  {10.3847/0004-637X/824/2/144}, \href
  {https://ui.adsabs.harvard.edu/abs/2016ApJ...824..144F} {824, 144}

\bibitem[\protect\citeauthoryear{{Fleury}, {Larena}  \& {Uzan}}{{Fleury}
  et~al.}{2021}]{Fleury++21}
{Fleury} P.,  {Larena} J.,   {Uzan} J.-P.,  2021, \mn@doi [\jcap]
  {10.1088/1475-7516/2021/08/024}, \href
  {https://ui.adsabs.harvard.edu/abs/2021JCAP...08..024F} {2021, 024}

\bibitem[\protect\citeauthoryear{{Gilman}, {Birrer}, {Treu}, {Nierenberg}  \&
  {Benson}}{{Gilman} et~al.}{2019}]{Gilman++19}
{Gilman} D.,  {Birrer} S.,  {Treu} T.,  {Nierenberg} A.,   {Benson} A.,  2019,
  \mn@doi [\mnras] {10.1093/mnras/stz1593}, \href
  {https://ui.adsabs.harvard.edu/abs/2019MNRAS.487.5721G} {487, 5721}

\bibitem[\protect\citeauthoryear{{Gilman}, {Birrer}, {Nierenberg}, {Treu}, {Du}
   \& {Benson}}{{Gilman} et~al.}{2020}]{Gilman++20}
{Gilman} D.,  {Birrer} S.,  {Nierenberg} A.,  {Treu} T.,  {Du} X.,   {Benson}
  A.,  2020, \mn@doi [\mnras] {10.1093/mnras/stz3480}, \href
  {https://ui.adsabs.harvard.edu/abs/2020MNRAS.491.6077G} {491, 6077}

\bibitem[\protect\citeauthoryear{{Gilman}, {Bovy}, {Treu}, {Nierenberg},
  {Birrer}, {Benson}  \& {Sameie}}{{Gilman} et~al.}{2021}]{Gilman++21}
{Gilman} D.,  {Bovy} J.,  {Treu} T.,  {Nierenberg} A.,  {Birrer} S.,  {Benson}
  A.,   {Sameie} O.,  2021, \mn@doi [\mnras] {10.1093/mnras/stab2335}, \href
  {https://ui.adsabs.harvard.edu/abs/2021MNRAS.507.2432G} {507, 2432}

\bibitem[\protect\citeauthoryear{{Gilman}, {Benson}, {Bovy}, {Birrer}, {Treu}
  \& {Nierenberg}}{{Gilman} et~al.}{2022}]{Gilman++22}
{Gilman} D.,  {Benson} A.,  {Bovy} J.,  {Birrer} S.,  {Treu} T.,   {Nierenberg}
  A.,  2022, \mn@doi [\mnras] {10.1093/mnras/stac670}, \href
  {https://ui.adsabs.harvard.edu/abs/2022MNRAS.512.3163G} {512, 3163}

\bibitem[\protect\citeauthoryear{{Gilman}, {Zhong}  \& {Bovy}}{{Gilman}
  et~al.}{2023}]{Gilman++23}
{Gilman} D.,  {Zhong} Y.-M.,   {Bovy} J.,  2023, \mn@doi [\prd]
  {10.1103/PhysRevD.107.103008}, \href
  {https://ui.adsabs.harvard.edu/abs/2023PhRvD.107j3008G} {107, 103008}

\bibitem[\protect\citeauthoryear{{Hao}, {Mao}, {Deng}, {Xia}  \& {Wu}}{{Hao}
  et~al.}{2006}]{Hao++06}
{Hao} C.~N.,  {Mao} S.,  {Deng} Z.~G.,  {Xia} X.~Y.,   {Wu} H.,  2006, \mn@doi
  [\mnras] {10.1111/j.1365-2966.2006.10545.x}, \href
  {https://ui.adsabs.harvard.edu/abs/2006MNRAS.370.1339H} {370, 1339}

\bibitem[\protect\citeauthoryear{{He} et~al.,}{{He} et~al.}{2022}]{He++22}
{He} Q.,  et~al., 2022, \mn@doi [\mnras] {10.1093/mnras/stac191}, \href
  {https://ui.adsabs.harvard.edu/abs/2022MNRAS.511.3046H} {511, 3046}

\bibitem[\protect\citeauthoryear{{He} et~al.,}{{He} et~al.}{2023}]{He++23}
{He} Q.,  et~al., 2023, \mn@doi [\mnras] {10.1093/mnras/stac2779}, \href
  {https://ui.adsabs.harvard.edu/abs/2023MNRAS.518..220H} {518, 220}

\bibitem[\protect\citeauthoryear{{Hezaveh} et~al.,}{{Hezaveh}
  et~al.}{2016}]{Hezaveh++16}
{Hezaveh} Y.~D.,  et~al., 2016, \mn@doi [\apj] {10.3847/0004-637X/823/1/37},
  \href {https://ui.adsabs.harvard.edu/abs/2016ApJ...823...37H} {823, 37}

\bibitem[\protect\citeauthoryear{{Hsueh}, {Enzi}, {Vegetti}, {Auger},
  {Fassnacht}, {Despali}, {Koopmans}  \& {McKean}}{{Hsueh}
  et~al.}{2020}]{Hsueh++20}
{Hsueh} J.~W.,  {Enzi} W.,  {Vegetti} S.,  {Auger} M.~W.,  {Fassnacht} C.~D.,
  {Despali} G.,  {Koopmans} L.~V.~E.,   {McKean} J.~P.,  2020, \mn@doi [\mnras]
  {10.1093/mnras/stz3177}, \href
  {https://ui.adsabs.harvard.edu/abs/2020MNRAS.492.3047H} {492, 3047}

\bibitem[\protect\citeauthoryear{{Jiang}, {Dekel}, {Freundlich}, {van den
  Bosch}, {Green}, {Hopkins}, {Benson}  \& {Du}}{{Jiang}
  et~al.}{2021}]{Jiang++21}
{Jiang} F.,  {Dekel} A.,  {Freundlich} J.,  {van den Bosch} F.~C.,  {Green}
  S.~B.,  {Hopkins} P.~F.,  {Benson} A.,   {Du} X.,  2021, \mn@doi [\mnras]
  {10.1093/mnras/staa4034}, \href
  {https://ui.adsabs.harvard.edu/abs/2021MNRAS.502..621J} {502, 621}

\bibitem[\protect\citeauthoryear{{Keeley}, {Nierenberg}, {Gilman}, {Birrer},
  {Benson}  \& {Treu}}{{Keeley} et~al.}{2023}]{Keeley++23}
{Keeley} R.~E.,  {Nierenberg} A.~M.,  {Gilman} D.,  {Birrer} S.,  {Benson} A.,
   {Treu} T.,  2023, \mn@doi [\mnras] {10.1093/mnras/stad2251}, \href
  {https://ui.adsabs.harvard.edu/abs/2023MNRAS.524.6159K} {524, 6159}

\bibitem[\protect\citeauthoryear{{Kim}, {Peter}  \& {Hargis}}{{Kim}
  et~al.}{2018}]{Kim++18}
{Kim} S.~Y.,  {Peter} A. H.~G.,   {Hargis} J.~R.,  2018, \mn@doi [\prl]
  {10.1103/PhysRevLett.121.211302}, \href
  {https://ui.adsabs.harvard.edu/abs/2018PhRvL.121u1302K} {121, 211302}

\bibitem[\protect\citeauthoryear{{Koekemoer} et~al.,}{{Koekemoer}
  et~al.}{2007}]{Koekemoer++07}
{Koekemoer} A.~M.,  et~al., 2007, \mn@doi [\apjs] {10.1086/520086}, \href
  {https://ui.adsabs.harvard.edu/abs/2007ApJS..172..196K} {172, 196}

\bibitem[\protect\citeauthoryear{{Koopmans}, {Browne}  \& {Jackson}}{{Koopmans}
  et~al.}{2004}]{Koopmans++04}
{Koopmans} L.~V.~E.,  {Browne} I.~W.~A.,   {Jackson} N.~J.,  2004, \mn@doi
  [\nar] {10.1016/j.newar.2004.09.047}, \href
  {https://ui.adsabs.harvard.edu/abs/2004NewAR..48.1085K} {48, 1085}

\bibitem[\protect\citeauthoryear{{Laroche}, {Gilman}, {Li}, {Bovy}  \&
  {Du}}{{Laroche} et~al.}{2022}]{Laroche++22}
{Laroche} A.,  {Gilman} D.,  {Li} X.,  {Bovy} J.,   {Du} X.,  2022, \mn@doi
  [\mnras] {10.1093/mnras/stac2677}, \href
  {https://ui.adsabs.harvard.edu/abs/2022MNRAS.517.1867L} {517, 1867}

\bibitem[\protect\citeauthoryear{{Lovell}}{{Lovell}}{2020}]{Lovell++20}
{Lovell} M.~R.,  2020, \mn@doi [\apj] {10.3847/1538-4357/ab982a}, \href
  {https://ui.adsabs.harvard.edu/abs/2020ApJ...897..147L} {897, 147}

\bibitem[\protect\citeauthoryear{{Ludlow}, {Bose}, {Angulo}, {Wang},
  {Hellwing}, {Navarro}, {Cole}  \& {Frenk}}{{Ludlow}
  et~al.}{2016}]{Ludlow++16}
{Ludlow} A.~D.,  {Bose} S.,  {Angulo} R.~E.,  {Wang} L.,  {Hellwing} W.~A.,
  {Navarro} J.~F.,  {Cole} S.,   {Frenk} C.~S.,  2016, \mn@doi [\mnras]
  {10.1093/mnras/stw1046}, \href
  {https://ui.adsabs.harvard.edu/abs/2016MNRAS.460.1214L} {460, 1214}

\bibitem[\protect\citeauthoryear{{Mansfield}, {Darragh-Ford}, {Wang}, {Nadler},
  {Diemer}  \& {Wechsler}}{{Mansfield} et~al.}{2024}]{Mansfield++23}
{Mansfield} P.,  {Darragh-Ford} E.,  {Wang} Y.,  {Nadler} E.~O.,  {Diemer} B.,
   {Wechsler} R.~H.,  2024, \mn@doi [\apj] {10.3847/1538-4357/ad4e33}, \href
  {https://ui.adsabs.harvard.edu/abs/2024ApJ...970..178M} {970, 178}

\bibitem[\protect\citeauthoryear{{McKean} et~al.,}{{McKean}
  et~al.}{2007}]{McKean++07}
{McKean} J.~P.,  et~al., 2007, \mn@doi [\mnras]
  {10.1111/j.1365-2966.2007.11744.x}, \href
  {https://ui.adsabs.harvard.edu/abs/2007MNRAS.378..109M} {378, 109}

\bibitem[\protect\citeauthoryear{{Minor}, {Gad-Nasr}, {Kaplinghat}  \&
  {Vegetti}}{{Minor} et~al.}{2021}]{Minor++21}
{Minor} Q.,  {Gad-Nasr} S.,  {Kaplinghat} M.,   {Vegetti} S.,  2021, \mn@doi
  [\mnras] {10.1093/mnras/stab2247}, \href
  {https://ui.adsabs.harvard.edu/abs/2021MNRAS.507.1662M} {507, 1662}

\bibitem[\protect\citeauthoryear{{Mondino}, {Tsantilas}, {Taki}, {Van Tilburg}
  \& {Weiner}}{{Mondino} et~al.}{2024}]{Mondino++23}
{Mondino} C.,  {Tsantilas} A.,  {Taki} A.-M.,  {Van Tilburg} K.,   {Weiner} N.,
   2024, \mn@doi [\mnras] {10.1093/mnras/stae1017}, \href
  {https://ui.adsabs.harvard.edu/abs/2024MNRAS.531..632M} {531, 632}

\bibitem[\protect\citeauthoryear{{More}, {Diemer}  \& {Kravtsov}}{{More}
  et~al.}{2015}]{More++15}
{More} S.,  {Diemer} B.,   {Kravtsov} A.~V.,  2015, \mn@doi [\apj]
  {10.1088/0004-637X/810/1/36}, \href
  {https://ui.adsabs.harvard.edu/abs/2015ApJ...810...36M} {810, 36}

\bibitem[\protect\citeauthoryear{{Nadler} et~al.,}{{Nadler}
  et~al.}{2021}]{Nadler++21}
{Nadler} E.~O.,  et~al., 2021, \mn@doi [\prl] {10.1103/PhysRevLett.126.091101},
  \href {https://ui.adsabs.harvard.edu/abs/2021PhRvL.126i1101N} {126, 091101}

\bibitem[\protect\citeauthoryear{{Nadler} et~al.,}{{Nadler}
  et~al.}{2023a}]{Nadler++23b}
{Nadler} E.~O.,  et~al., 2023a, \mn@doi [\apj] {10.3847/1538-4357/acb68c},
  \href {https://ui.adsabs.harvard.edu/abs/2023ApJ...945..159N} {945, 159}

\bibitem[\protect\citeauthoryear{{Nadler}, {Yang}  \& {Yu}}{{Nadler}
  et~al.}{2023b}]{Nadler++23}
{Nadler} E.~O.,  {Yang} D.,   {Yu} H.-B.,  2023b, \mn@doi [\apjl]
  {10.3847/2041-8213/ad0e09}, \href
  {https://ui.adsabs.harvard.edu/abs/2023ApJ...958L..39N} {958, L39}

\bibitem[\protect\citeauthoryear{{Nadler}, {Gluscevic}, {Driskell}, {Wechsler},
  {Moustakas}, {Benson}  \& {Mao}}{{Nadler} et~al.}{2024}]{Nadler++24}
{Nadler} E.~O.,  {Gluscevic} V.,  {Driskell} T.,  {Wechsler} R.~H.,
  {Moustakas} L.~A.,  {Benson} A.,   {Mao} Y.-Y.,  2024, \mn@doi [\apj]
  {10.3847/1538-4357/ad3bb1}, \href
  {https://ui.adsabs.harvard.edu/abs/2024ApJ...967...61N} {967, 61}

\bibitem[\protect\citeauthoryear{{Navarro}, {Frenk}  \& {White}}{{Navarro}
  et~al.}{1997}]{NFW}
{Navarro} J.~F.,  {Frenk} C.~S.,   {White} S. D.~M.,  1997, \mn@doi [\apj]
  {10.1086/304888}, \href
  {https://ui.adsabs.harvard.edu/abs/1997ApJ...490..493N} {490, 493}

\bibitem[\protect\citeauthoryear{{Nierenberg}, {Treu}, {Wright}, {Fassnacht}
  \& {Auger}}{{Nierenberg} et~al.}{2014}]{Nierenberg++14}
{Nierenberg} A.~M.,  {Treu} T.,  {Wright} S.~A.,  {Fassnacht} C.~D.,   {Auger}
  M.~W.,  2014, \mn@doi [\mnras] {10.1093/mnras/stu862}, \href
  {https://ui.adsabs.harvard.edu/abs/2014MNRAS.442.2434N} {442, 2434}

\bibitem[\protect\citeauthoryear{{Nierenberg} et~al.,}{{Nierenberg}
  et~al.}{2017}]{Nierenberg++17}
{Nierenberg} A.~M.,  et~al., 2017, \mn@doi [\mnras] {10.1093/mnras/stx1400},
  \href {https://ui.adsabs.harvard.edu/abs/2017MNRAS.471.2224N} {471, 2224}

\bibitem[\protect\citeauthoryear{{Nierenberg} et~al.,}{{Nierenberg}
  et~al.}{2020}]{Nierenberg++20}
{Nierenberg} A.~M.,  et~al., 2020, \mn@doi [\mnras] {10.1093/mnras/stz3588},
  \href {https://ui.adsabs.harvard.edu/abs/2020MNRAS.492.5314N} {492, 5314}

\bibitem[\protect\citeauthoryear{{Nierenberg} et~al.,}{{Nierenberg}
  et~al.}{2024}]{Nierenberg++23}
{Nierenberg} A.~M.,  et~al., 2024, \mn@doi [\mnras] {10.1093/mnras/stae499},
  \href {https://ui.adsabs.harvard.edu/abs/2024MNRAS.530.2960N} {530, 2960}

\bibitem[\protect\citeauthoryear{{Nightingale} et~al.,}{{Nightingale}
  et~al.}{2024}]{Nightingale++24}
{Nightingale} J.~W.,  et~al., 2024, \mn@doi [\mnras] {10.1093/mnras/stad3694},
  \href {https://ui.adsabs.harvard.edu/abs/2024MNRAS.52710480N} {527, 10480}

\bibitem[\protect\citeauthoryear{{O'Riordan} \& {Vegetti}}{{O'Riordan} \&
  {Vegetti}}{2024}]{ORiordan++24}
{O'Riordan} C.~M.,  {Vegetti} S.,  2024, \mn@doi [\mnras]
  {10.1093/mnras/stae153}, \href
  {https://ui.adsabs.harvard.edu/abs/2024MNRAS.528.1757O} {528, 1757}

\bibitem[\protect\citeauthoryear{{Oguri} \& {Marshall}}{{Oguri} \&
  {Marshall}}{2010}]{Oguri++10}
{Oguri} M.,  {Marshall} P.~J.,  2010, \mn@doi [\mnras]
  {10.1111/j.1365-2966.2010.16639.x}, \href
  {https://ui.adsabs.harvard.edu/abs/2010MNRAS.405.2579O} {405, 2579}

\bibitem[\protect\citeauthoryear{{Oh}, {Nierenberg}, {Gilman}  \&
  {Birrer}}{{Oh} et~al.}{2024}]{Oh++24}
{Oh} M. S.~H.,  {Nierenberg} A.,  {Gilman} D.,   {Birrer} S.,  2024, \mn@doi
  [arXiv e-prints] {10.48550/arXiv.2404.17124}, \href
  {https://ui.adsabs.harvard.edu/abs/2024arXiv240417124O} {p. arXiv:2404.17124}

\bibitem[\protect\citeauthoryear{{Planck Collaboration} et~al.,}{{Planck
  Collaboration} et~al.}{2020}]{PlanckCosmo}
{Planck Collaboration} et~al., 2020, \mn@doi [\aap]
  {10.1051/0004-6361/201833910}, \href
  {https://ui.adsabs.harvard.edu/abs/2020A&A...641A...6P} {641, A6}

\bibitem[\protect\citeauthoryear{{Powell}, {Vegetti}, {McKean}, {Spingola},
  {Stacey}  \& {Fassnacht}}{{Powell} et~al.}{2022}]{Powell++22}
{Powell} D.~M.,  {Vegetti} S.,  {McKean} J.~P.,  {Spingola} C.,  {Stacey}
  H.~R.,   {Fassnacht} C.~D.,  2022, \mn@doi [\mnras] {10.1093/mnras/stac2350},
  \href {https://ui.adsabs.harvard.edu/abs/2022MNRAS.516.1808P} {516, 1808}

\bibitem[\protect\citeauthoryear{{Powell}, {Vegetti}, {McKean}, {White},
  {Ferreira}, {May}  \& {Spingola}}{{Powell} et~al.}{2023}]{Powell++23}
{Powell} D.~M.,  {Vegetti} S.,  {McKean} J.~P.,  {White} S. D.~M.,  {Ferreira}
  E. G.~M.,  {May} S.,   {Spingola} C.,  2023, \mn@doi [\mnras]
  {10.1093/mnrasl/slad074}, \href
  {https://ui.adsabs.harvard.edu/abs/2023MNRAS.524L..84P} {524, L84}

\bibitem[\protect\citeauthoryear{{Schmidt} et~al.,}{{Schmidt}
  et~al.}{2023}]{Schmidt++23}
{Schmidt} T.,  et~al., 2023, \mn@doi [\mnras] {10.1093/mnras/stac2235}, \href
  {https://ui.adsabs.harvard.edu/abs/2023MNRAS.518.1260S} {518, 1260}

\bibitem[\protect\citeauthoryear{{Schneider}, {Smith}, {Macci{\`o}}  \&
  {Moore}}{{Schneider} et~al.}{2012}]{Schneider++12}
{Schneider} A.,  {Smith} R.~E.,  {Macci{\`o}} A.~V.,   {Moore} B.,  2012,
  \mn@doi [\mnras] {10.1111/j.1365-2966.2012.21252.x}, \href
  {https://ui.adsabs.harvard.edu/abs/2012MNRAS.424..684S} {424, 684}

\bibitem[\protect\citeauthoryear{{Seng{\"u}l}, {Dvorkin}, {Ostdiek}  \&
  {Tsang}}{{Seng{\"u}l} et~al.}{2022}]{Sengul+22}
{Seng{\"u}l} A.~{\c{C}}.,  {Dvorkin} C.,  {Ostdiek} B.,   {Tsang} A.,  2022,
  \mn@doi [\mnras] {10.1093/mnras/stac1967}, \href
  {https://ui.adsabs.harvard.edu/abs/2022MNRAS.515.4391S} {515, 4391}

\bibitem[\protect\citeauthoryear{{S{\'e}rsic}}{{S{\'e}rsic}}{1963}]{Sersic63}
{S{\'e}rsic} J.~L.,  1963, Boletin de la Asociacion Argentina de Astronomia La
  Plata Argentina, \href
  {https://ui.adsabs.harvard.edu/abs/1963BAAA....6...41S} {6, 41}

\bibitem[\protect\citeauthoryear{{Shajib} et~al.,}{{Shajib}
  et~al.}{2019}]{Shajib++19}
{Shajib} A.~J.,  et~al., 2019, \mn@doi [\mnras] {10.1093/mnras/sty3397}, \href
  {https://ui.adsabs.harvard.edu/abs/2019MNRAS.483.5649S} {483, 5649}

\bibitem[\protect\citeauthoryear{{Shajib} et~al.,}{{Shajib}
  et~al.}{2020}]{Shajib++20}
{Shajib} A.~J.,  et~al., 2020, \mn@doi [\mnras] {10.1093/mnras/staa828}, \href
  {https://ui.adsabs.harvard.edu/abs/2020MNRAS.494.6072S} {494, 6072}

\bibitem[\protect\citeauthoryear{{Sheth}, {Mo}  \& {Tormen}}{{Sheth}
  et~al.}{2001}]{ST01}
{Sheth} R.~K.,  {Mo} H.~J.,   {Tormen} G.,  2001, \mn@doi [\mnras]
  {10.1046/j.1365-8711.2001.04006.x}, \href
  {https://ui.adsabs.harvard.edu/abs/2001MNRAS.323....1S} {323, 1}

\bibitem[\protect\citeauthoryear{{Slone}, {Jiang}, {Lisanti}  \&
  {Kaplinghat}}{{Slone} et~al.}{2023}]{Slone++23}
{Slone} O.,  {Jiang} F.,  {Lisanti} M.,   {Kaplinghat} M.,  2023, \mn@doi
  [\prd] {10.1103/PhysRevD.107.043014}, \href
  {https://ui.adsabs.harvard.edu/abs/2023PhRvD.107d3014S} {107, 043014}

\bibitem[\protect\citeauthoryear{{Spergel} \& {Steinhardt}}{{Spergel} \&
  {Steinhardt}}{2000}]{Spergel++00}
{Spergel} D.~N.,  {Steinhardt} P.~J.,  2000, \mn@doi [\prl]
  {10.1103/PhysRevLett.84.3760}, \href
  {https://ui.adsabs.harvard.edu/abs/2000PhRvL..84.3760S} {84, 3760}

\bibitem[\protect\citeauthoryear{{Spingola}, {McKean}, {Auger}, {Fassnacht},
  {Koopmans}, {Lagattuta}  \& {Vegetti}}{{Spingola}
  et~al.}{2018}]{Spingola++18}
{Spingola} C.,  {McKean} J.~P.,  {Auger} M.~W.,  {Fassnacht} C.~D.,  {Koopmans}
  L.~V.~E.,  {Lagattuta} D.~J.,   {Vegetti} S.,  2018, \mn@doi [\mnras]
  {10.1093/mnras/sty1326}, \href
  {https://ui.adsabs.harvard.edu/abs/2018MNRAS.478.4816S} {478, 4816}

\bibitem[\protect\citeauthoryear{{Springel} et~al.,}{{Springel}
  et~al.}{2008}]{Springel++08}
{Springel} V.,  et~al., 2008, \mn@doi [\mnras]
  {10.1111/j.1365-2966.2008.14066.x}, \href
  {https://ui.adsabs.harvard.edu/abs/2008MNRAS.391.1685S} {391, 1685}

\bibitem[\protect\citeauthoryear{{St{\"u}cker}, {Angulo}, {Hahn}  \&
  {White}}{{St{\"u}cker} et~al.}{2022}]{Stucker++22}
{St{\"u}cker} J.,  {Angulo} R.~E.,  {Hahn} O.,   {White} S. D.~M.,  2022,
  \mn@doi [\mnras] {10.1093/mnras/stab3078}, \href
  {https://ui.adsabs.harvard.edu/abs/2022MNRAS.509.1703S} {509, 1703}

\bibitem[\protect\citeauthoryear{{Tulin}, {Yu}  \& {Zurek}}{{Tulin}
  et~al.}{2013}]{Tulin++13}
{Tulin} S.,  {Yu} H.-B.,   {Zurek} K.~M.,  2013, \mn@doi [\prd]
  {10.1103/PhysRevD.87.115007}, \href
  {https://ui.adsabs.harvard.edu/abs/2013PhRvD..87k5007T} {87, 115007}

\bibitem[\protect\citeauthoryear{{Van Tilburg}, {Taki}  \& {Weiner}}{{Van
  Tilburg} et~al.}{2018}]{VanTillburg++18}
{Van Tilburg} K.,  {Taki} A.-M.,   {Weiner} N.,  2018, \mn@doi [\jcap]
  {10.1088/1475-7516/2018/07/041}, \href
  {https://ui.adsabs.harvard.edu/abs/2018JCAP...07..041V} {2018, 041}

\bibitem[\protect\citeauthoryear{{Vegetti}, {Koopmans}, {Auger}, {Treu}  \&
  {Bolton}}{{Vegetti} et~al.}{2014}]{Vegetti++14}
{Vegetti} S.,  {Koopmans} L.~V.~E.,  {Auger} M.~W.,  {Treu} T.,   {Bolton}
  A.~S.,  2014, \mn@doi [\mnras] {10.1093/mnras/stu943}, \href
  {https://ui.adsabs.harvard.edu/abs/2014MNRAS.442.2017V} {442, 2017}

\bibitem[\protect\citeauthoryear{{Vegetti}, {Despali}, {Lovell}  \&
  {Enzi}}{{Vegetti} et~al.}{2018}]{Vegetti++18}
{Vegetti} S.,  {Despali} G.,  {Lovell} M.~R.,   {Enzi} W.,  2018, \mn@doi
  [\mnras] {10.1093/mnras/sty2393}, \href
  {https://ui.adsabs.harvard.edu/abs/2018MNRAS.481.3661V} {481, 3661}

\bibitem[\protect\citeauthoryear{{Vegetti} et~al.,}{{Vegetti}
  et~al.}{2023}]{Vegetti++23}
{Vegetti} S.,  et~al., 2023, \mn@doi [arXiv e-prints]
  {10.48550/arXiv.2306.11781}, \href
  {https://ui.adsabs.harvard.edu/abs/2023arXiv230611781V} {p. arXiv:2306.11781}

\bibitem[\protect\citeauthoryear{{Wagner-Carena}, {Aalbers}, {Birrer},
  {Nadler}, {Darragh-Ford}, {Marshall}  \& {Wechsler}}{{Wagner-Carena}
  et~al.}{2023}]{Wagner-Carena++23}
{Wagner-Carena} S.,  {Aalbers} J.,  {Birrer} S.,  {Nadler} E.~O.,
  {Darragh-Ford} E.,  {Marshall} P.~J.,   {Wechsler} R.~H.,  2023, \mn@doi
  [\apj] {10.3847/1538-4357/aca525}, \href
  {https://ui.adsabs.harvard.edu/abs/2023ApJ...942...75W} {942, 75}

\bibitem[\protect\citeauthoryear{{Yang} \& {Yu}}{{Yang} \&
  {Yu}}{2021}]{Yang++21}
{Yang} D.,  {Yu} H.-B.,  2021, \mn@doi [\prd] {10.1103/PhysRevD.104.103031},
  \href {https://ui.adsabs.harvard.edu/abs/2021PhRvD.104j3031Y} {104, 103031}

\bibitem[\protect\citeauthoryear{{Yang}, {Nadler}  \& {Yu}}{{Yang}
  et~al.}{2023}]{Yang++23}
{Yang} D.,  {Nadler} E.~O.,   {Yu} H.-B.,  2023, \mn@doi [\apj]
  {10.3847/1538-4357/acc73e}, \href
  {https://ui.adsabs.harvard.edu/abs/2023ApJ...949...67Y} {949, 67}

\bibitem[\protect\citeauthoryear{{Zentner} \& {Bullock}}{{Zentner} \&
  {Bullock}}{2003}]{Zentner++03}
{Zentner} A.~R.,  {Bullock} J.~S.,  2003, \mn@doi [\apj] {10.1086/378797},
  \href {https://ui.adsabs.harvard.edu/abs/2003ApJ...598...49Z} {598, 49}

\makeatother
\end{thebibliography}
		
		\appendix
		\section{The imaging data likelihood}
		\label{app:A}
		\begin{figure*}
			\includegraphics[trim=0.5cm 0cm 0.5cm
			1cm,width=0.485\textwidth]{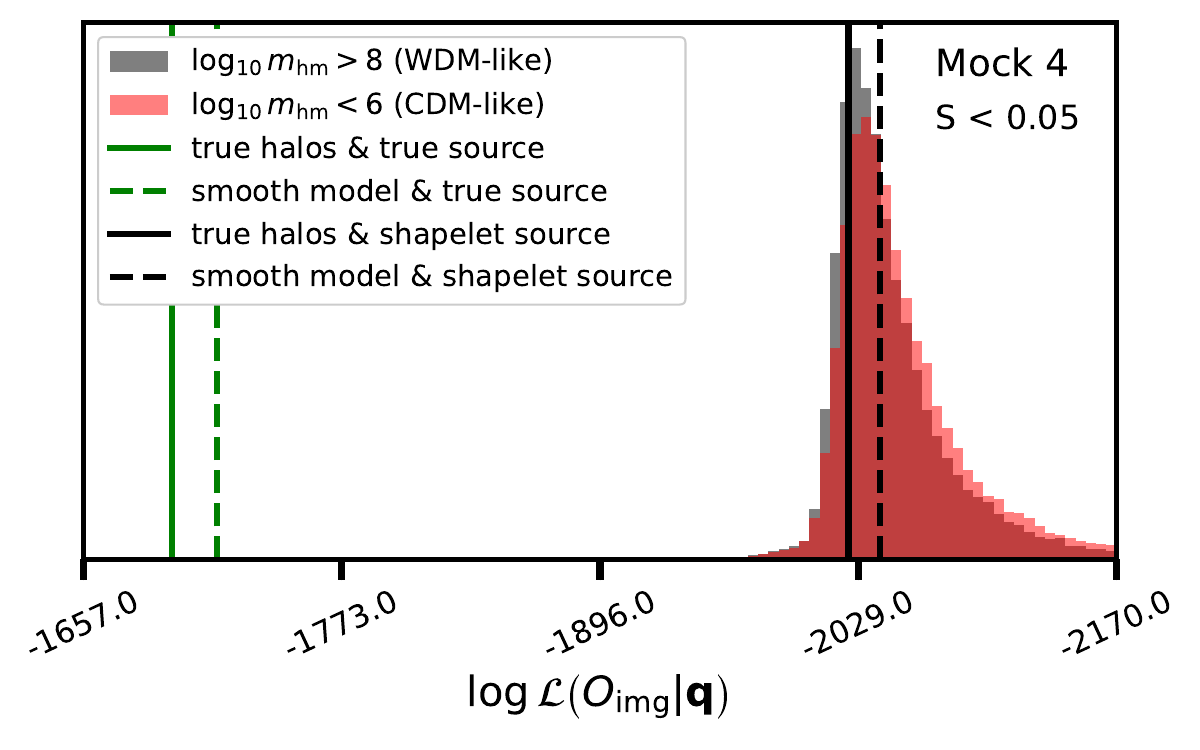}
			\includegraphics[trim=0.5cm 0cm 0.5cm
			1cm,width=0.485\textwidth]{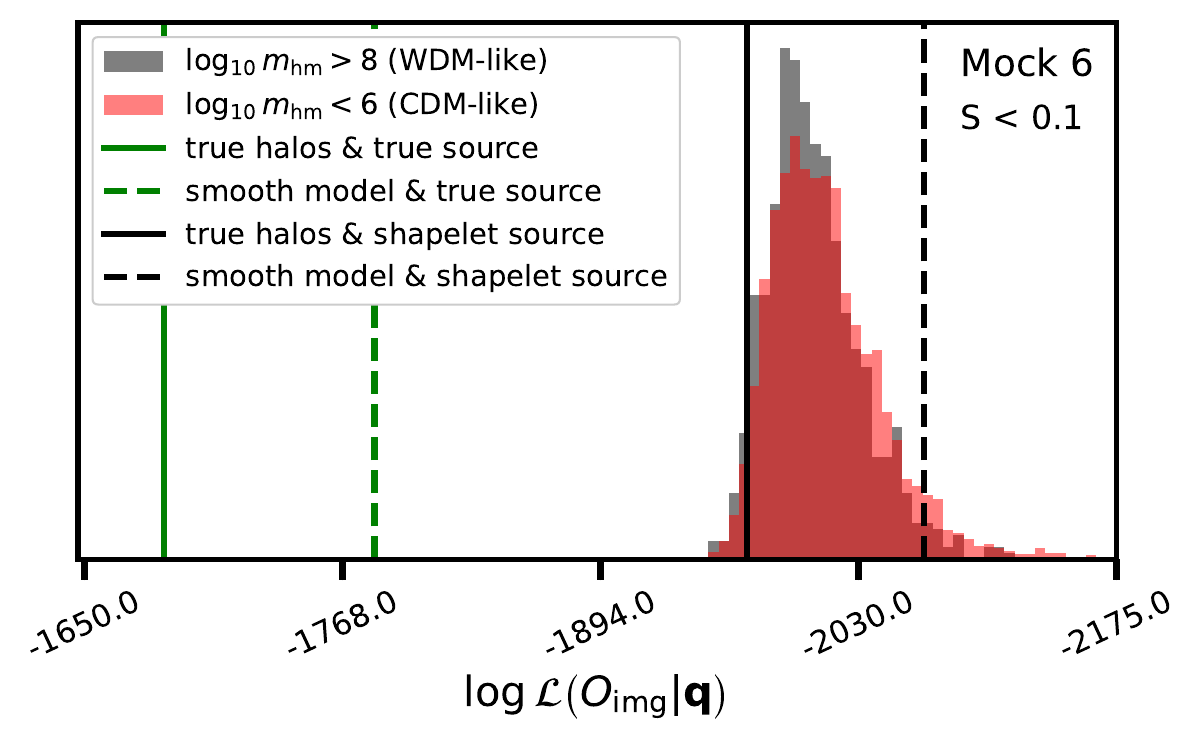}
			\includegraphics[trim=0.5cm 0.cm 0.5cm
			0cm,width=0.485\textwidth]{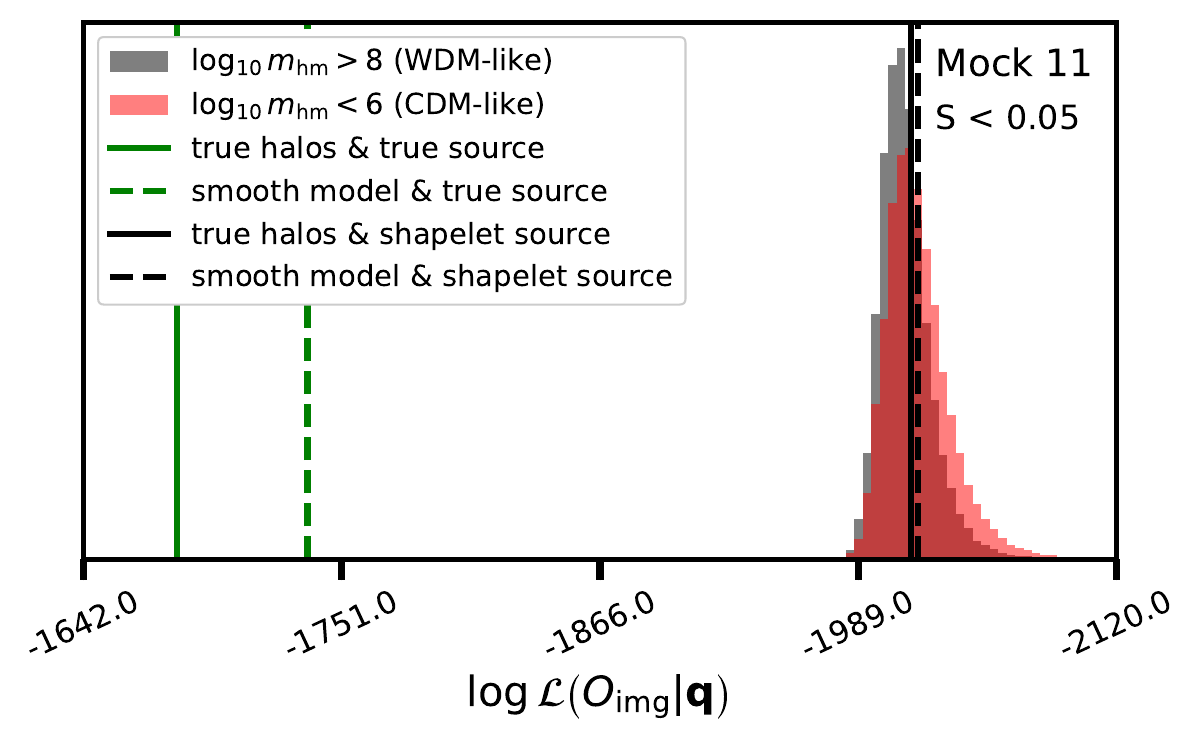}
			\includegraphics[trim=0.5cm 0cm 0.5cm
			0cm,width=0.485\textwidth]{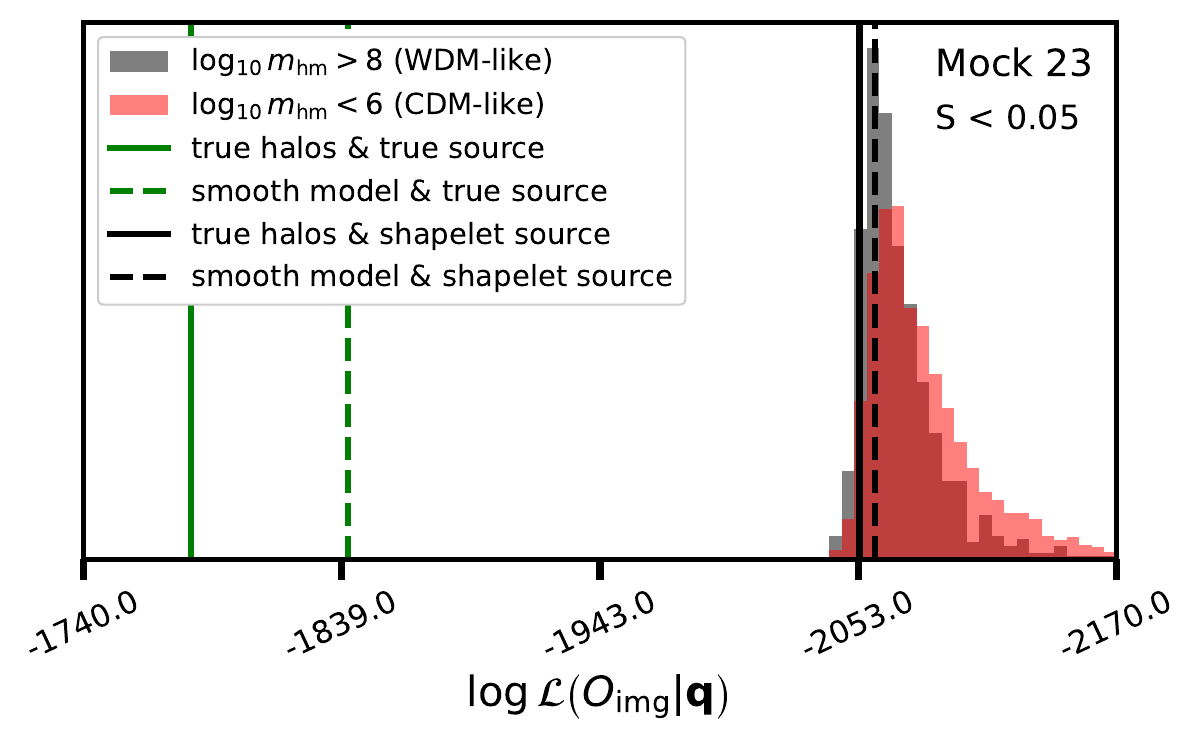}
			\caption{\label{fig:imagedatastats} The log-likelihood of the imaging data inferred for the four mock case study lenses after down-selecting on realizations that fit the observed flux ratios, as indicated by the threshold applied to the $S$ statistic in each panel. The gray and red histograms shows values of the log-likelihood for realizations with $m_{\rm{hm}}$ above $10^8 M_{\odot}$ and below $10^6 M_{\odot}$, respectively. Vertical bars represent the log-likelihoods from fits to the mock data when portions of the lens and source light model are known perfectly. Green vertical bars represent the log-likelihood obtained with perfect knowledge of the source. Black vertical bars represent the log-likelihood computed with imperfect knowledge of the source. Solid lines correspond to the log-likelihood obtained with knowledge of the ``true'' population of substructure, while dashed lines correspond to a smooth lens model fit to the mock lens.}
		\end{figure*}
		As discussed in Section \ref{ssec:inferencesummaries}, our strategy for incorporating the imaging data likelihood is one in which the imaging data alone does not constrain dark matter hyper-parameters. This requirement differs from the likelihood function relevant for gravitational imaging of individual halos \citep[e.g.][]{Vegetti++14,Powell++22,Powell++23}, in which one explicitly uses imaging data to characterize the mass and position of a dark substructure. Obtaining a reliable likelihood of substructure properties from the imaging data requires a careful calibration of the sensitivity function of the lensed arc and various systematics associated with the lens and source light models \citep{Vegetti++14,Cao++22,Despali++22,He++23,ORiordan++24}. The strategies to calibrate the sensitivity function and contend with systematic uncertainties in current gravitational imaging studies with single-halo models do not necessarily carry over to our analysis methods because we perform the lens mass and source light reconstruction with full populations of subhalos and line-of-sight halos. 
		
		Figure \ref{fig:imagedatastats} shows the imaging data likelihoods derived in our analysis. The four panels show the distribution of log-likelihoods derived from fits to the imaging data for the four case study mock lenses. Vertical bars represent varying degrees of knowledge regarding the true population of halos and the true structure of the lensed source, as indicated by the figure legend. The black and red distributions show the log-likelihoods obtained for WDM realizations ($m_{\rm{hm}}>10^8 M_{\odot}$) and CDM-like realizations ($m_{\rm{hm}}<10^6 M_{\odot}$) that match the flux ratios, as indicated by the tolerance threshold for acceptance based on the flux ratio summary statistic $S$ (Equation \ref{eqn:frstat}). 
		
		Several features apparent in Figure \ref{fig:imagedatastats} dissuade us from using the imaging data likelihood to directly constrain the properties of substructure. First, we see that randomly generated populations of halos, i.e. the ``wrong'' substructure models, sometimes result in better fits to the imaging data than the ``correct'' population of halos when we have imperfect knowledge of the source. Second, models with fewer halos (shown in red) systematically improve the imaging data likelihood relative to models with many halos (black). Both of these features could arise from degeneracies between small-scale structure in the lens mass distribution and small-scale features in the source light. These considerations motivate the inclusion of the importance sampling weights in Equation \ref{eqn:importanceweights2}, which one can interpret as an adjustment of the prior volume associated with the source light model such that all dark matter models are equally likely when constrained by imaging data alone.   
		
		We have experimented with re-running the analysis for some mock systems with shapelets having $n_{\rm{max}}=20$. This drives the $\chi^2_{\rm{DOF}}$ to values $\lesssim 1$, which implies a certain degree of over-fitting. However, the systematic bias in which models with fewer halos are preferred by the imaging data persists, and we do not obtain significantly tighter constraints on the main deflector mass profile. 
		
		\section{The effect of multipoles on substructure constraints}
		\label{app:B}
		Shortly after this work appeared, \citet{Cohen++24} (hereafter C24) claimed that the presence of multipole perturbations to galaxy density profiles -- in particular, the $m=3$ and $m=4$ moments that we considered in this work -- preclude inferences of substructure properties from quadruply-imaged quasars. C24 draw this conclusion from a series of lens modeling experiements in which they fit a model that includes only $m=3$ and $m=4$ multipole perturbations to the image positions and flux ratios of mock lenses perturbed by substructure. C24 speculate that a minor difference in the modeling of the position angle of the $m=4$ term, $\phi_4$, explains the discrepancy between their findings and the constraints on substructure properties we obtain from analyzing only the image positions and flux ratios of 25 mock lenses (shown by the black posterior in Figure \ref{fig:inference}); we fix $\phi_4$ to the position angle of the underlying elliptical power law profile, while C24 allow it to vary freely. 
		
\begin{figure}
			\includegraphics[trim=0cm 0cm 0cm
			0cm,width=0.45\textwidth]{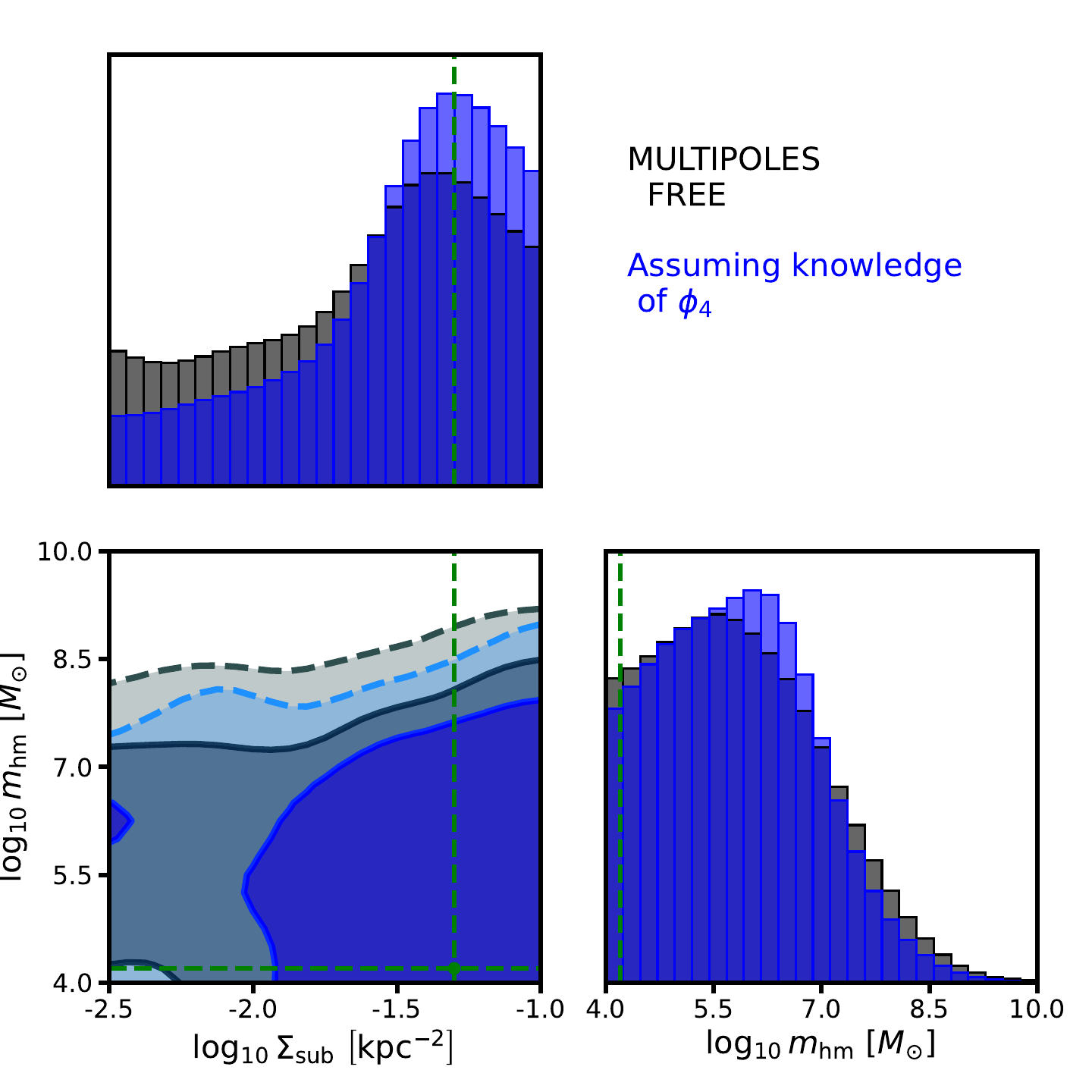}
			\caption{\label{fig:multipoleeffect} The posterior distribution for $\Sigma_{\rm{sub}}$ and $m_{\rm{hm}}$ obtained from modeling only image positions and flux ratios for the 25 lenses created with a CDM ground truth. The black posterior shows the result of analyzing the mock data while allowing the orientation of the $m=4$ multipole perturbation to vary freely, and the blue posterior shows the effect of adding importance weights that enforce alignment with the position angle of the underlying elliptical power law profile. }
		\end{figure}

		Using the methods discussed in this paper, we can investigate the claims by C24 regarding the effect of multipole perturbations in substructure inferences. To begin, we repeat the inference presented in Section \ref{sec:results} assigning the same degree of model flexibility to the multipole terms as advocated by C24. Specifically, we sample $a_3$, $a_4$, and $\phi_3$ from the priors described in Section \ref{ssec:massprofilemaindeflector}, but now allow $\phi_4$ to vary freely between $-\pi/8$ and $\pi/8$. Figure \ref{fig:multipoleeffect} shows the result of analyzing the 25 mock lenses with a CDM ground truth using the more flexible prior on $\phi_4$. We emphasize that the mocks in our sample have non-zero amplitudes of $a_3$ and $a_4$, and thus our simulations do not have priors centered on the ground truth. 
		
		After marginalizing over $a_3$, $a_4$, $\phi_3$, and $\phi_4$, we obtain the black posterior distribution shown in Figure \ref{fig:multipoleeffect}. The blue posterior distribution in Figure \ref{fig:multipoleeffect} includes importance sampling weights that enforce alignment between the $m=4$ term and the underlying elliptical power law profile to isolate the effect of a freely-varying $\phi_4$ angle on the constraints. Assigning the same degree of flexibility to the lens model as advocated by C24, we see no evidence for a systematic bias in the inferred model parameters, or a degeneracy between multipoles and halos that would preclude the constraints shown in Figure \ref{fig:multipoleeffect}. 
		
		To understand these results in the context of the claims by C24, we note that for most problems there exists some other model besides the one under consideration that can fit the dataset. C24 consider a model in which only multipoles can resolve flux ratio anomalies, and compute the required properties of the multipole terms in this scenario. However, to conclude that models with substructure are indistinguishable from lens models that include only multipoles, one must actually calculate the likelihood function and demonstrate that the data cannot distinguish the models statistically when both dark matter halos and multipole terms are present in the lens model. C24 fundamentally cannot make statements regarding relative likelihoods, and therefore the constraining power of the data, because they do not include substructure in the model they use to analyze mock data. 
		
		Models including substructure are strongly preferred from the perspective of Bayesian model selection, both in our sample of mock lenses and the four cases considered by C24. For the 25 mocks in this paper and the 4 generated by C24 we generate 600,000 possible lens models that include both dark matter substructure and multipole perturbations, and 600,000 lens models that include only multipole perturbations. We draw samples of $a_3$, $a_4$, $\phi_3$, and $\phi_4$ from the same prior distributions for each calculation, with the amplitudes and orientations of these terms allowed to vary freely, and we assign the same priors for $\Sigma_{\rm{sub}}$ and $m_{\rm{hm}}$ as discussed in Section \ref{ssec:darkmattermodels}. For the 29 systems in consideration, we perform a Bayesian model comparison by evaluating the posterior odds given the image positions and flux ratios
		\begin{equation}
			\frac{p\left(M_1 | \boldsymbol{O_{\rm{img}}}, \boldsymbol{O_{\rm{f}}}\right) }{p\left(M_2 | \boldsymbol{O_{\rm{img}}}, \boldsymbol{O_{\rm{f}}}\right)} = \frac{p\left(M_1\right)}{p\left(M_2\right)} \frac{\mathcal{L}\left(\boldsymbol{O_{\rm{img}}}, \boldsymbol{O_{\rm{f}}} | M_1\right)}{\mathcal{L}\left(\boldsymbol{O_{\rm{img}}}, \boldsymbol{O_{\rm{f}}}| M_2\right)}. 
		\end{equation}
		Here, $M_1$, represents the model that includes both substructure and multipoles, and $M_2$ represents the model that only includes multipoles. The first term on the right represents the ratio of our prior beliefs regarding the probability of $M_1$ and $M_2$, and the second term is the Bayes factor. For this test, we will ignore the multiple lines of evidence pointing towards the existence of dark matter substructure and assume that both models are equally likely. We can approximate the Bayes factor from the number of accepted samples generated under the Approximate Bayesian Computing approach outlined in Section \ref{sec:lensmodeling}. We use a stringent acceptance threshold $\epsilon < 0.01$ for these calculations to match the flux ratios precisely. 
		
\begin{figure}
			\includegraphics[trim=0cm 0cm 0cm
			0cm,width=0.45\textwidth]{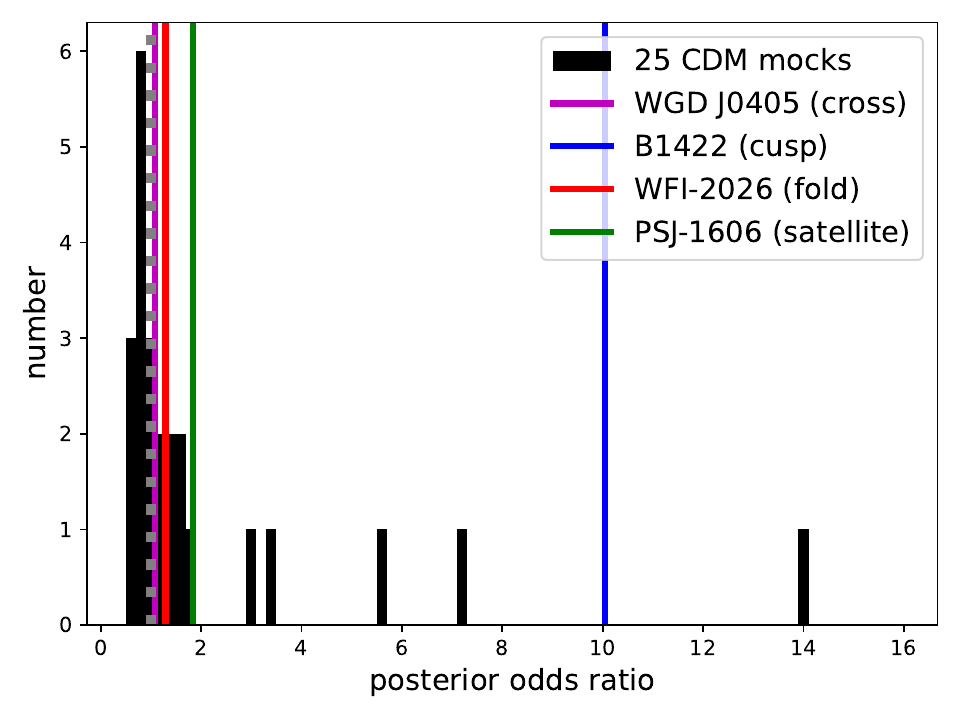}
			\caption{\label{fig:posteriorodds} The posterior odds ratios, or Bayes factors (Equation B1), for the twenty-five mock lenses in our sample (black histogram), and the four mock lenses considered by C24 (colored vertical bars). A number greater than one indicates a Bayesian preference for a model that includes dark matter substructure and  multipole perturbations, while a number less than one indicates a model with only multipoles is preferred. The dashed grey vertical bar marks a Bayes factor of one.}
		\end{figure}

		Figure \ref{fig:posteriorodds} shows the distribution of the posterior odds for each mock lens, with numbers greater than one indicating that a model with substructure and multipoles is preferred relative to a model that only includes multipoles. The 25 mocks we generate with a CDM ground truth are represented by the black histogram, and the odds for the four cases considered C24 are marked with vertical bars. One in four of the mocks considered by C24 exhibits a notable Bayesian preference (odds > 2) for substructure in the lens model, compared with one in five of the mocks we generate. The four mocks considered by C24 have a joint Bayes factor (the product of the individual Bayes factors) of 26, and for the 25 mocks we consider with $a_3$, $a_4$, $\phi_3$, and $\phi_4$ allowed to vary freely, we obtain a joint Bayes factor of 1140. From Figure \ref{fig:posteriorodds}, we conclude that the mock lenses considered by C24, which they use to support the claim that quadruply-imaged quasars cannot constrain substructure properties, actually exhibit a Bayesian preference for substructure in the lens model consistent with what one expects in CDM. 
		
		The reason we can constrain substructure properties, even with multipoles included in the lens model with the same degree of flexibility as advocated by C24, is that models including substructure reproduce the observed data more frequently than lens models that include only multipoles. As a result, the ``multipoles-only'' explanation for the data is heavily disfavored from a Bayesian standpoint. Accounting for lens models that include both substructure and multipole terms requires the careful evaluation of the integral in Equation \ref{eqn:likelihoodsingle}, as outlined in this paper and in previous work \citep{Gilman++19,Gilman++20}. These tests emphasize the importance of a rigorous statistical treatment of the problem in order to validate statements regarding the constraining power of certain datasets over different models, and the importance of validating modeling assumptions through tests on realistic datasets. 

		\bsp	
		\label{lastpage}
	\end{document}